 \documentclass[final,5p,times,twocolumn,authoryear]{elsarticle}

\usepackage[authoryear]{natbib}
\usepackage{graphicx}
\usepackage{subcaption}
\usepackage{amssymb,amsmath}
\usepackage{pdflscape}

\usepackage{lipsum}
\usepackage{epsfig}
\usepackage{epstopdf}
\usepackage[utf8]{inputenc}
\usepackage{newunicodechar}
\usepackage{textgreek}
\usepackage{physics}
\usepackage{inputenc}

\usepackage{afterpage}

\usepackage{amsthm}

\usepackage{multirow}
\usepackage{array}
\usepackage{url} 
\usepackage{xcolor}
\usepackage{adjustbox}
\usepackage{soul}
\usepackage{orcidlink}
\usepackage{multirow}
\usepackage{array}

\usepackage{longtable}%\usepackage{siunitx}
\usepackage{enumitem}
\usepackage{threeparttable}

\usepackage[utf8]{inputenc}
\usepackage{textcomp}
\DeclareUnicodeCharacter{2009}{\,}

\journal{High Energy Astrophysics}

\begin{document}

\begin{frontmatter}


%% \cortext[cor1]{Zahir Shah}

%\title{Transition Blazar S5 1803+784: Multi-Wavelength Variability and Emission Mechanisms}
%\title{Tempo-spectral modeling of blazars and constraining the Lorentz-factor and emission site of $\gamma$-rays}
\title{Broadband Spectral Modeling of Blazars: Constraining the Lorentz Factor and Gamma-Ray Emission Site}

%\author[first]{Javaid Tantry}
%\affiliation[first]{{Department of Physics, University of Kashmir,}, city={Srinagar},
%           postcode={190006}, 
%            state={Kashmir},
%            country={India}}

%\author[second]{Zahir Shah}
%\affiliation[second]{{Department of Physics, Central University of Kashmir,},
%            addressline={Ganderbal}, 
%            city={Varanasi},
%           postcode={191131}, 
%            state={Kashmir},
%            country={India}}  

%\author[first]{Naseer Iqbal}

%%%%%%%%%%%%%%%%%%%%%%%%%%%%%%%%%%%%%%%%   Demo

%\author[first]{Javaid Tantry\,\orcidlink{0009-0002-5150-3604}}
%\ead{javaidtantray9@gmail.com}

\author[first]{Ajay Sharma\,\orcidlink{0000-0002-5221-0822}}
\ead{ajjjkhoj@gmail.com}

\author[second]{Aishwarya Sarath\,\orcidlink{0009-0004-8256-8093}}
\ead{aishwarya.sarath@mail.udp.cl}

\author[first]{Sakshi Chaudhary}
\ead{gurjarsakshi51@gmail.com}

\author[third]{Debanjan Bose\,\orcidlink{0000-0003-1071-5854}}
\ead{debanjan.tifr@gmail.com}

%\affiliation[first]{{Department of Physics,}, {University of Kashmir,}, 
%            city={Srinagar},
%            postcode={190006}, 
%            state={Kashmir},
 %           country={India}}

\affiliation[first]{{S N Bose National Centre for Basic Sciences,},addressline={Block JD, Salt Lake}, city={Kolkata},
           postcode={700106}, 
            state={West Bengal},
            country={India}}

\affiliation[second]{{Instituto de Estudios Astrofísicos, Facultad de Ingeniería y Ciencias, Universidad Diego Portales,},addressline={Av. Ejército Libertador 441}, city={Santiago}, country={Chile}}

\affiliation[third]{{Department of Physics, Central University of Kashmir,}, 
            addressline={Ganderbal}, 
            postcode={191131}, 
            state={Kashmir},
            country={India}} 

\cortext[cor1]{Corresponding author: Ajay Sharma, Debanjan Bose }

%%%%%%%%%%%%%%%%%%%%%%%%%%%%%%%%%%%%%%%%%%%%%

\begin{abstract}

We present a comprehensive temporal and spectral analysis of a few blazars using multi-wavelength observations. Rapid flux variations are quantified via the doubling/halving timescale method, revealing the shortest variability timescales of a few hours in $\gamma$-ray emissions. The broadband fractional variability is systematically computed and examined as a function of frequency, displaying a characteristic double-hump structure akin to the typical spectral energy distribution (SED) of blazars. To distinguish between different emission states, we utilize the Bayesian block algorithm, which effectively identifies distinct flux states for targeted spectral modeling. A one-zone leptonic emission framework is employed to model the broadband emission during these states. The minimum Doppler factors are estimated based on the shortest variability timescales observed in the $\gamma$-ray emissions. Under the external Compton scenario, we constrain the location of the gamma-ray emitting region and the Lorentz factor using three physical conditions: the upper limit on the jet collimation parameter, $\Gamma \theta < 1$; the upper limit on the synchrotron self-Compton contribution, $L_{\mathrm{SSC}} \lesssim L_X$; and the observational constraint on the cooling break energy, $E_{\mathrm{cool, obs}} \lesssim 100$ MeV.

\end{abstract} 


\begin{keyword}
galaxies: active \sep galaxies:  blazar \sep jets \sep radiation mechanisms: non-thermal - gamma-rays \sep galaxies:  Jets; Active

%% keywords here, in the form: keyword \sep keyword, up to a maximum of 6 keywords

\end{keyword}

\end{frontmatter}

%\tableofcontents



\section{Introduction}
\label{introduction}

Active galactic nuclei (AGNs) are among the most powerful and energetic objects in the universe, powered by the accretion of matter onto supermassive black holes (SMBHs) with masses ranging from $10^6$ to $10^{10} \ \rm{M}_{\odot}$ \citep{soƚtan1982masses}. Blazars, a subclass of AGNs, are luminous objects that host relativistic jets that are closely aligned with our line of sight—typically within a few degrees ($<5^{\circ}$) \citep{ghisellini1993relativistic, urry1995unified, blandford2019relativistic}. These objects exhibit non-thermal emission that spans the entire electromagnetic (EM) spectrum, from radio waves to very high energy (VHE; $>$100 GeV) $\gamma$-rays \citep{urry1995unified, ulrich1997variability, padovani2017active}. Based on the strength and presence of broad emission lines in their optical spectra, blazars are further classified into two classes: BL Lacertae objects (BL Lacs), which have weak or absent emission lines, and flat-spectrum radio quasars (FSRQs), which exhibit prominent broad lines with equivalent widths greater than (EW)$> 5 \ \text{\AA}$ \citep{giommi2012simplified}.

Blazars are also known for their extreme variability across the EM spectrum, with timescales ranging from minutes to years \citep{ulrich1997variability, aharonian2007exceptional, raiteri2013awakening, goyal2022multiwavelength}. Their broadband spectral energy distributions (SEDs), constructed using simultaneous multi-wavelength observations, serve as powerful diagnostics of the particle acceleration mechanisms and emission processes acting within jets. Typically, the SEDs of blazars exhibit a characteristic two-hump structure \citep{marscher1980relativistic, konigl1981relativistic, fossati1998unifying, abdo2010spectral}. The low-energy hump, extending from radio to optical/UV and sometimes into X-rays, is attributed to synchrotron radiation from relativistic electrons (or positrons) travelling in magnetic field within jet \citep{urry1982pks}.

The high-energy hump, peaking at MeV to TeV energy range, is generally explained by two competing scenarios. In leptonic models, it arises from inverse Compton (IC) scattering, either via synchrotron self-Compton (SSC) processes—where the synchrotron photons are upscattered by the same electrons that produced them \citep{ghisellini1985inhomogeneous, maraschi1992jet, sikora2009constraining}—or via external Compton (EC) processes, where electrons upscatter external photons originating from the broad-line region (BLR), accretion disk, or dusty torus \citep{dermer1992high, dermer1993model, sikora1994comptonization}. Alternatively, in hadronic models, the high-energy emission results from proton-proton or proton-photon interactions within the jet \citep{mannheim1993proton, aharonian2000tev}.

Several theoretical models have been proposed in the literature to explain the particle acceleration and emission mechanisms operating within blazar jets. Identifying the appropriate model responsible for the observed non-thermal emission critically depends on localizing the emission region along the jet. To uncover the underlying cause of gamma-ray flaring activity, it is essential to determine the gamma-ray emission site relative to the central supermassive black hole (SMBH). This is particularly important at high energies, where external radiation fields play a significant role in the production of gamma-rays. At small distances from the SMBH (e.g., $r \gtrsim 0.01$ pc) \citep{ghisellini1996origin}, the dominant external radiation field is the thermal emission from the accretion disk \citep{dermer2002transformation}. Around $r \sim 0.1$ pc, the broad-line region (BLR) becomes the primary source of external photons \citep{sikora1994comptonization}. At larger distances ($r\gtrsim1$ pc), the dominant external radiation field arises from the infrared emission of the dusty torus (DT) \citep{blazejowski2000comptonization}.\par
PKS 1424-41 is a flat-spectrum radio quasar (FSRQ; \citealt{ajello20173fhl}) located at a redshift of $z = 1.522$, with a supermassive black hole (SMBH) mass estimated to be $\sim 4.5 \times 10^9 \ M_{\odot}$ \citep{fan2004black}. This blazar has exhibited significant flaring activity, particularly during the period 2008–2014 \citep{buson2014unusual}, and its broadband emission has been modeled by several studies \citep{celotti2008power, paliya2017general, abhir2021multi}. Notably, the source showed recurrent sub-flares between 2013 and 2019, indicative of quasi-periodic oscillations (QPOs) \citep{chen2024transient, sharma2025exploring}, and exhibited most intense flaring activity during 2022–2023.\par
PKS 0736+01 was first detected by the Parkes radio telescope \citep{day1966parkes} and later identified as a blazar \citep{lister2005mojave, lister2009mojave}. The source is located at a redshift of $z = 0.189$ \citep{ho2009magellan}. Broad emission lines in its optical–UV spectrum were reported by \citealt{baldwin1975spectrophotometry} and \citealt{malkan1986ultraviolet}, confirming its classification as an active galactic nucleus. Very high energy (VHE) $\gamma$-ray emission from this source was detected by H.E.S.S. during the period February 15–23, 2015 \citep{abdalla2020hess}. The observed monoscopic and stereoscopic spectra were both well described by a power-law model of the form $\mathrm{dN/dE = N_0 (E/E_0)^{-\alpha}}$. The corresponding photon indices were estimated to be $\alpha = 3.1 \pm 0.3_{\mathrm{stat}} \pm 0.2_{\mathrm{syst}}$ (monoscopic) and $\alpha = 4.2 \pm 0.8_{\mathrm{stat}} \pm 0.2_{\mathrm{syst}}$ (stereoscopic). The normalization fluxes at 200 GeV were found to be $N_0 = (1.0 \pm 0.2_{\mathrm{stat}} \pm 0.3_{\mathrm{syst}}) \times 10^{-10} \ \mathrm{cm^{-2} \ s^{-1} \ TeV^{-1}}$ and $(1.1 \pm 0.3_{\mathrm{stat}} \pm 0.2_{\mathrm{syst}}) \times 10^{-10} \ \mathrm{cm^{-2} \ s^{-1} \ TeV^{-1}}$ for the monoscopic and stereoscopic spectra, respectively. A year-timescale QPO in the $\gamma$-ray emission from the source has also been identified \citep{sharma2025exploring}.\par
PKS 0208-512 was identified in the Parkes radio survey \citep{bolton1964parkes}, located at redshfit $z=1.003$ \citep{healey2008cgrabs}. The broadband SEDs have been modeled by \citep{khatoon2022temporal}.\par
PKS 0035-252 is a FSRQ, located at redshift $z=0.49$ and has the black hole mass of $log(M)\sim7.18 M_{\odot}$ \citep{pei2022estimation}.

The paper is summarized as follows: Section \ref{sec:data_reduction_sec2} covers multi-wavelength observations of Fermi-LAT, Swift-XRT, and UVOT, and their data reduction procedures. In Section \ref{sec:doubling}, we perform temporal analysis, including fractional variability amplitude estimation, flare symmetry analysis, and doubling/halving timescale estimation, using multi-wavelength data of blazars in our sample. A detailed discussion on the correlation analysis between different wavebands is provided in Section~\ref{sec:correlation}. Section \ref{sec:sec4_SED_modeling} focuses on broadband spectral energy distributions (SEDs) analysis. In Section \ref{sec:JET_power}, we assess the jet power from SED modeling. In section \ref{sec:min_Doppler_factor}, we estimated the minimum Doppler factor utilizing the fastest variability timescale in $\gamma$-ray emissions. Section \ref{sec:constraint_r_Gamma} presents the constraints on $\gamma$-ray emission site $r$ and the bulk Lorentz factor $\Gamma$, and we summarise the findings and discussion in Section \ref{sec:discussion}.

\begin{table*}
\setlength{\extrarowheight}{7pt}
\setlength{\tabcolsep}{5pt}
\centering
\begin{threeparttable}
\caption{The general information of blazars in our sample.}
\label{tab:source_sample}
\begin{tabular}{c c c c c c c c c}
\hline
\hline
Source  & 4FGL association & R.A. & Decl. & Redshift & Class & log $M_{BH}$  & \multicolumn{2}{c}{Reference}  \\
\cline{8-9}
 & & & & & & & $z$ & $M_{BH}$\\
 & & & & ($z$) & & units of $(M_{\odot})$ \\
(1) & (2) & (3) & (4) & (5) & (6) & (7) & (8) & (9)\\
[+2pt]
\hline
PKS 0736+01 & J0739.2+0137 &  114.820 & 1.622 & 0.189 & FSRQ & 8--8.7 &  (a) & (a,b,f) \\ 
PKS 1424-41 & J1427.9-4206 & 216.98 & -42.10 & 1.522 & FSRQ & 9.65 & (c) & (d) \\
%S2 0109+22 & J0112.1+2245 & 18.02 & 22.75 & 0.36 & BL Lac & 8.6 & (f) & (e)\\
%PKS 0244-470 & J0245.9-4650 & 41.49 & -46.84 & 1.385 & FSRQ & 8.48, 8.32 & $\sim$225 & 8 & (h) & (g) & (h) \\
%PKS 0405-385 & J0407.0-3826 & 61.76 & -38.43 & 1.285 & FSRQ & 8.7 & $\sim$1037 & 5 & (i) & (j) & (j)\\
PKS 0208-512 & J0210.7-5101 & 32.69 & -51.02 & 1.003 & FSRQ & 8.84 &  (e) & (e) \\
PKS 0035-252 & J0038.2-2459 & 9.56 & -24.99 & 0.49 & FSRQ & 7.18 & (g) & (g) \\
[+5pt]
\hline
\end{tabular}

\begin{tablenotes} % Add the text note here
\small
\item Note: Column (1): blazar name; Column (2): blazar name in the Fermi-LAT $4^{th}$ catalog; Column (3 - 4): coordinate of blazar; Column (5): redshift; Column (6): blazar type; Column (7): black hole mass in units of $M_{\odot}$; Column (8-9): references -- (a) \citep{abdalla2020hess}, (b) \citep{xiong2014intrinsic}, (c) \citep{chen2024transient}, (d) \citep{fan2004black}, (e) \citep{ghisellini2010general}, (f) \citep{zhang2024fundamental}, and (g) \citep{pei2022estimation}.

\end{tablenotes}
\end{threeparttable}
\end{table*}

\section{Multi-wavelength Observations } \label{sec:data_reduction_sec2}
\subsection{Fermi-LAT observation} \label{sec:fermilat}

The Fermi Gamma-ray Space Telescope, launched by NASA on June 11, 2008, onboard two instruments: the Large Area Telescope (LAT) and the Gamma-ray Burst Monitor (GBM). Together, they enable comprehensive gamma-ray observations across a wide energy range, from a few keV to 500 GeV. The Fermi-LAT, a pair-conversion gamma-ray detector, is designed to explore high-energy gamma rays from $\sim$20 MeV to 500 GeV. It provides a wide field of view ($>$2 sr ), covering about 20$\%$ of the entire sky. Since its launch, Fermi-LAT has conducted all-sky surveys every three hours, providing near-continuous observations of $\gamma$-ray emissions from astrophysical sources \citep{atwood2009large}.\par
We collected the Fermi-LAT observations of blazars in our sample, Table~\ref{tab:source_sample}, spanning from August 5, 2008 (MJD 54683) to April 1, 2025 (MJD 60766). Further details for each source are provided in Table~\ref{tab:source_sample}. During the data download procedure, we chose the energy range of 0.1-300 GeV with Pass8 class events (evclass==128, evtype==3) recommended by the Fermi-LAT collaboration from a region of interest (ROI) with a radius of $10^{\circ}$ centered at the source coordinate. The analysis of $\gamma$-ray data was carried out using the standard point-source analysis procedures provided by the \textit{Fermi} Science Tools package (v11r05p3), made available by the Fermi Science Support Center. To minimize contamination from the Earth's limb, we applied a zenith angle cut of $> 90^\circ$. The good time intervals (GTIs) were selected using the standard filter expression: \texttt{(DATA\_QUAL > 0) \&\& (LAT\_CONFIG == 1)}. This ensures that only high-quality observational data were considered. We used the \texttt{GTLTCUBE} and \texttt{GTEXPOSURE} tools to compute the integrated livetime as a function of sky position and off-axis angle, and the corresponding exposure map, respectively. Galactic and extragalactic diffuse background emissions were modeled using the files \texttt{gll\_iem\_v07.fits}\footnote{\label{fermi}\url{https://fermi.gsfc.nasa.gov/ssc/data/access/lat/BackgroundModels.html}} and \texttt{iso\_P8R3\_SOURCE\_V3\_v1.txt}\footref{fermi}, respectively.

To generate the source model, we used the \texttt{make4FGLxml.py} script, which creates an XML file containing the spectral models and spatial locations of sources in the region of interest. The unbinned likelihood analysis was performed using the \texttt{GTLIKE} tool \citep{cash1979parameter, mattox1996likelihood}, employing the instrument response function (IRF) \texttt{P8R3\_SOURCE\_V3} and the generated XML model.

The significance of the target source was evaluated using the \texttt{GTTSMAP} tool, which calculates the Test Statistic (TS), defined as:

\[
\mathrm{TS} = 2 \Delta \log(\text{likelihood}) = -2 \log\left( \frac{L_0}{L} \right),
\]

where $L$ and $L_0$ are the maximum likelihood values of the model with and without the point source at the target location, respectively. The TS value provides a measure of source significance and is approximately related to the detection significance via $\mathrm{TS} \sim \sigma^2$ \citep{mattox1996likelihood}.
\par
We adopted a criterion with TS$\ge$9 for data points in the light curve, and weekly binned light curves of sources (Table~\ref{tab:source_sample}) are generated using Fermipy\footnote{\url{https://fermipy.readthedocs.io/en/latest/}}. The resulting weekly binned $\gamma$-ray light curve is shown in Figure \ref{Fig-LC_PKS1424}, \ref{Fig-LC_PKS0736}, \ref{Fig-LC_PKS0208}, and \ref{Fig-LC_PKS0035}.\par

\subsection{Swift-XRT}
\textit{Swift} is a space-based observatory equipped with the Burst Alert Telescope (BAT), X-ray Telescope (XRT), and Ultraviolet/Optical Telescope (UVOT) \citep{burrows2005swift}, enabling broad electromagnetic coverage from optical–ultraviolet to X-rays. The Swift-XRT is sensitive to soft X-rays in the energy range of 0.3--10~keV. We processed the XRT data in photon counting (PC) mode using the \texttt{xrtpipeline}\footnote{\url{https://www.swift.ac.uk/analysis/xrt/xrtpipeline.php}} (version 0.13.5) with the calibration database \texttt{CALDB} (version 20190910) to generate cleaned event files. The tool \texttt{xselect} (v2.5b)\footnote{\url{https://heasarc.gsfc.nasa.gov/docs/software/lheasoft/ftools/xselect/index.html}} was used to extract both the source and background spectra.

For the source region, a circular aperture with a radius of 40 arcsec centered on the source was used, while an offset circular region of radius 80 arcsec was selected for the background. The auxiliary response files (ARFs) were created using the \texttt{xrtmkarf}\footnote{\url{https://www.swift.ac.uk/analysis/xrt/arfs.php}} tool. To prepare the spectra for $\chi^2$ statistics, we grouped the data using the \texttt{grppha}\footnote{\url{https://heasarc.gsfc.nasa.gov/ftools/caldb/help/grppha.txt}} tool, ensuring a minimum of 20 counts per bin.

Spectral analysis was carried out in the 0.3--10.0~keV range using the \texttt{XSPEC}\footnote{\url{https://heasarc.gsfc.nasa.gov/xanadu/xspec/}} software (v12.13.0c) \citep{1996ASPC..101...17A}. During fitting with a power-law model, the neutral hydrogen column densities were fixed at: $N_H = 7.7 \times 10^{20}\ \mathrm{cm^{-2}}$ for PKS 1424$-$41 \citep{refId0}, $N_H = 7.8 \times 10^{20}\ \mathrm{cm^{-2}}$ for PKS 0736+01 \citep{dickey1990hi}, $N_H = 1.61 \times 10^{20}\ \mathrm{cm^{-2}}$ for PKS 0208$-$512\footnote{\label{nH_col}\url{https://heasarc.gsfc.nasa.gov/cgi-bin/Tools/w3nh/w3nh.pl}}, and $N_H = 1.24 \times 10^{20}\ \mathrm{cm^{-2}}$ for PKS 0035$-$252\footref{nH_col}, while all other model parameters were allowed to vary freely.
\\

\subsection{Swift-UVOT}
%\subsection{\emph{Swift}-UVOT}
Swift-UVOT provides optical–ultraviolet photometric observations through six filters: U (3465~\AA), V (5468~\AA), B (4392~\AA), UVW1 (2600~\AA), UVM2 (2246~\AA), and UVW2 (1928~\AA) \citep{roming2005}. The Swift-UVOT telescope \citep{burrows2005swift} has observed all the sources across different flux states used in both temporal and spectral analyses, see Sect.~\ref{sec:doubling}, ~\ref{sec:correlation}, and ~\ref{sec:sec4_SED_modeling}. The data for each source were retrieved from the HEASARC archive and processed using the \texttt{HEASoft} software package\footnote{\url{https://heasarc.gsfc.nasa.gov/docs/software/heasoft/}} (version 6.26.1).

For each filter, individual images from multiple observations were processed and co-added using the \texttt{UVOTIMSUM} tool\footnote{\url{https://www.swift.ac.uk/analysis/uvot/image.php}}, and source photometry was performed using \texttt{UVOTSOURCE}\footnote{\url{https://heasarc.gsfc.nasa.gov/lheasoft/help/uvotsource.html}}, which provides observed magnitudes.

To correct for Galactic extinction, we applied reddening values $E(B-V)$ obtained from \cite{schlafly2011measuring}\footnote{\url{https://irsa.ipac.caltech.edu/applications/DUST/}}: $E(B-V)=0.1055$ for PKS 1424$-$41, $E(B-V)=0.1164$ for PKS 0736+01, $E(B-V)=0.0174$ for PKS 0208$-$512, and $E(B-V)=0.0129$ for PKS 0035$-$252, with $A_V / E(B-V) = 3.1$. The extinction-corrected magnitudes were then converted to flux densities using the zero-point magnitudes provided in \cite{poole2008photometric}.

\begin{figure*}
    \centering
    \includegraphics[width=0.85\textwidth]{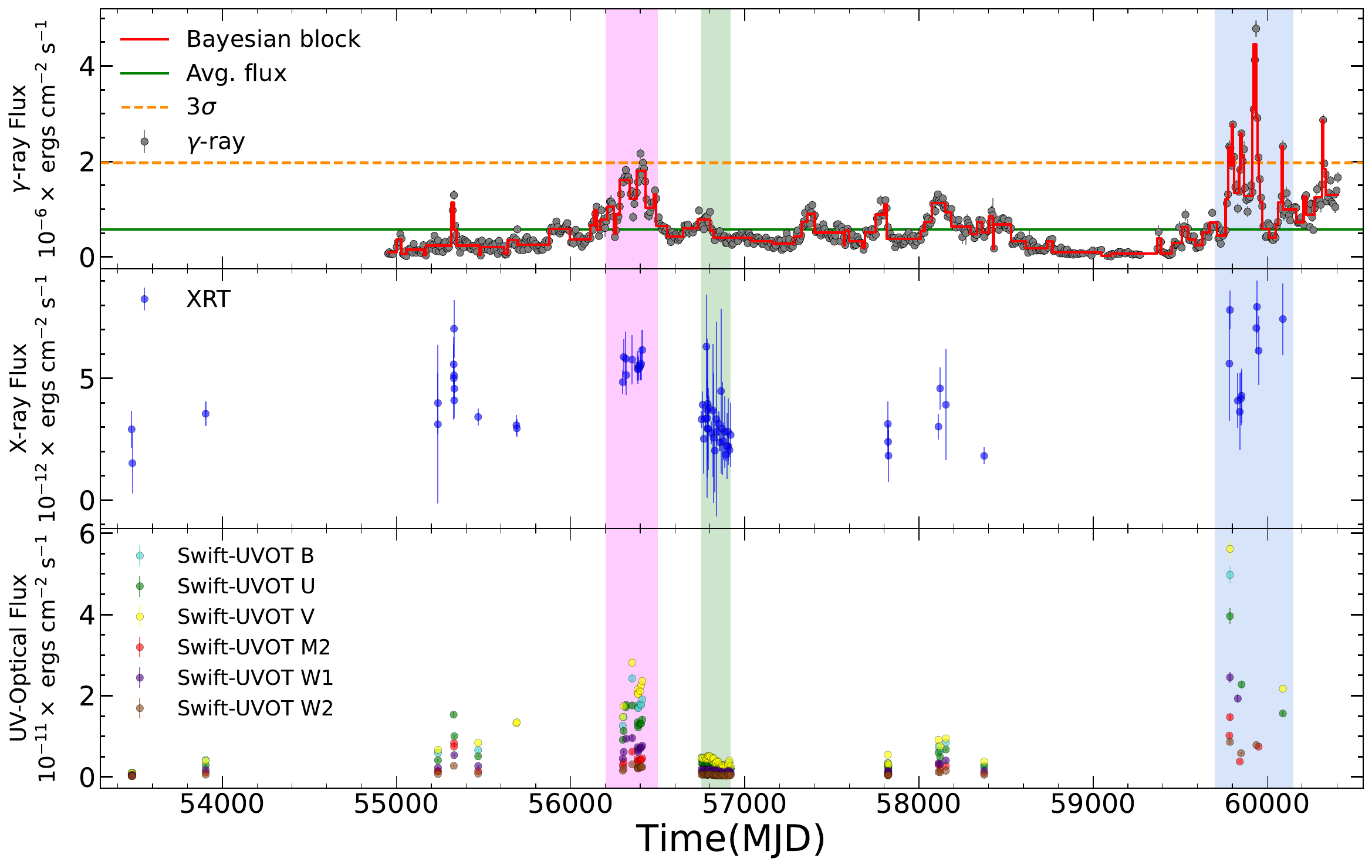}
    \vspace{1pt}
    \caption{Figure presents the multi-wavelength light curves of PKS 1424$-$41 spanning from MJD 54952 to 60405. The top panel shows the \textit{Fermi}-LAT $\gamma$-ray flux (in grey) expressed in units of $10^{-6} \ \mathrm{erg \ cm^{-2} \ s^{-1}}$, overlaid with Bayesian blocks (in red). The solid green and dashed orange lines represent the average $\gamma$-ray flux and the $3\sigma$ flux threshold, respectively. The middle panel displays the \textit{Swift}-XRT X-ray flux in the $0.3$--$10$~keV energy range, in units of $10^{-12} \ \mathrm{erg \ cm^{-2} \ s^{-1}}$. The bottom panel shows the optical and ultraviolet fluxes obtained from \textit{Swift}-UVOT observations, expressed in units of $10^{-11} \ \mathrm{erg \ cm^{-2} \ s^{-1}}$. The vertical shaded regions mark distinct flux states of the blazar: Flare 1 (magenta), Flare 2 (light blue), and the quiescent state (green).}
    \label{Fig-LC_PKS1424}    
\end{figure*}

\begin{figure*}
    \centering
    \includegraphics[width=0.85\textwidth]{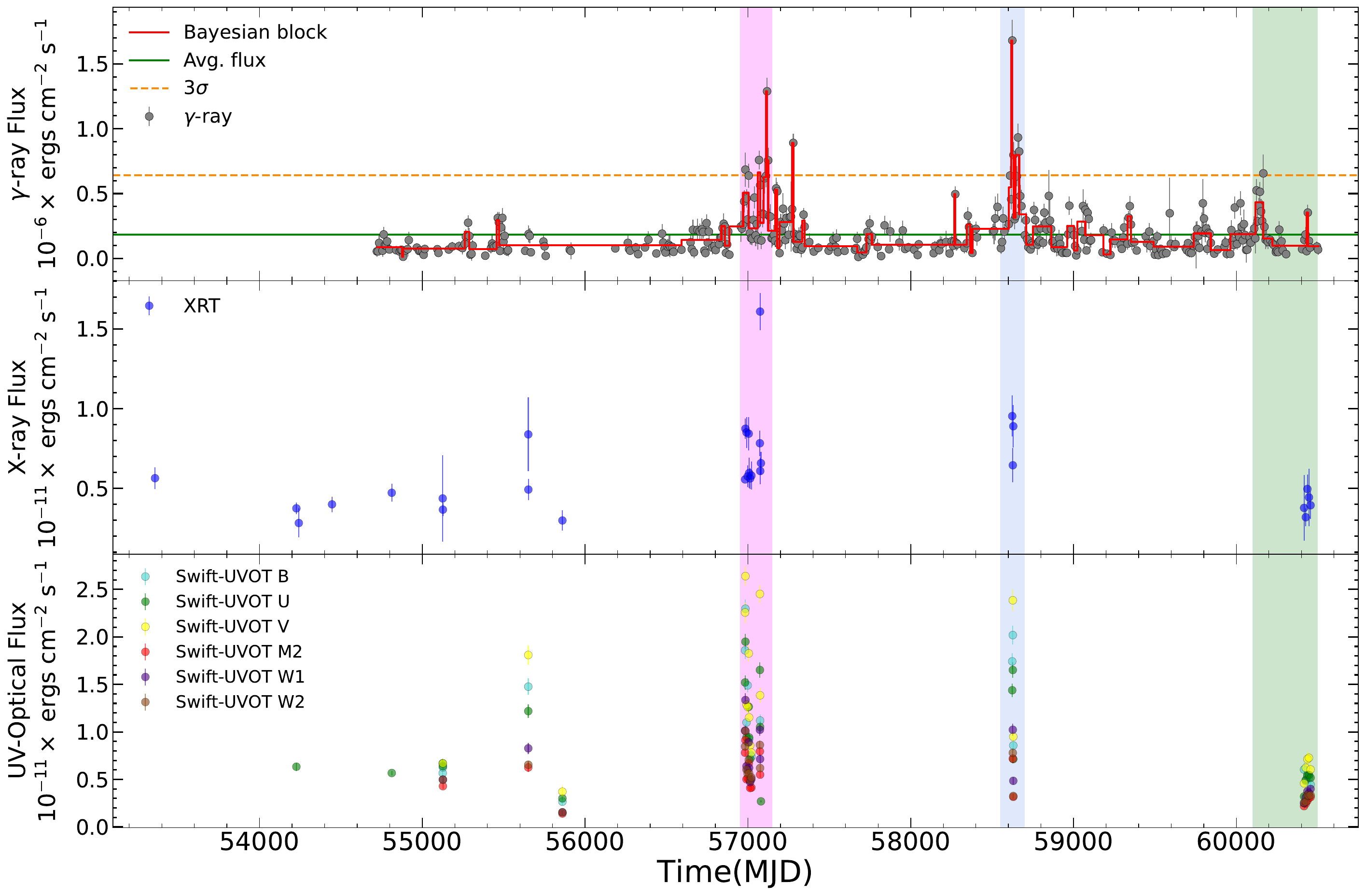}
    \caption{The multi-wavelength light curves of the blazar PKS 0736+01 are shown, with notations and conventions similar to those used in Figure~\ref{Fig-LC_PKS1424}. }
    \label{Fig-LC_PKS0736}    
\end{figure*}

\begin{figure*}
    \centering
    \includegraphics[width=0.88\textwidth]{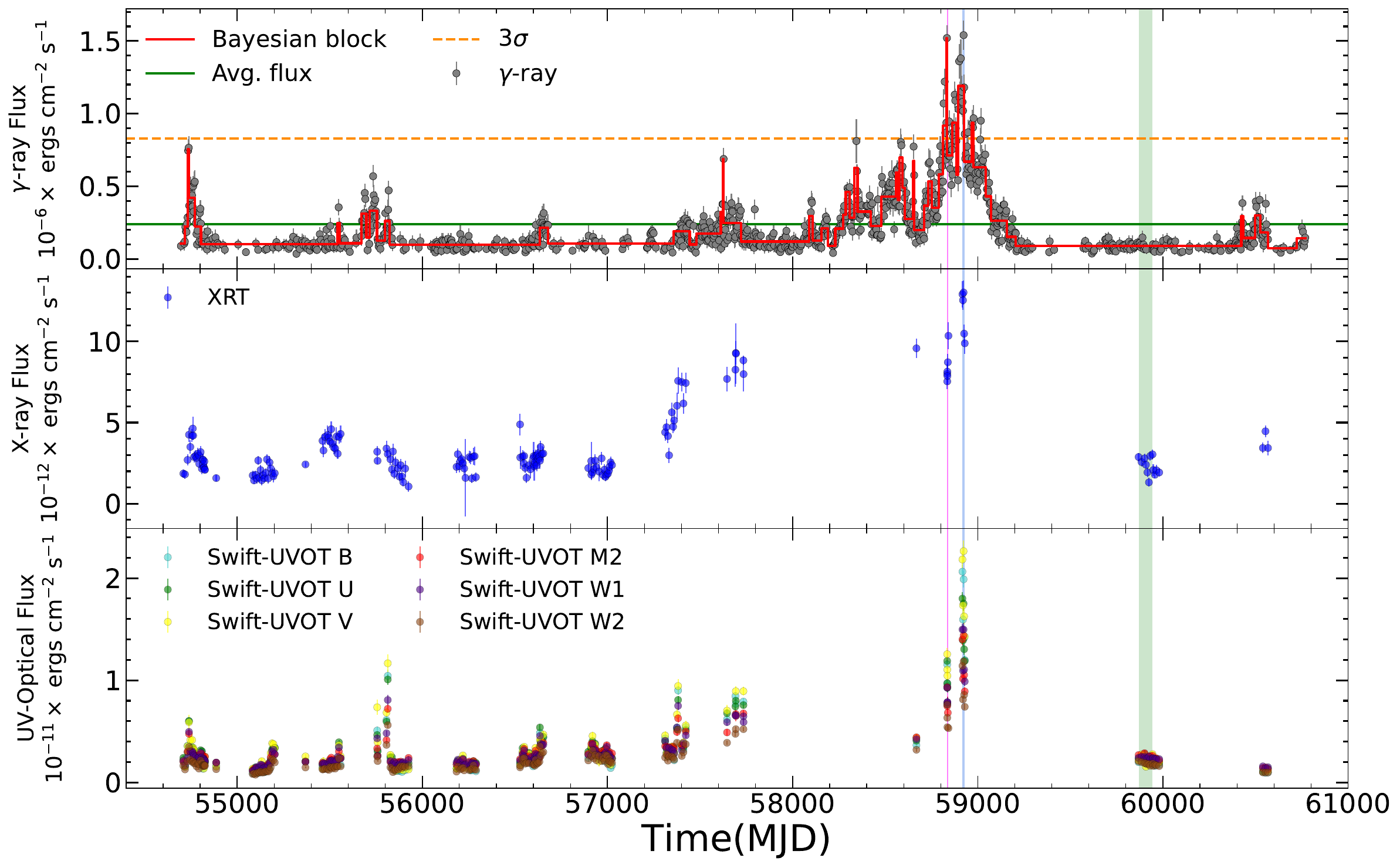}
    \caption{The multi-wavelength light curves of the blazar PKS 0208-512 are shown, with notations and conventions similar to those used in Figure~\ref{Fig-LC_PKS1424}. }
    \label{Fig-LC_PKS0208}    
\end{figure*}

\begin{figure*}
    \centering
    \includegraphics[width=0.84\textwidth]{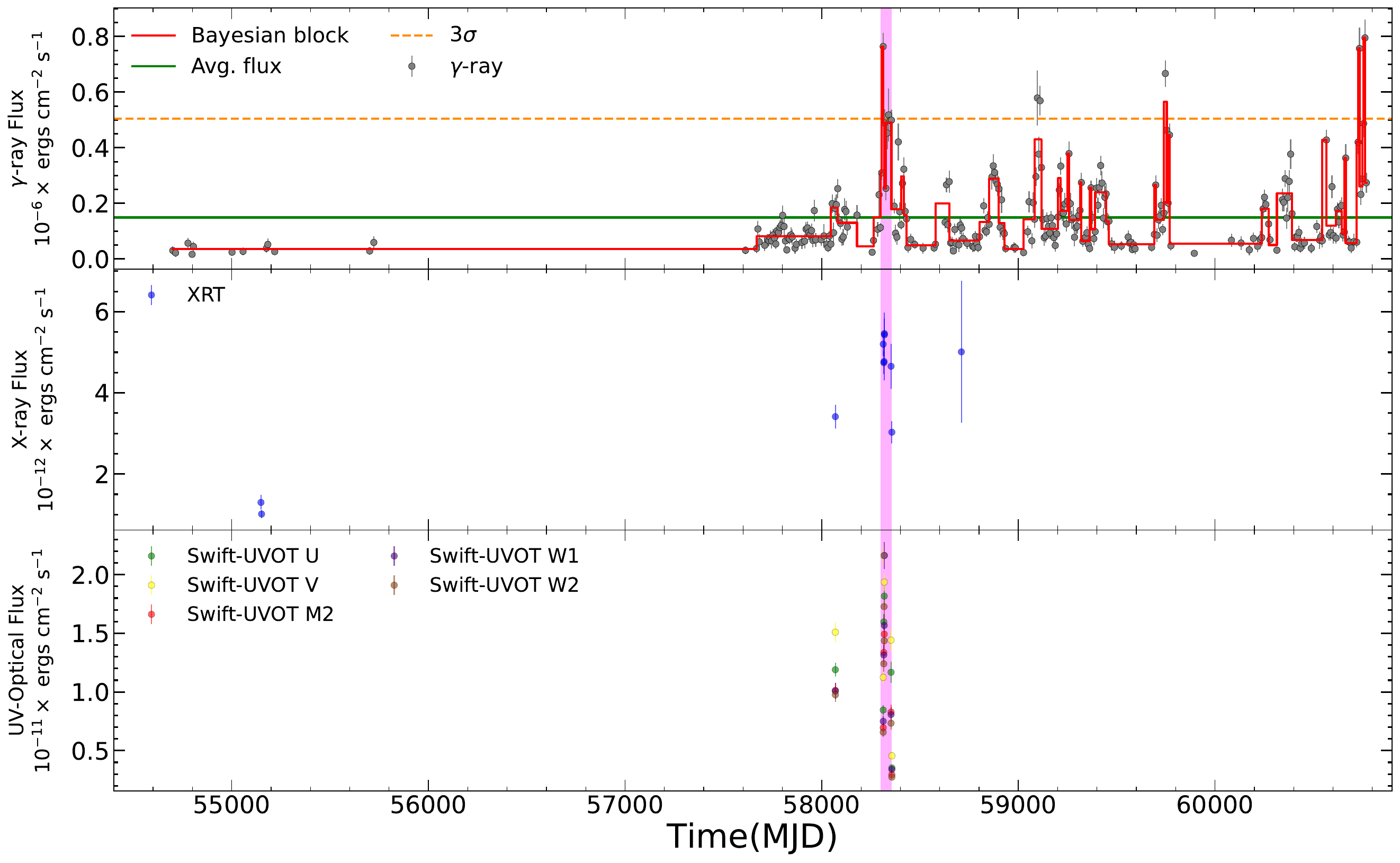}
    \caption{The multi-wavelength light curves of the blazar PKS 0035-252 are displayed, with notations and conventions similar to those used in Figure~\ref{Fig-LC_PKS1424}. }
    \label{Fig-LC_PKS0035}    
\end{figure*}

%\section{Temporal modeling}\label{sec:sec3_temporal_modeling}
%Figure \ref{Fig-LC_FLUX_distribution} presents the 7-day binned $\gamma$-ray light curves for all blazars in our sample. These light curves display clear flux modulations, including both low- and high-flux states, as well as indications of long-term and transient quasi-periodic oscillations. In the left column, the logarithm of the normalized flux is plotted over time, while the right column shows the corresponding flux distributions, also in logarithmic scale, fitted with Gaussian functions. A detailed discussion on the QPO analysis, including the methodologies used and key findings, is provided in Section \ref{sec:QPO_search}. Additionally, we define the high-flux state in the $\gamma$-ray light curve of the blazar PKS 1424-41 as periods when the flux exceeds the average $\gamma$-ray flux level. PKS 1424-41 exhibits two prominent outbursts during the intervals MJD 56299–56411 and MJD 59782–60089. The first is classified as a semi-outburst, while the second is a major flare, with flux surpassing $4 \times 10^{-6} \ \mathrm{ph \ cm^{-2} \ s^{-1}}$. Similar flaring activity is also observed in the Swift X-ray and UVOT bands. A quiescent state is identified during the interval MJD 56751–56919.\par
%For the blazar PKS 0736+01, two major outbursts are identified during the periods MJD 56982–57080 and MJD 58623–58630, with flux levels exceeding $1 \times 10^{-6} \ \mathrm{ph \ cm^{-2} \ s^{-1}}$, and a relatively low-flux state is observed during the interval MJD 60000–60300.

\begin{table*}
\setlength{\extrarowheight}{7pt}
\setlength{\tabcolsep}{10pt}
\centering
\caption{Fractional variability amplitude $F_{var}$ obtained in different energy bands.}

\begin{tabular}{l c c c c}

%\multicolumn{3}{c}{\textbf{Data set (2008--2024)}}\\
\hline 
& PKS 1424-41 & PKS 0736 +01 & PKS 0208-512 & PKS 0035-252 \\
\cline{2-5} 
Energy band  & $\rm F_{var}$ & $\rm F_{var}$ & $\rm F_{var}$ & $\rm F_{var}$\\
[+2pt]
\hline
Fermi $\gamma$-ray (\mbox{0.1-300\,GeV}) & 1.15$\pm$0.007 & 0.95$\pm$0.02  & 0.89$\pm$0.007 & 0.87$\pm$0.012\\  
Swift X-ray (0.3-10 keV) & 0.16$\pm$0.06 & 0.31$\pm$0.05  & 0.64$\pm$0.01 & 0.36$\pm$0.04\\
Swift UVOT-W2 & 1.36$\pm$0.014 & 0.42$\pm$0.014  & 0.74$\pm$0.005 & 0.49$\pm$0.022\\
Swift UVOT-M2 & 1.20$\pm$0.014 & 0.43$\pm$0.016 & 0.67$\pm$0.006 & 0.45$\pm$0.027\\
Swift UVOT-W1 & 1.26$\pm$0.012 & 0.48$\pm$0.015 & 0.72$\pm$0.005 & 0.52$\pm$0.025\\
Swift UVOT-U & 1.07 $\pm$ 0.009 & 0.52$\pm$0.012 & 0.83$\pm$0.005 & 0.44$\pm$0.022\\
Swift UVOT-B & 1.17 $\pm$ 0.01 & 0.57$\pm$0.014 & 1.01$\pm$0.006 & --\\
Swift UVOT-V & 1.10 $\pm$ 0.009 & 0.56$\pm$0.013 & 1.01$\pm$0.008 & 0.41$\pm$0.02\\
[+4pt]
\hline
\end{tabular}
\label{tab:Fvar}
\end{table*}

\section{Temporal variability study}\label{sec:doubling}
In this study, we used multi-wavelength observations to conduct a temporal variability analysis, which included estimating the fractional variability amplitude, modeling flares using the sum of exponential functions (SOEs), and determining the shortest variability timescales through the flux doubling/halving method. Based on these results, we further estimated the size and location of the emission region. A detailed discussion is presented below.
\subsection{Fractional variability}
We utilized the multi-wavelength emissions from the sources in our sample to estimate the variability amplitude. To measure the level of intrinsic variability in the source, the fractional variability amplitude $\left( F_{var} \right)$ is defined as the square root of excess variance $\left( \sigma_{XS} \right)$, which is derived by subtracting the contribution of measurement uncertainties on flux from the variance of the light curve \citep{sharma2024probing}. The mathematical expression for $ F_{var}$ is given as:

\begin{equation}
    F_{var} = \sqrt{\frac{S^2 - \overline{\sigma_{err}^2}}{\overline{x}^2}}
\end{equation}

Here, $S^2$ represents the variance of the light curve, defined as $\frac{1}{N - 1} \sum_{i=1}^N \left( \langle x \rangle - x_i\right)^2$, $ \overline{x}^2$ denotes the average flux, and $\overline{\sigma_{err}^2}$ represents the mean square error, which is defined as $\frac{1}{N} \sum_{i=1}^{N} \sigma_{err, \ i}^2$. The uncertainty on the excess variance is defined as

\begin{equation}
    err(\sigma_{NXS}^2) = \sqrt{\left(\sqrt{\frac{2}{N}} \cdot \frac{\overline{\sigma_{err}^2}}{\overline{x}^2}\right)^2 + \left(\sqrt{\frac{\overline{\sigma_{err}^2}}{N}} \cdot \frac{2F_{var}}{\overline{x}} \right)^2 } 
\end{equation}
Here, the total number of data points in the light curve is represented by N, and the error on $F_{var}$ is given as

\begin{equation}
    \Delta F_{var} = \sqrt{F_{var}^2 + err\left( \sigma_{NXS}^2 \right)} - F_{var}
\end{equation}

\begin{figure*}
    \centering
    \includegraphics[width=0.43\textwidth]{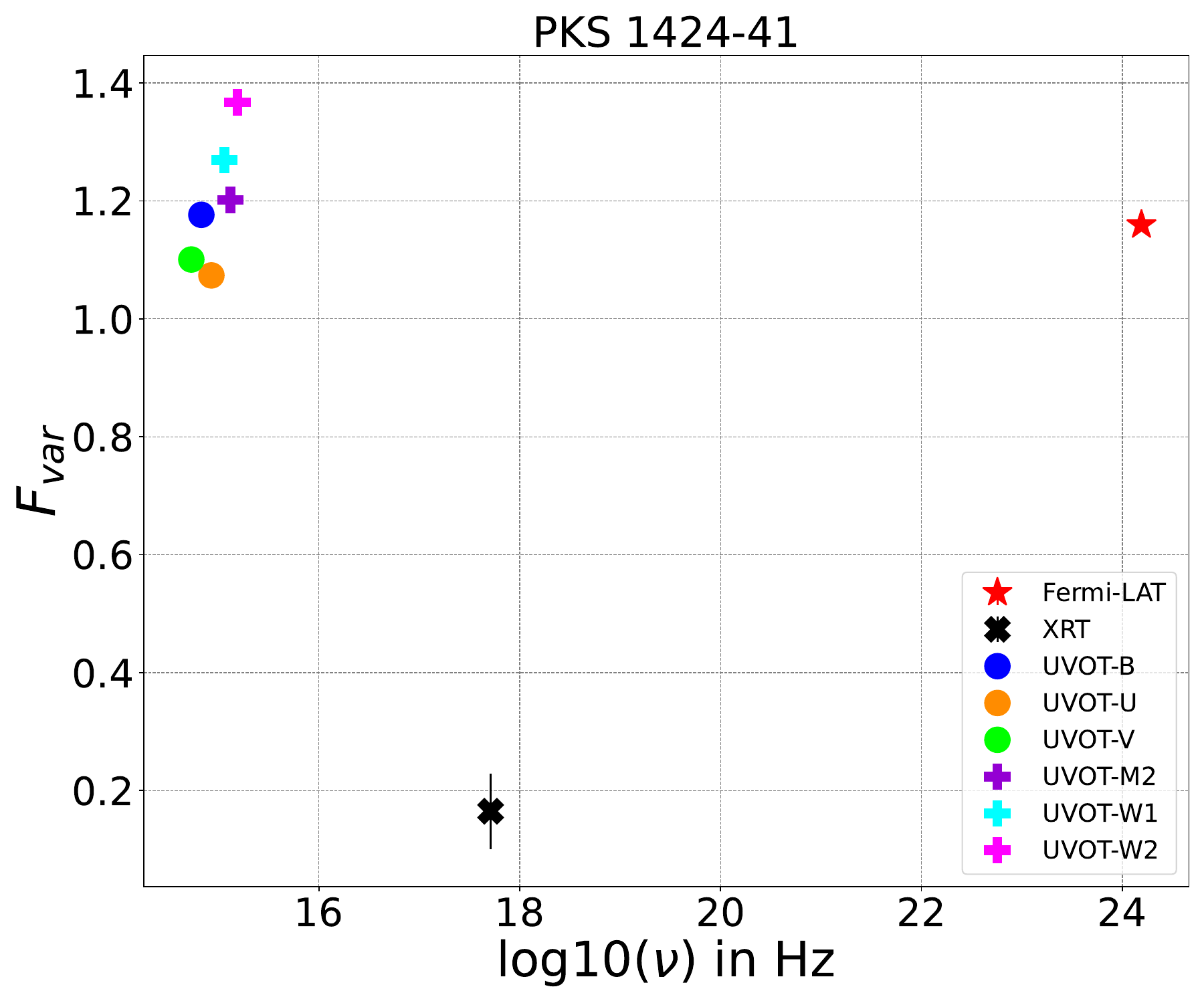}
    \hspace{1pt}
    \includegraphics[width=0.43\textwidth]{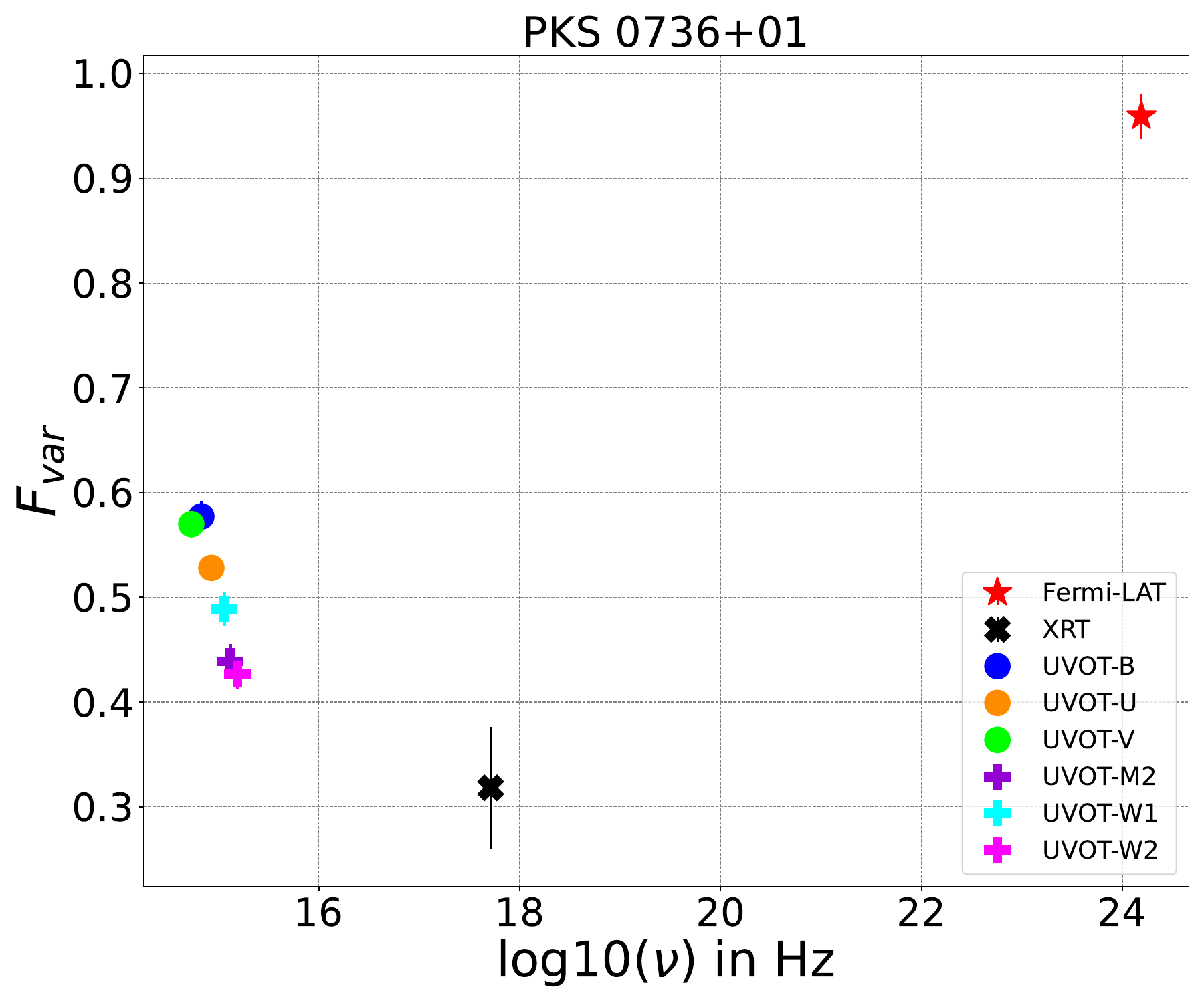}
    \vspace{1pt}
    \includegraphics[width=0.43\textwidth]{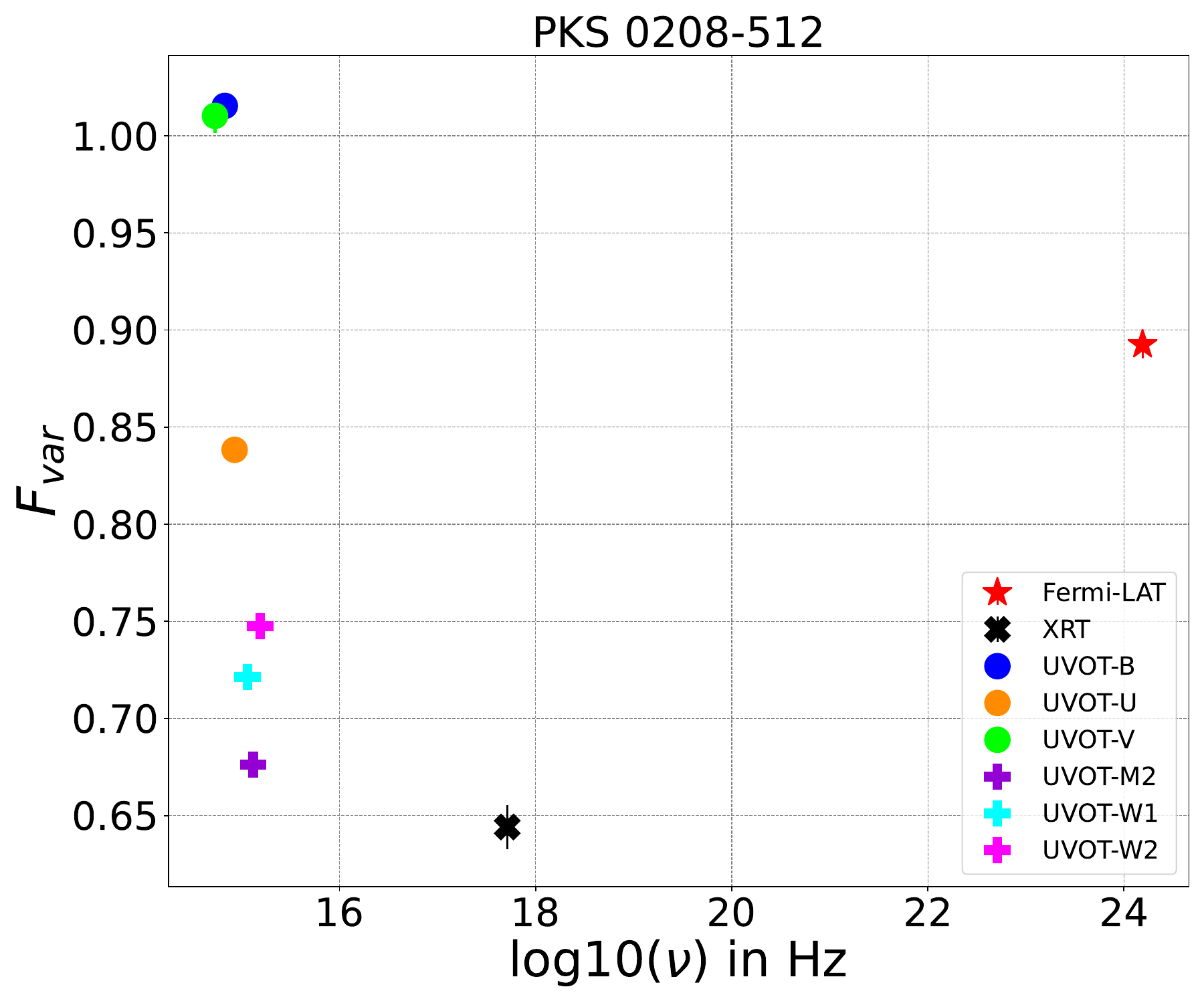}
    \hspace{1pt}
    \includegraphics[width=0.43\textwidth]{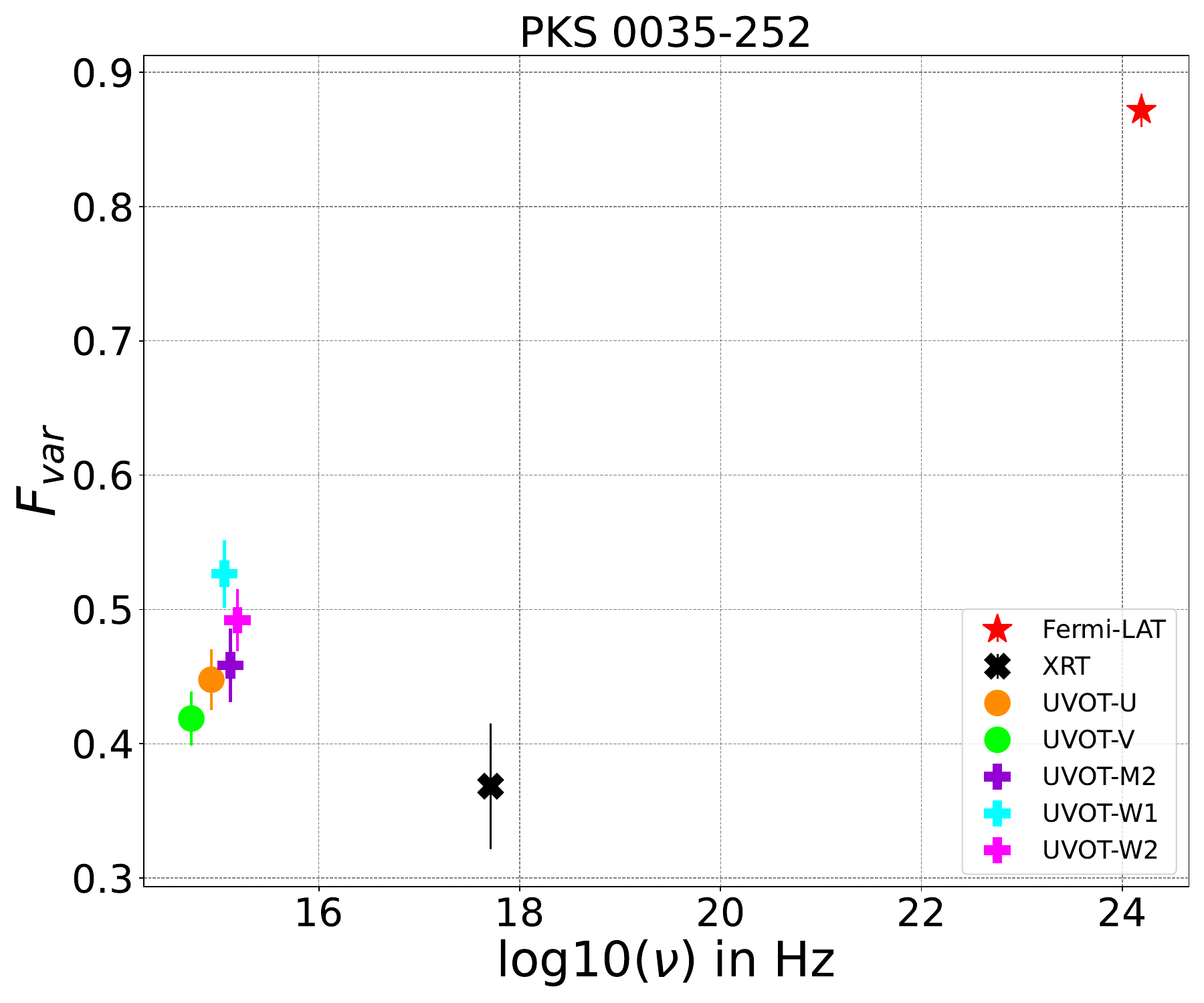}
    \caption{Fractional variability amplitude estimated using the multi-wavelength data of blazars. The top left panel presents fractional variability amplitudes of PKS 1424-41, while the top right, bottom left, and bottom right panels correspond to PKS 0736+01, PKS 0208-512, and PKS 0035-252, respectively. }
    \label{Fig-Fvar}    
\end{figure*}

The fractional variability amplitudes were calculated across all observed wavebands and are presented in Table~\ref{tab:Fvar} and Figure~\ref{Fig-Fvar}. For the blazar PKS 1424-41, a higher variability amplitude is observed in the $\gamma$-ray, UV, and optical bands compared to the X-ray band. In the case of PKS 0736+01, although the X-ray variability is relatively low, the UV and optical bands show slightly higher variability amplitudes, with the $\gamma$-ray band exhibiting the strongest variability overall for this source. PKS 0208–512 exhibits stronger variability in the optical bands, while its X-ray variability amplitude is found to be the lowest. The fractional variability analysis of PKS 0035–252 reveals that the $\gamma$-ray variability amplitude is the highest compared to the X-ray and optical–UV bands. The variability pattern resembles the shape of the broadband SED as seen in many case studies \citep{sharma2024probing, tantry2025study}.

\begin{table*}
\setlength{\extrarowheight}{7pt}
\setlength{\tabcolsep}{0.9pt}
\centering
\caption{The fastest variability time-scale calculated in $\gamma$-ray emissions. }

\begin{tabular}{| c | c | c | c | c | c | c | c | c | c | c | c |}

\hline
Source  & $\mathrm{T_{start} (t_1)}$ & $\mathrm{Flux (t_1)}$  & $\mathrm{T_{end} (t_2)}$ & $\mathrm{Flux (t_2)}$ & $\mathrm{t_d}$ & $\mathrm{{t_{d,z}}}$ & Significance & R/D & Size & Location\\
& MJD & ($10^{-7} \mathrm{ph \ cm^{-2} \ s^{-1}}$) & MJD & ($10^{-7} \mathrm{ph \ cm^{-2} \ s^{-1}}$) & Hr & Hr & & & $\times 10^{15} cm$ & $\times 10^{16} cm$\\ 
[+5pt]

(1) & (2) & (3) & (4) & (5) & (6) & (7) & (8) & (9) & (10) & (11) \\
\hline
PKS 1424-41 & 59935.38 & 28.88 $\pm$ 8.64 & 59935.63 & 75.24 $\pm$ 9.70 & 3.0 $\pm$ 0.9 & 1.19 $\pm$ 0.35 & 5.36 & R & 1.97 & 6.05\\
PKS 0736+01 & 58619.50 & 5.03 $\pm$ 1.66 & 58620.50 & 26.46 $\pm$ 3.35 & 2.73 $\pm$ 0.63 & 2.29 $\pm$ 0.53 & 12.87 & R & 4.21 & 1.43\\
PKS 0208-512 & 58793.50 & 3.17 $\pm$ 0.9 & 58794.50 & 10.20 $\pm$ 1.46 & 9.83 $\pm$ 2.65 & 4.90 $\pm$ 1.33 & 7.78 & R & 13.8 & 72.8\\
[+5pt]
\hline
\end{tabular}

\label{tab:Fastest_variability}
\end{table*}

\subsection{Flare fitting with exponential function}

The prominent and sub-prominent flares observed in the $\gamma$-ray light curves are analyzed using a sum of exponential functions. This approach enables the extraction of the rise and decay timescales associated with each distinct peak in the light curves and provides insight into the symmetry of the flare. The functional form of the model is expressed as \citep{abdo2010gamma}:

\begin{equation}
f(t) = a_0 + \sum_{i=1}^{N} 2a_i \left[\exp \left( \frac{T_i - t}{T_{r,i}} \right) + \exp \left( \frac{t - T_i}{T_{d,i}} \right) \right]^{-1}
\label{eq:flare_fitting}
\end{equation}

Here, $a_0$ denotes the baseline flux level, $a_i$ are the amplitude coefficients of the individual flares, and $T_i$ represents the peak time of the $i$-th flare. The parameters $T_{r,i}$ and $T_{d,i}$ correspond to the characteristic rise and decay timescales, respectively. For symmetric flare, $T_{r} = T_{d}$.

\begin{figure*}
    \centering
    \begin{subfigure}[b]{0.8\textwidth}
        \includegraphics[width=\textwidth]{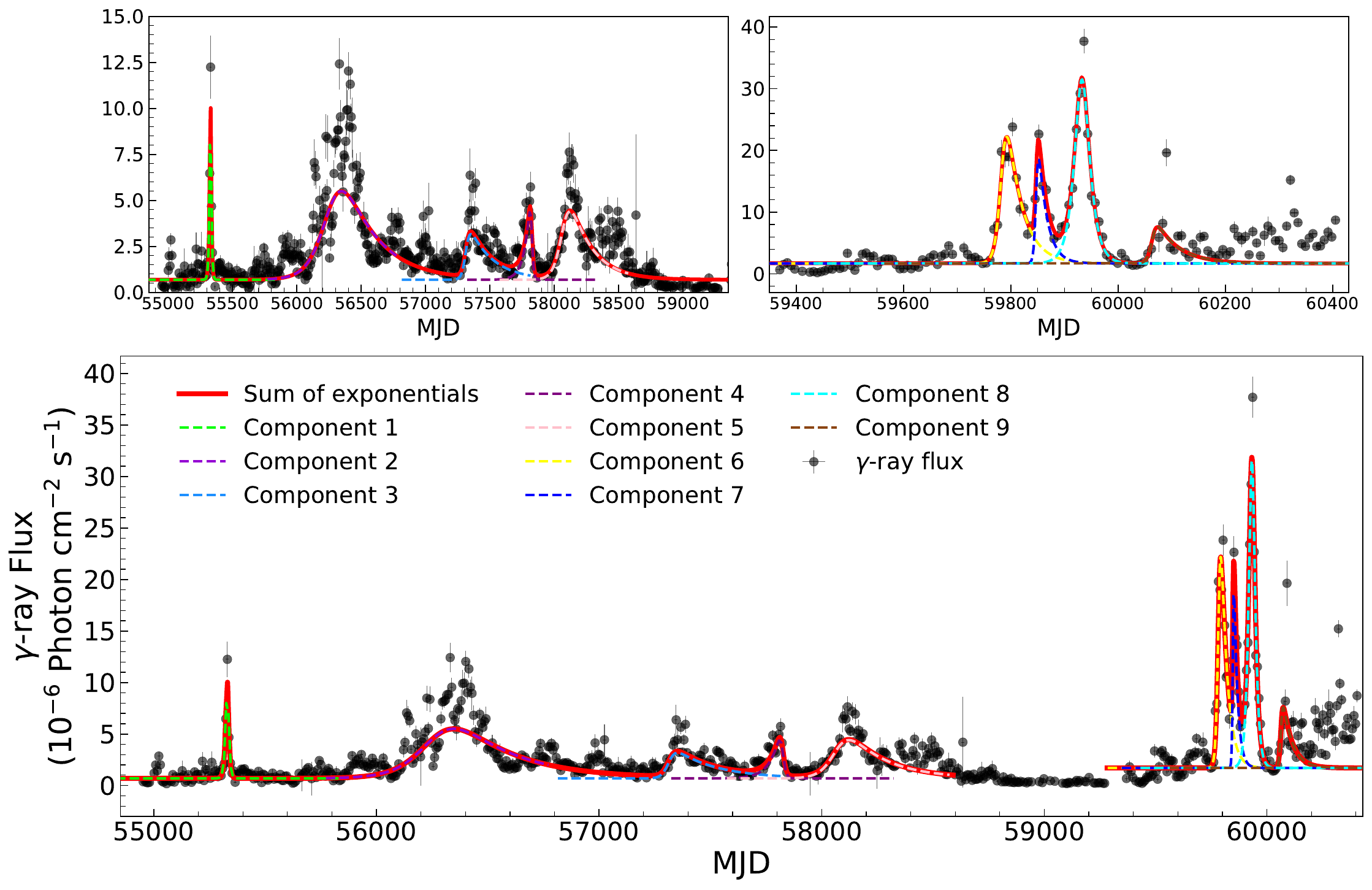}
        \caption{\small PKS 1424-41}
        \label{fig:flare_a}
    \end{subfigure}
    \vspace{1pt}

    \begin{subfigure}[b]{0.8\textwidth}
        \includegraphics[width=\textwidth]{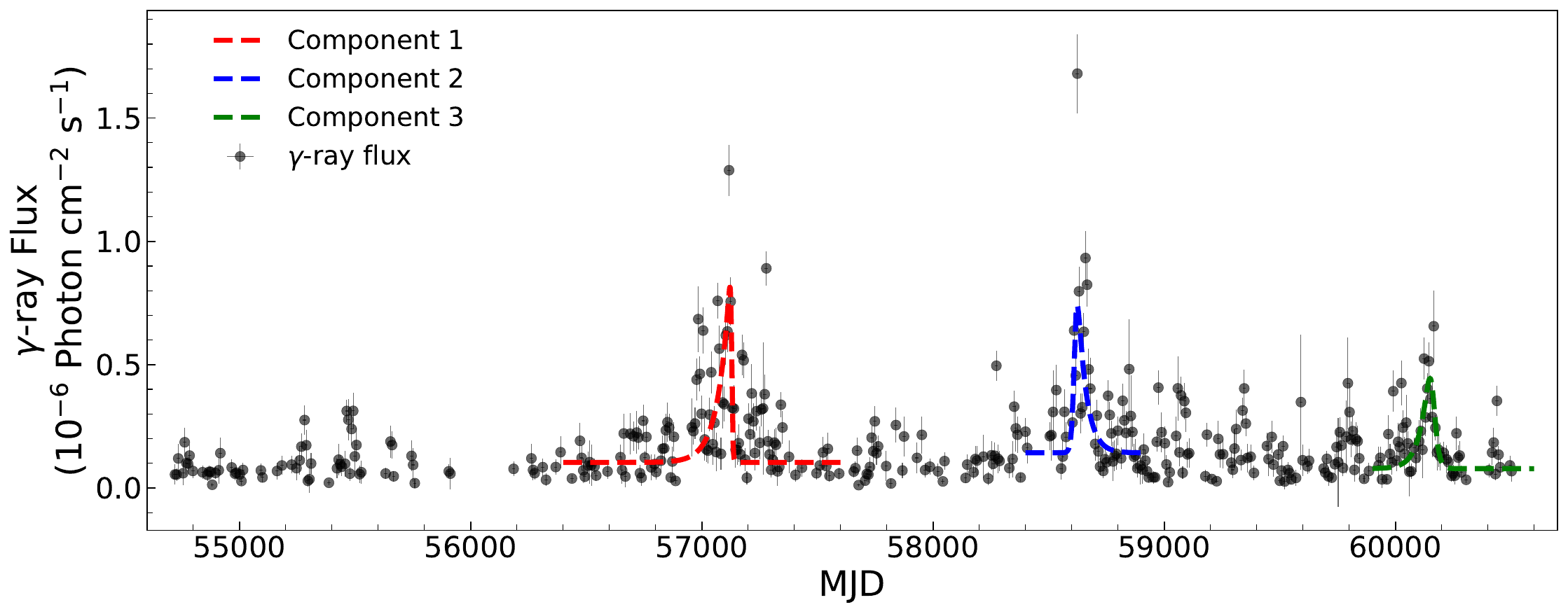}
        \caption{\small PKS 0736+01}
        \label{fig:flare_b}
    \end{subfigure}
    \vspace{1pt}

    \begin{subfigure}[b]{0.8\textwidth}
        \includegraphics[width=\textwidth]{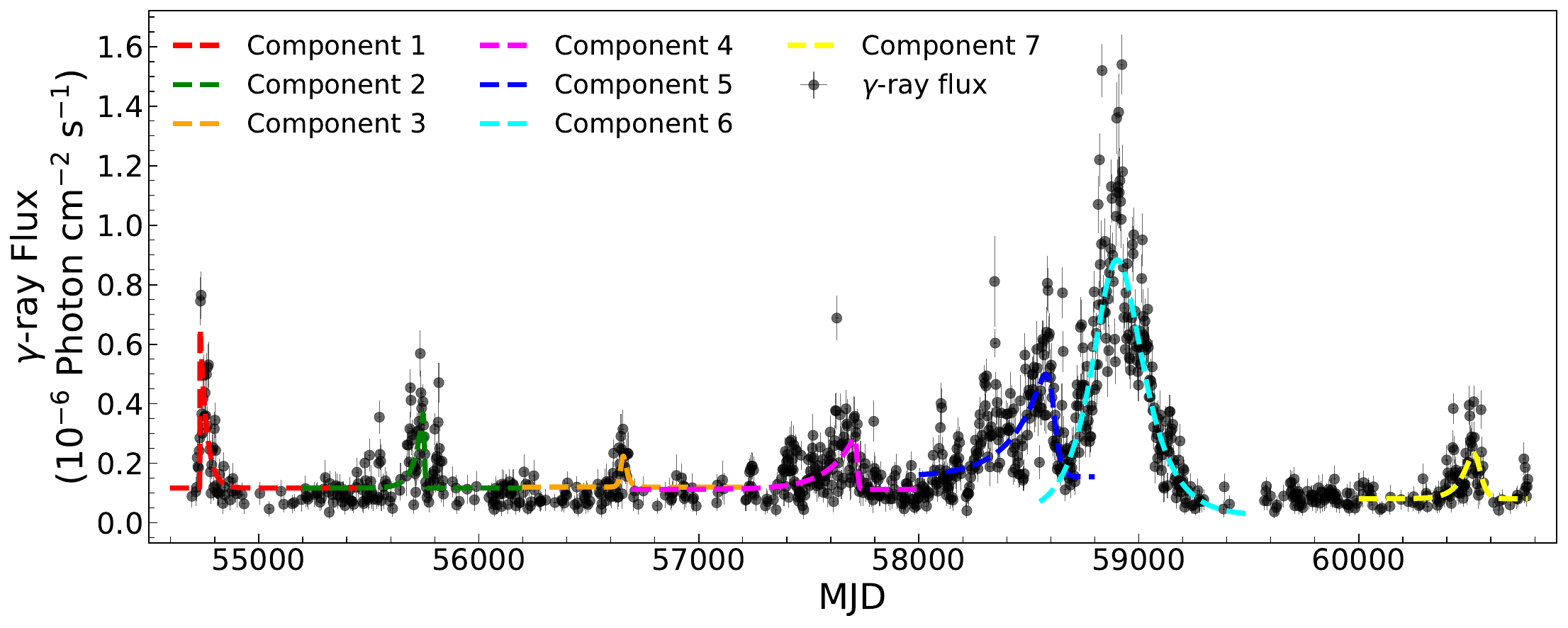}
        \caption{\small PKS 0208-512}
        \label{fig:flare_c}
    \end{subfigure}

    \caption{Flare modeling of $\gamma$-ray emissions using a sum of exponentials (SOE) function. Sub-figures (a), (b), and (c) show SOE-fitted $\gamma$-ray light curves for PKS 1424-41, PKS 0736+01, and PKS 0208-512, respectively. See Table~\ref{tab:flare_fitting_results} for the derived timescales.}
    \label{Fig-Flare_fitting}    
\end{figure*}

To quantify the degree of flare symmetry, we define the symmetry parameter $\zeta = \frac{T_d - T_r}{T_d + T_r}$. The value of $|\zeta|$ provides a measure of asymmetry: flares with $|\zeta| < 0.3$ are considered markedly symmetric, those with $0.3 < |\zeta| < 0.7$ are moderately asymmetric, and flares with $0.7 < |\zeta| < 1$ are classified as markedly asymmetric. Equation~\ref{eq:flare_fitting} serves as the basis for characterizing the temporal profiles of the observed flares. The corresponding fit parameters and graphical representations are presented in Table~\ref{tab:flare_fitting_results} and Figure~\ref{Fig-Flare_fitting}, respectively.

\subsection{Doubling/halving timescale}
The fastest variability timescales in the $\gamma$-ray light curves were estimated as the time required for the flux to either double or halve, and were calculated only for those light curves where the maximum flux exceeds  $10^{-6} \ \mathrm{ph\ cm^{-2}\ s^{-1}}$ (for bright flares). The expression used for determining the doubling or halving timescale is given by:

\begin{equation}
F(t_2) = F(t_1) \times 2^{\left(\frac{t_2 - t_1}{T_d}\right)}
\label{eq:doubling_halving}
\end{equation}

Here, $F(t_1)$ and $F(t_2)$ are the fluxes measured at times $t_1$ and $t_2$, respectively, and $T_d$ is the characteristic timescale associated with flux doubling or halving.

Equation~\ref{eq:doubling_halving} was applied to consecutive flux measurements in the light curves %that satisfy the criterion of test statistic $TS \ge 9$. 
and only those variations with a statistical significance of at least $3\sigma$ were considered. The significance was computed as the ratio of the flux difference $F(t_2) - F(t_1)$ to the uncertainty in $F(t_1)$. The derived doubling/halving timescales ($T_d$) were then converted into variability timescales using the relation $t_{\mathrm{var}} = \ln(2) \times T_d$.

The fastest variability timescales identified are: $\sim$3.0 hours with a significance of $5.36\sigma$ for PKS 1424-41, $\sim$2.73 hours with $12.87\sigma$ for PKS 0736+01, and $\sim$9.83 hours with $7.78\sigma$ for PKS 0208-512. These results are summarized in Table~\ref{tab:Fastest_variability}.\par

Based on the derived fastest variability timescales in the $\gamma$-ray emission from blazars, the size of the emitting region$-$assumed to be a spherical blob$-$can be constrained using the following relation:

\begin{equation}
R < \frac{c \ t_{\mathrm{var}} \ \delta}{1+z}
\label{eq:R}
\end{equation}

Here, $R$ is the radius of the emitting region, $c$ is the speed of light, $t_{\mathrm{var}}$ is the observed variability timescale, $\delta$ is the Doppler factor, $\delta \sim \Gamma$, and $z$ is the redshift of the source.

Furthermore, the distance of the emission region from the central SMBH, denoted as $R_H$, can be estimated by:

\begin{equation}
R_H \sim \frac{2 \ c \ t_{\mathrm{var}} \ \delta^2}{1+z}
\end{equation}

Using these expressions, the size and location of the emission region were estimated for blazars. For PKS 1424$-$41, the emitting region has a radius of $< 1.97 \times 10^{15}$ cm and is located at $\sim 6.05 \times 10^{16}$ cm from the SMBH. Similarly, for PKS 0736+01, the corresponding values are $<4.21 \times 10^{15}$ cm and $\sim 1.43 \times 10^{17}$ cm, while for PKS 0208$-$512, the estimated size and distance are $<1.38 \times 10^{16}$ cm and $\sim7.28 \times 10^{17}$ cm, respectively.

%\begin{figure*}
%    \centering
%    \includegraphics[width=0.8\textwidth]{Flare_fitting_PKS1424.pdf}
%    \vspace{1pt}
%    \includegraphics[width=0.8\textwidth]{Flare_fitting_PKS0736+01.pdf}
%    \vspace{1pt}
%    \includegraphics[width=0.8\textwidth]{Flare_fitting_PKS0208.pdf}
%    \caption{Flare modeling of $\gamma$-ray emissions from blazars PKS 1424-41 and PKS 0736+01 using a sum of exponentials (SOE) function. The second panel shows the 7-day binned $\gamma$-ray light curve of PKS 1424-41 fitted with SOE, highlighting the estimated rise and decay trend with timescales (Table~\ref{tab:flare_fitting_results}) of both bright and semi-bright flares. A zoomed-in view of the fitted light curve of PKS 1424-41 is shown in the two sub-figures in the first row of the figure. The third and fourth panels display the SOE-fitted weekly binned $\gamma$-ray light curves of PKS 0736+01 and PKS 0208-512. }
%    \label{Fig-Flare_fitting}    
%\end{figure*}

\begin{table*}%{ccccccc}
%\tablewidth{50pt}
\centering
\setlength{\extrarowheight}{8pt}
\setlength{\tabcolsep}{4pt}
\caption{The results of the LAT flare profile fitting for PKS 1424-41, with flare timescales measured in days, are as follows: for components 1 to 5, the base flux is (6.92 $\pm$ 0.39)$\times 10^{-7} (\mathrm{ph \ cm^{-2} \ s^{-1}})$ while for components 6 to 9, the base flux is 1.70 $\pm$ 0.21 $\times 10^{-6} (\mathrm{ph \ cm^{-2} \ s^{-1}})$. For PKS 037+01, the base flux values for the three components are (1.03 $\pm$ 0.13)$\times 10^{-7} (\mathrm{ph \ cm^{-2} \ s^{-1}})$, (1.42 $\pm$ 0.47)$\times 10^{-7} (\mathrm{ph \ cm^{-2} \ s^{-1}})$, and (0.78 $\pm$ 0.11)$\times 10^{-7} (\mathrm{ph \ cm^{-2} \ s^{-1}})$, respectively. Similarly, for PKS 0208-512, the base flux values for all seven components are, respectively: (1.16 $\pm$ 0.032)$\times 10^{-7} (\mathrm{ph \ cm^{-2} \ s^{-1}})$,(1.16 $\pm$ 0.034)$\times 10^{-7} (\mathrm{ph \ cm^{-2} \ s^{-1}})$, (1.19 $\pm$ 0.055)$\times 10^{-7} (\mathrm{ph \ cm^{-2} \ s^{-1}})$, (1.12 $\pm$ 0.037)$\times 10^{-7} (\mathrm{ph \ cm^{-2} \ s^{-1}})$, (0.23 $\pm$ 0.19)$\times 10^{-7} (\mathrm{ph \ cm^{-2} \ s^{-1}})$, (1.74 $\pm$ 0.17)$\times 10^{-7} (\mathrm{ph \ cm^{-2} \ s^{-1}})$, (0.23 $\pm$ 0.19)$\times 10^{-7} (\mathrm{ph \ cm^{-2} \ s^{-1}})$, (5.50 $\pm$ 0.24)$\times 10^{-8} (\mathrm{ph \ cm^{-2} \ s^{-1}})$, and (8.05 $\pm$ 0.38)$\times 10^{-8} (\mathrm{ph \ cm^{-2} \ s^{-1}})$.
\label{tab:flare_fitting_results}}
\begin{tabular}{llccccl}
%\tablehead{
\hline
\hline
Source & Component & Amplitude ($F_0$) &  $t_{peak}$ & $t_{rise}$ & $t_{decay}$ & Nature of flare ($\zeta$) \\
 & & ($10^{-7} \ \mathrm{ph \ cm^{-2} \ s^{-1}}$) &  (MJD) & (days) & (days) & \\
\hline
\multirow{9}{*}{PKS 1424-41} & Component 1 & 9.53 $\pm$  0.27 & 55330.14 & 6.33 $\pm$ 2.16 & 5.26 $\pm$ 1.95 & -0.09 (Symmetric) \\
& Component 2 & 4.20 $\pm$ 0.30 &  56272.62 & 87.78 $\pm$ 10.60 & 269.18 $\pm$ 25.88 & 0.50 (Moderatly asymmetric)\\
& Component 3 & 1.64 $\pm$ 0.27 &  57315.27 & 16.58 $\pm$ 8.36 & 187.18 $\pm$ 43.73 & 0.83 (Asymmetric)\\
& Component 4 & 2.81 $\pm$ 0.54 &  57823.71 & 45.61 $\pm$ 12.66 & 7.57 $\pm$ 4.13 & -0.71 (Asymmetric)\\
& Component 5 & 2.99 $\pm$ 0.32 &  58074.09 & 39.21 $\pm$ 9.06 & 175.20 $\pm$ 25.78 & 0.63 (Moderatly asymmetric)\\
& Component 6 & 16.62 $\pm$ 2.53 &  59784.49 & 6.57 $\pm$ 1.4 & 28.51 $\pm$ 7.72 & 0.62 (Moderatly asymmetric)\\
& Component 7 & 12.40 $\pm$ 3.38 &  59847.60 & 2.36 $\pm$ 0.24 & 16.34 $\pm$ 1.4 & 0.74 (Asymmetric)\\
& Component 8 & 29.83 $\pm$ 2.39 &  59932.12 & 12.06 $\pm$ 2.85 & 12.86 $\pm$ 1.79 & 0.03 (Symmetric)\\
& Component 9 & 4.09 $\pm$ 1.51 &  60064.00 & 4.78 $\pm$ 1.6 & 42.00 $\pm$ 0.9 & 0.79 (Asymmetric)\\
[1.5pt]
\hline
\multirow{3}{*}{PKS 0736+01} & Component 1 & 4.35 $\pm$  1.08 & 57128.94 & 50.01 $\pm$ 15.82 & 2.67 $\pm$ 0.29 & -0.89 (Asymmetric)\\
& Component 2 & 4.60 $\pm$ 1.49 & 58613.68 & 6.83 $\pm$ 1.82 & 6.70 $\pm$ 1.29 & 0.68 (Moderatly asymmetric)\\
& Component 3 & 3.22 $\pm$ 0.69 & 60159.79 & 37.40 $\pm$ 13.11 & 12.73 $\pm$ 2.04 & -0.49 (Moderatly asymmetric)\\
[1.5pt]
\hline
\multirow{7}{*}{PKS 0208-512} & Component 1 & 2.79 $\pm$  0.5 & 54732 & 0.15 $\pm$ 0.07 & 29.68 $\pm$ 5.23 & 0.98 (Asymmetric) \\
& Component 2 & 1.56 $\pm$ 0.38 &  55753 & 38.38 $\pm$ 12.9 & 2.44 $\pm$ 1.05 & -0.88 (Asymmetric)\\
& Component 3 & 0.99 $\pm$ 0.69 &  56653 & 8.40 $\pm$ 3.06 & 15.51 $\pm$ 2.1 & 0.29 (Symmetric)\\
& Component 4 & 0.99 $\pm$ 0.20 &  57720 & 119.7 $\pm$ 20 & 3.99 $\pm$ 1.40 & -0.93 (Asymmetric)\\
& Component 5 & 2.07 $\pm$ 0.16 &  58614 & 300.9 $\pm$ 50.5 & 9.22 $\pm$ 3.6 & -0.94 (Asymmetric)\\
& Component 6 & 8.58 $\pm$ 0.29 &  58894 & 96.13 $\pm$ 9.44 & 110.05 $\pm$ 13.51 & 0.06 (Symmetric)\\
& Component 7 & 1.36 $\pm$ 0.21 &  60535 & 51.09 $\pm$ 14.8 & 20.13 $\pm$ 10.7 & -0.43 (Moderatly asymmetric)\\
[1.5pt]
\hline
\end{tabular}
\end{table*}

\begin{figure}
    \centering
    \includegraphics[width=0.45\textwidth]{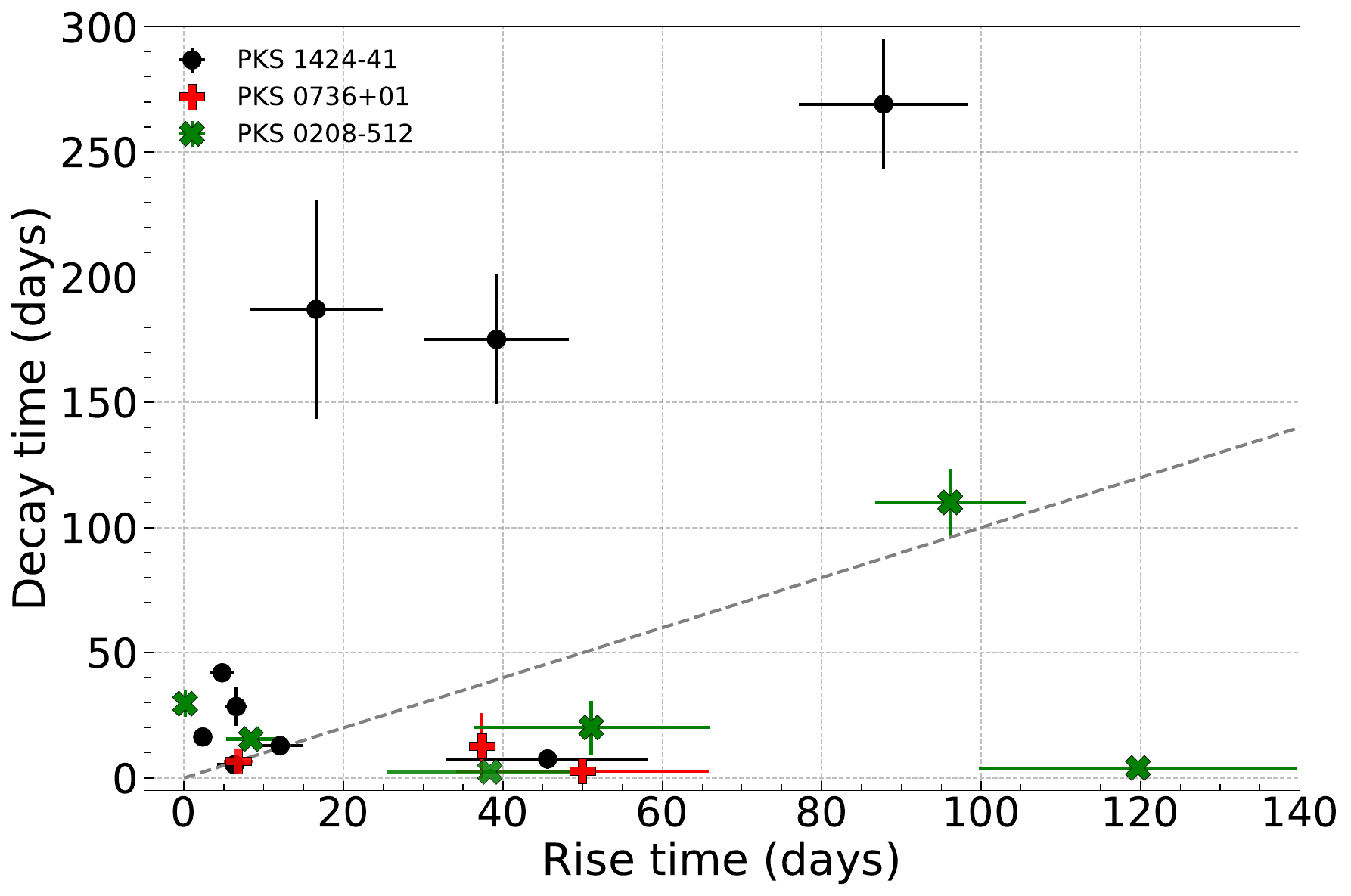}
    \caption{The rise and decay timescales obtained from the $\gamma$-ray light curve modeling using a sum of exponential functions. Black data points correspond to PKS 1424-41, while red and green points represent the rise and decay timescales of PKS 0736+01 and PKS 0208-512, respectively. The grey dashed line indicates equality between rise and decay timescales. }
    \label{Fig-Rise_decay_timescale}    
\end{figure}

\begin{figure}
    \centering
    \includegraphics[width=0.47\textwidth]{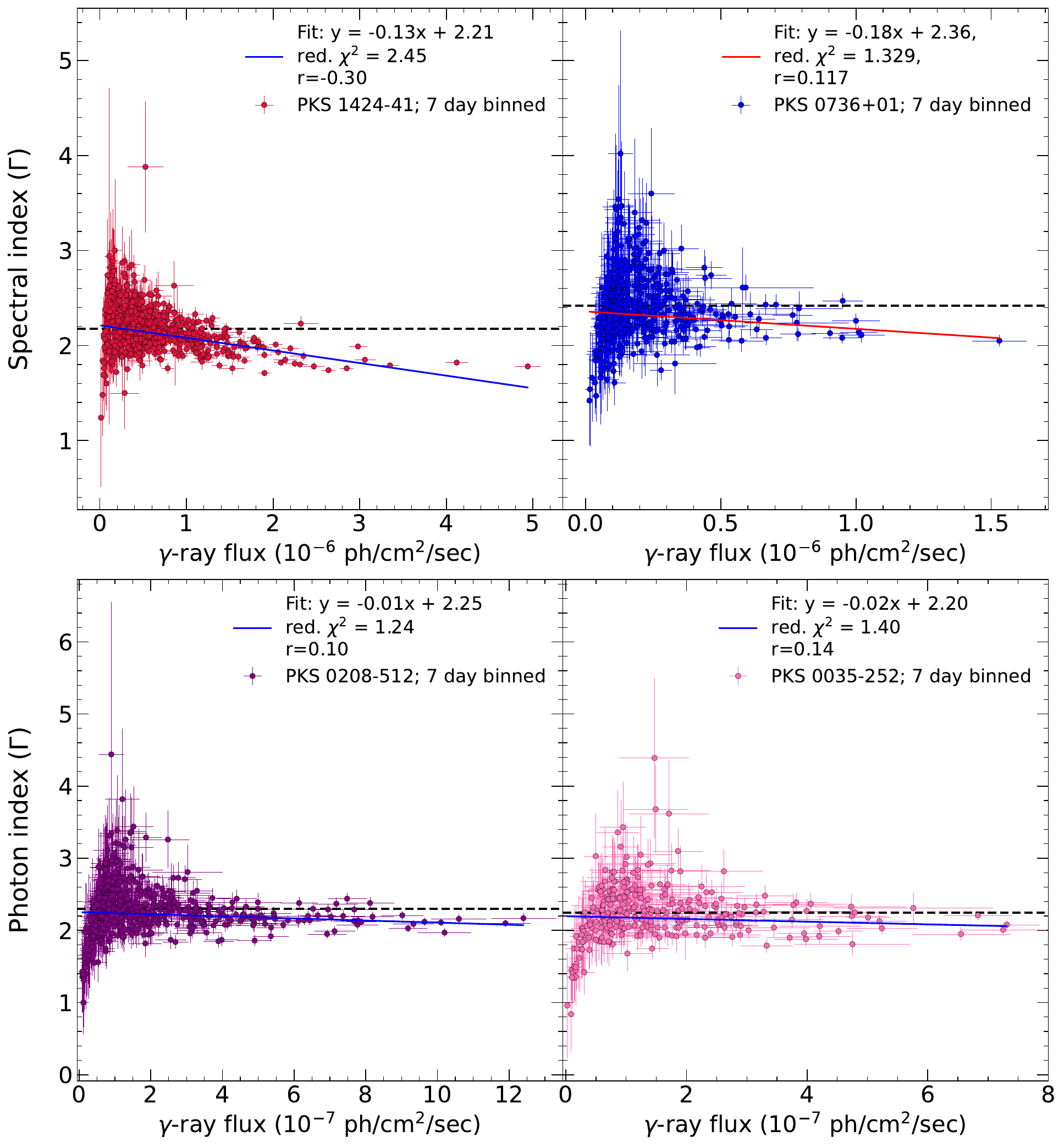}
    \caption{The variation in photon index as a function of $\gamma$-ray flux of PKS 1424-41, PKS 0736+01, PKS 0208-512, and PKS 0035-252. }
    \label{Fig-Index_flux_correlation}    
\end{figure}

\section{Correlation study}\label{sec:correlation}
\subsection{Flux-index correlation}
To investigate the correlation between $\gamma$-ray flux and corresponding photon index, we performed a correlation analysis using weekly binned $\gamma$-ray flux measurements and their corresponding photon indices for all sources in our sample. It is important to note that the photon index, denoted here by $\Gamma$, should not be confused with the Lorentz factor. We applied Spearman’s rank correlation method to assess the presence of any monotonic relationship between the two variables. The results, summarized in Table~\ref{tab:correlation_results} and shown in Figure~\ref{Fig-Index_flux_correlation}, indicate a weak positive correlation between the $\gamma$-ray flux and photon index across all sources, as reflected by the low Spearman correlation coefficients ($\rho$). Nevertheless, the very low p-values associated with the correlation coefficients suggest that the observed correlations are statistically significant and unlikely to have occurred by chance.

\begin{table}
\setlength{\extrarowheight}{4pt}
\setlength{\tabcolsep}{4pt}
\centering
\caption{The correlation between photon index and $\gamma$-ray flux of blazars.}
\begin{tabular}{l c c}

\multicolumn{3}{c}{\textbf{Correlation between $\gamma$-ray flux and index}}\\
\hline 
& \multicolumn{2}{c}{Spearman's rank correlation} \\
\cline{2-3} 
Source  & $\rho$ & p-value\\
[+2pt]
\hline
PKS 1424-41 & -0.30 & 2.42$\times 10^{-18}$\\  
PKS 0736+01 & 0.117 & 0.0098 \\
%S2 0109+22 & 0.362 & 2.69$\times 10^{-23}$\\
%PKS 0244-470 & 0.40 & 1.36$\times 10^{-11}$\\
%PKS 0405-385 & 0.39 & 3.51$\times 10^{-17}$\\
PKS 0208-512 & 0.10 & 0.012\\
PKS 0035-252 & 0.14 & 0.011\\
[+4pt]
\hline
\end{tabular}
\label{tab:correlation_results}
\end{table}
    
\subsection{Cross-correlation between different wave bands}\label{sec:ICCF}
A cross-correlation analysis between $\gamma$-rays and UV-optical band light curves was carried out by utilizing the interpolated cross-correlation function (ICCF: \cite{peterson1998uncertainties, peterson2004central}), which is one of the commonly used methods in the time-series analysis of AGNs. The ICCF method emerges as a powerful technique to estimate the time lag between two time series. The method uses the linear interpolation method to deal with unevenly sampled AGN light curves and calculate the cross-correlation coefficient as a function of the time lag for two time series:

\begin{equation}
    F_{CCF}(\tau)=\frac{1}{N} \sum_{i=1}^N \frac{\left[ L(t_i) - \bar{L} \right] \left[ C(t_i - \tau) - \bar{C} \right]}{\sigma_L \sigma_C}
\end{equation}

where N is the number of data points in the light curves, L and C. Each light curve has a corresponding mean value ($\bar{L}$ and $\bar{C}$ ) and uncertainty ($\sigma_L$ and $\sigma_C$).

The ICCF is evaluated for a time lag ($\tau$) in a range [-100,100] days with a searching step $\Delta \tau$, which should be smaller than the median sampling time of the light curves. We adopted $\Delta \tau$=7 days and used the public PYTHON version of the ICCF, PYCCF \citep{2018ascl.soft05032S} in this study. By applying the ICCF to the light curves, we estimated the cross-correlation and its corresponding centroid for time lags around the peak. We adopted the centroid of the CCF ($\tau_{cent}$) using only time lags with r$> 0.8 r_{max}$, where $r_{max}$ is the peak value of the CCF. The 1$\sigma$ confidence on the time lag is estimated using a model-independent Monte Carlo method. We estimated the time lags between $\gamma$-ray and UV-optical light curves based on CCCD for blazars PKS 1424-41 and PKS 0736+01, and the observed time lags are mentioned in Table~\ref{tab:ICCF} and Figures~\ref{Fig-CCF_PKS1424-41},~\ref{Fig-CCF_PKS0736}, and ~\ref{Fig-CCF_PKS0208}.

%\begin{table}
%\setlength{\extrarowheight}{4pt}
%\setlength{\tabcolsep}{4pt}
%\centering
%\caption{Fractional variability amplitude $F_{var}$ obtained in different energy bands. Columns 1: energy band 2: fractional variability amplitude with uncertainty.}

%\begin{tabular}{l c c c c}

%\hline 
%& \multicolumn{2}{c}{PKS 1424-41} & \multicolumn{2}{c}{PKS 0736 +01} \\
%\cline{2-5} 
%Light curves & CCCD & CCPD & CCCD & CCPD\\
% & Days & Days & Days & Days\\
%[+2pt]
%\hline
%$\gamma$-ray vs B band  & 58.88$_{-6.64}^{+5.09}$ & 58$_{-8}^{+6}$ &  54.73$_{-66.85}^{+15.31}$ & 54$_{-66}^{+16}$\\  
%$\gamma$-ray vs U band & -37.43$_{-29.76}^{+28.79}$ & -62$_{-8}^{+58}$ & 20.4$_{-34.28}^{+48.54}$ &  18$_{-30}^{+50}$ \\
%$\gamma$-ray vs V band & -38.97$_{-29.01}^{+18.86}$ & -64$_{-8}^{+46}$ & -64.27$_{-45.47}^{+6.61}$ &  -66$_{-50}^{+4}$\\
%$\gamma$-ray vs M2 band & -11.55$_{-52.3}^{+9.66}$ & -2$_{-64}^{+4}$ & 56.05$_{-69.05}^{+14.45}$ &  54$_{-66}^{+16}$\\
%$\gamma$-ray vs W1 band & -68.11$_{-3.1}^{+47.18}$ & -68$_{-6}^{+58}$ & 59.87$_{-59.05}^{+10.96}$ &  58$_{-58}^{+12}$\\
%$\gamma$-ray vs W2 band & -7.95$_{-57.53}^{+11.2}$ & -6$_{-60}^{+4}$ & 51.94$_{-65.83}^{+18.08}$ &  52$_{-64}^{+18}$\\
%[+4pt]
%\hline
%\end{tabular}
%\label{tab:ICCF}
%\end{table}

\begin{table*}
\setlength{\extrarowheight}{8pt}
\setlength{\tabcolsep}{10pt}
\centering
\caption{The Cross-correlation analysis between $\gamma$-ray and Swift-UVOT filters.}

\begin{tabular}{l c c c c c c}

%\multicolumn{5}{c}{\textbf{Data set (2008--2024)}}\\
\hline 
& \multicolumn{2}{c}{PKS 1424-41} & \multicolumn{2}{c}{PKS 0736 +01} & \multicolumn{2}{c}{PKS 0208-512} \\
\cline{2-7} 
Light curves & CCCD & CCPD & CCCD & CCPD & CCCD & CCPD\\
 & Days & Days & Days & Days & Days & Days\\
[+2pt]
\hline
$\gamma$-ray vs B band  & 58.88 & 58 &  54.73 & 54 & 6.43 & 1\\  
$\gamma$-ray vs U band & -37.43 & -62 & 20.4 &  18 & -9.38 & 1 \\
$\gamma$-ray vs V band & -38.97 & -64 & -64.27 &  -66 & 23.2 & 8\\
$\gamma$-ray vs M2 band & -11.55 & -2 & 56.05 &  54 & -2.17 & 1 \\
$\gamma$-ray vs W1 band & -68.11 & -68 & 59.87 &  58 & 0.8 & 1\\
$\gamma$-ray vs W2 band & -7.95 & -6 & 51.94 &  52 & 1.28 & 1\\
[+4pt]
\hline
\end{tabular}
\label{tab:ICCF}
\end{table*}

\begin{figure*}
    \centering
    \includegraphics[width=0.49\textwidth]{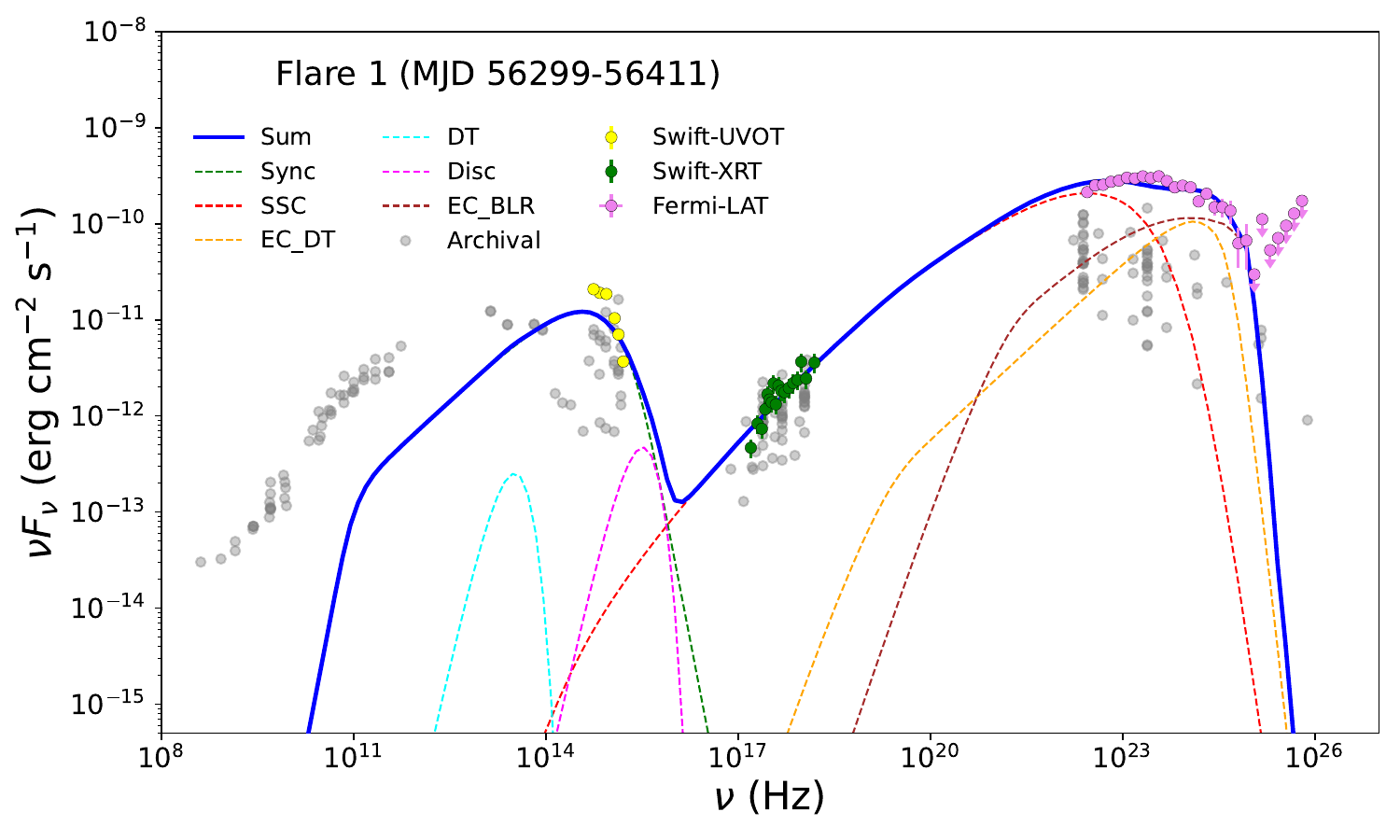} \hspace{1pt} 
    \includegraphics[width=0.49\textwidth]{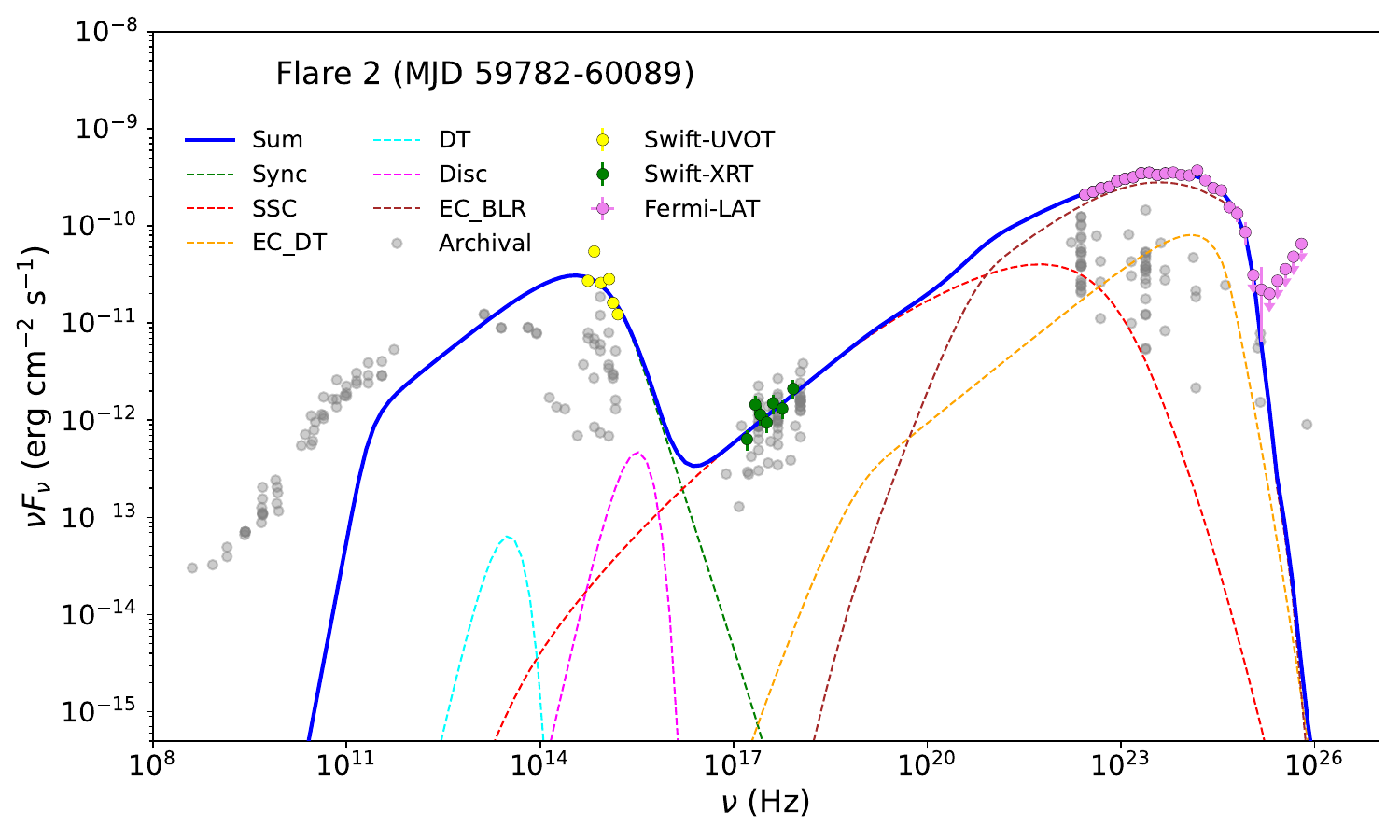}
    \vspace{1pt}
    \includegraphics[width=0.49\textwidth]{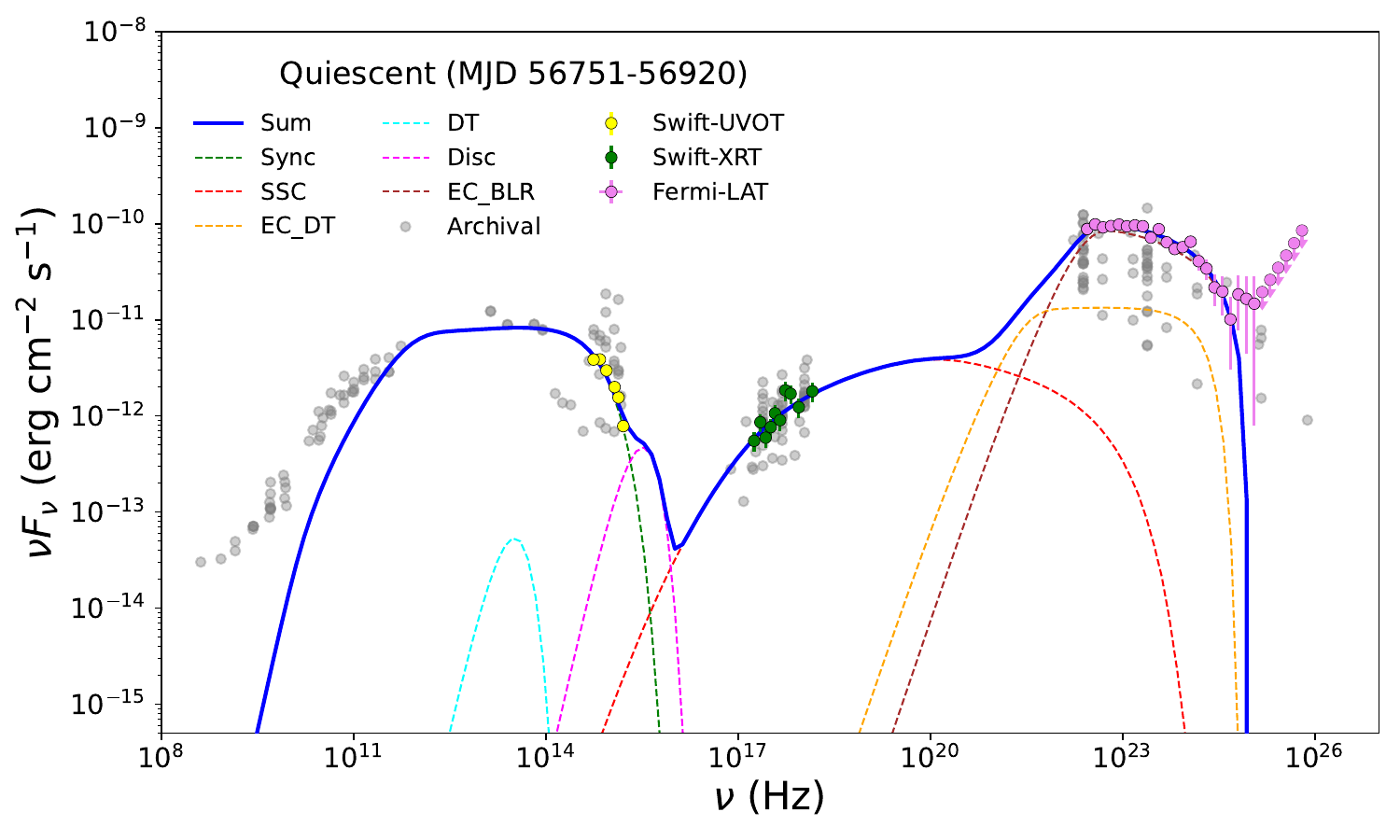}
    \caption{The broadband SEDs of Flare 1 and 2, and the quiescent states of PKS 1424-41 fitted with a one-zone leptonic model using JetSet. The data points and various lines in different colors are self-explanatory.  }
    \label{Fig-JETSET_PKS1424}    
\end{figure*}

\begin{figure*}
    \centering
    \includegraphics[width=0.49\textwidth]{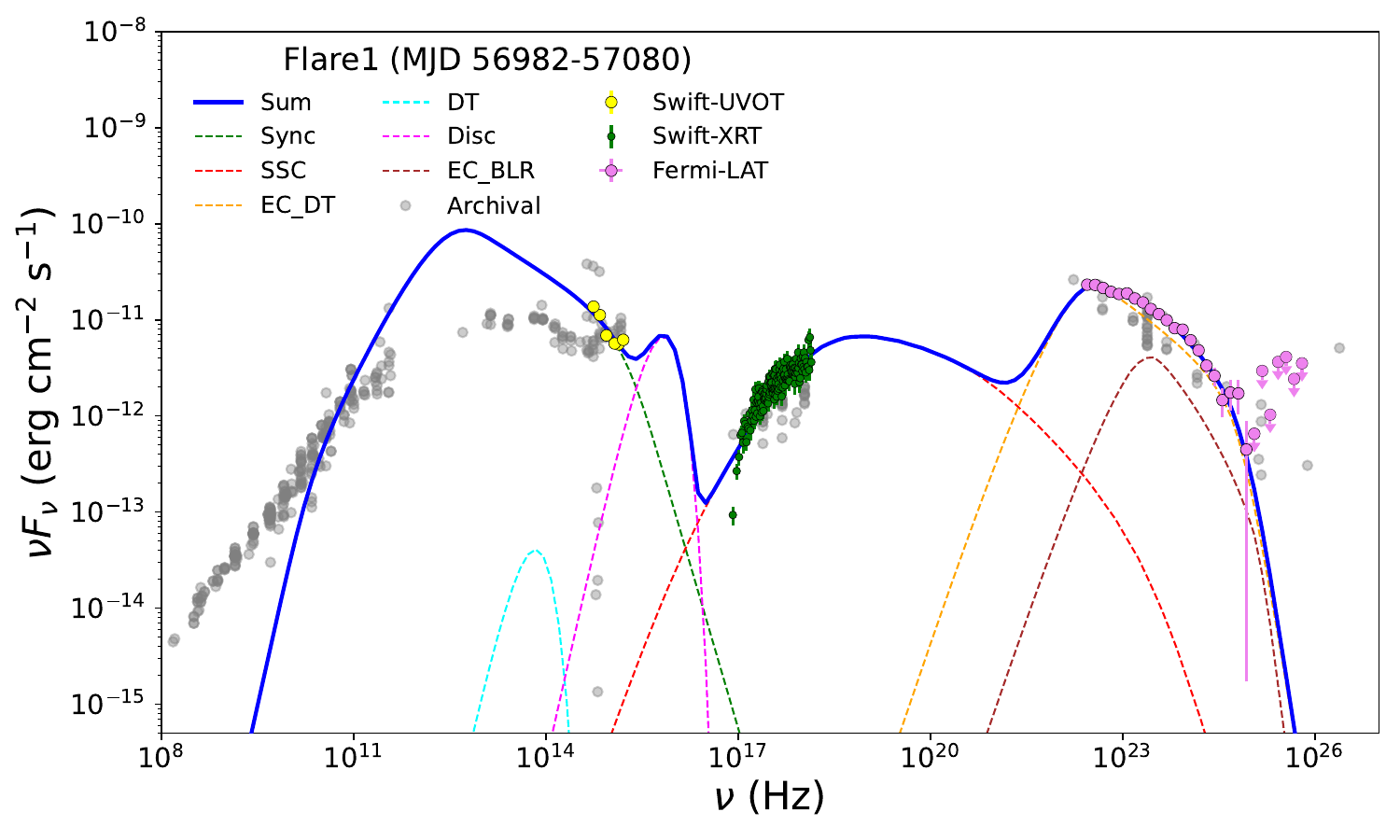}
    \hspace{1pt} 
    \includegraphics[width=0.49\textwidth]{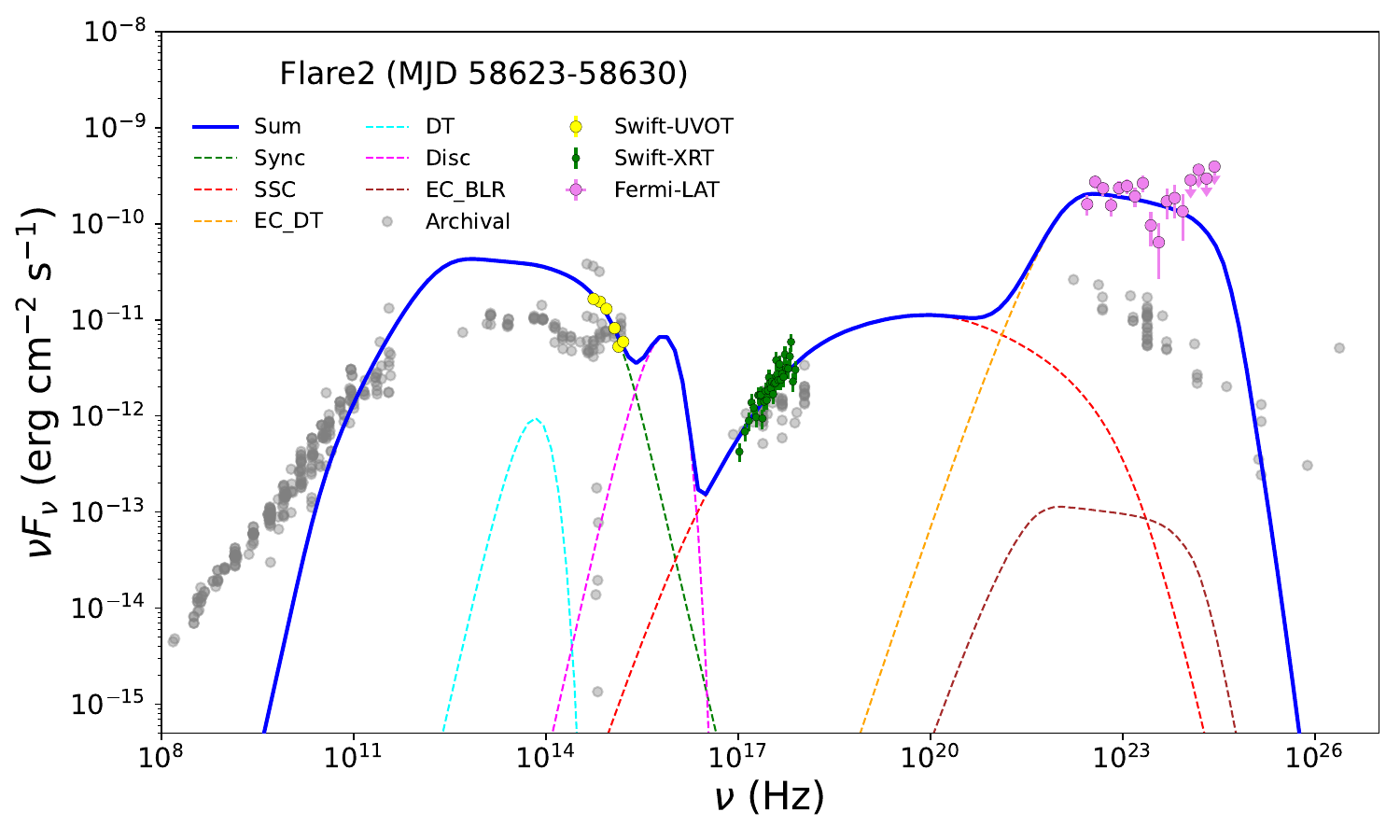}
    \vspace{1pt}
    \includegraphics[width=0.49\textwidth]{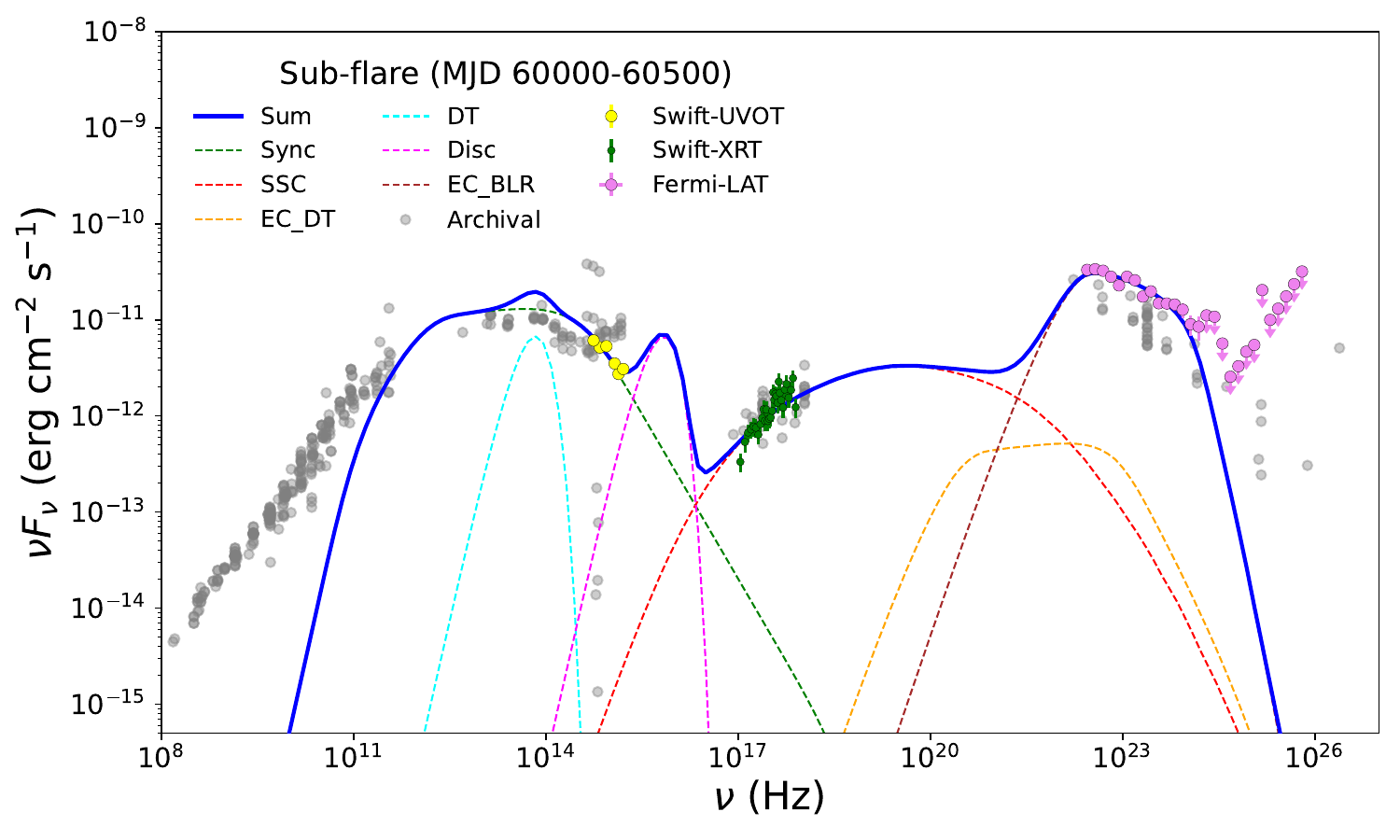}
    \caption{The broadband SEDs of Flare 1 and 2, and a sub-flare of PKS 0736+01 fitted with one-zone leptonic model. }
    \label{Fig-JETSET_PKS0736}    
\end{figure*}

\begin{figure*}
    \centering
    \includegraphics[width=0.49\textwidth]{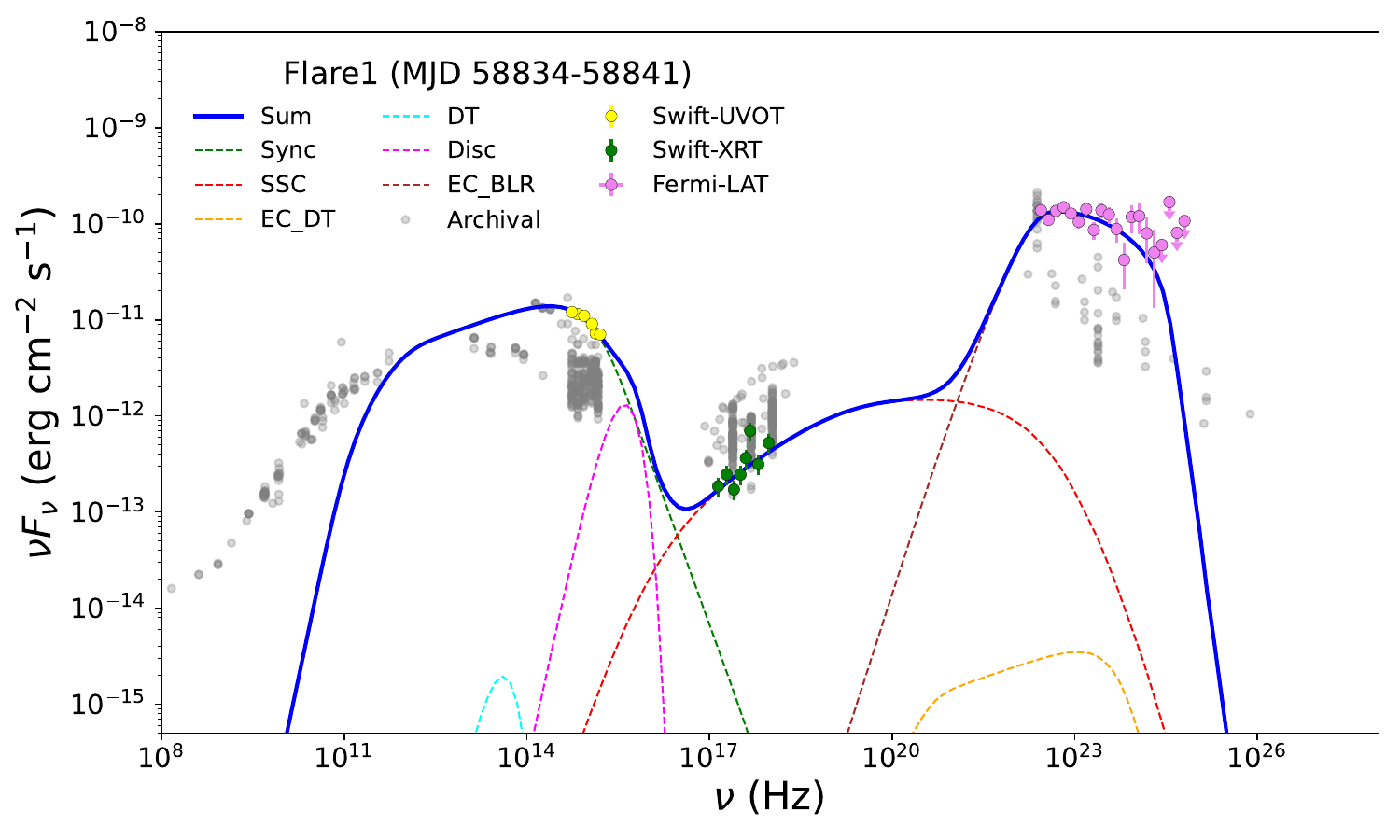}\hspace{1pt}
    \includegraphics[width=0.49\textwidth]{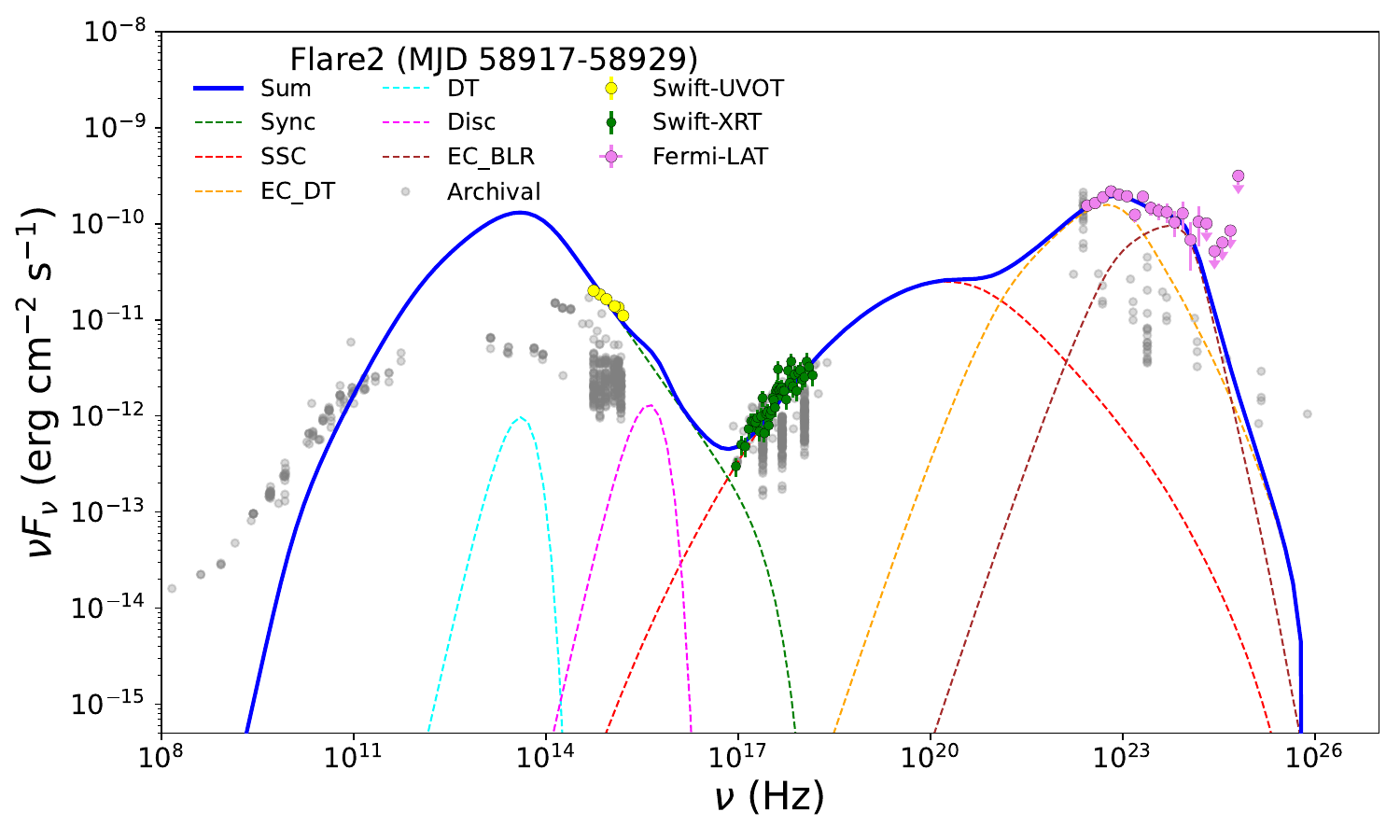}
    \vspace{1pt}
    \includegraphics[width=0.49\textwidth]{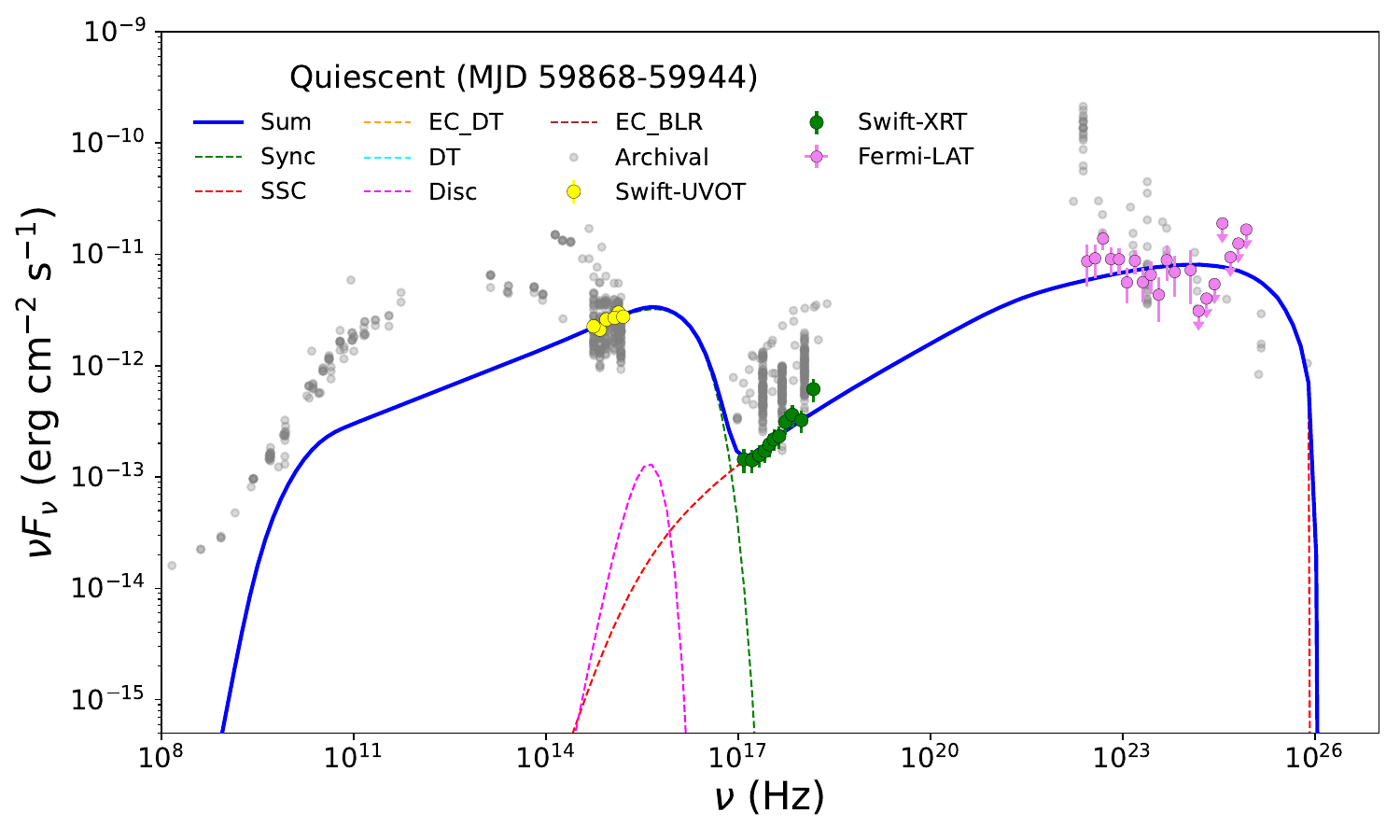}
    \caption{The broadband SEDs of Flare 1 and 2, and a low-flux state of PKS 0208-512 fitted with one-zone leptonic model. }
    \label{Fig-JETSET_PKS0208}    
\end{figure*}

\begin{figure*}
    \centering
    \includegraphics[width=0.49\textwidth]{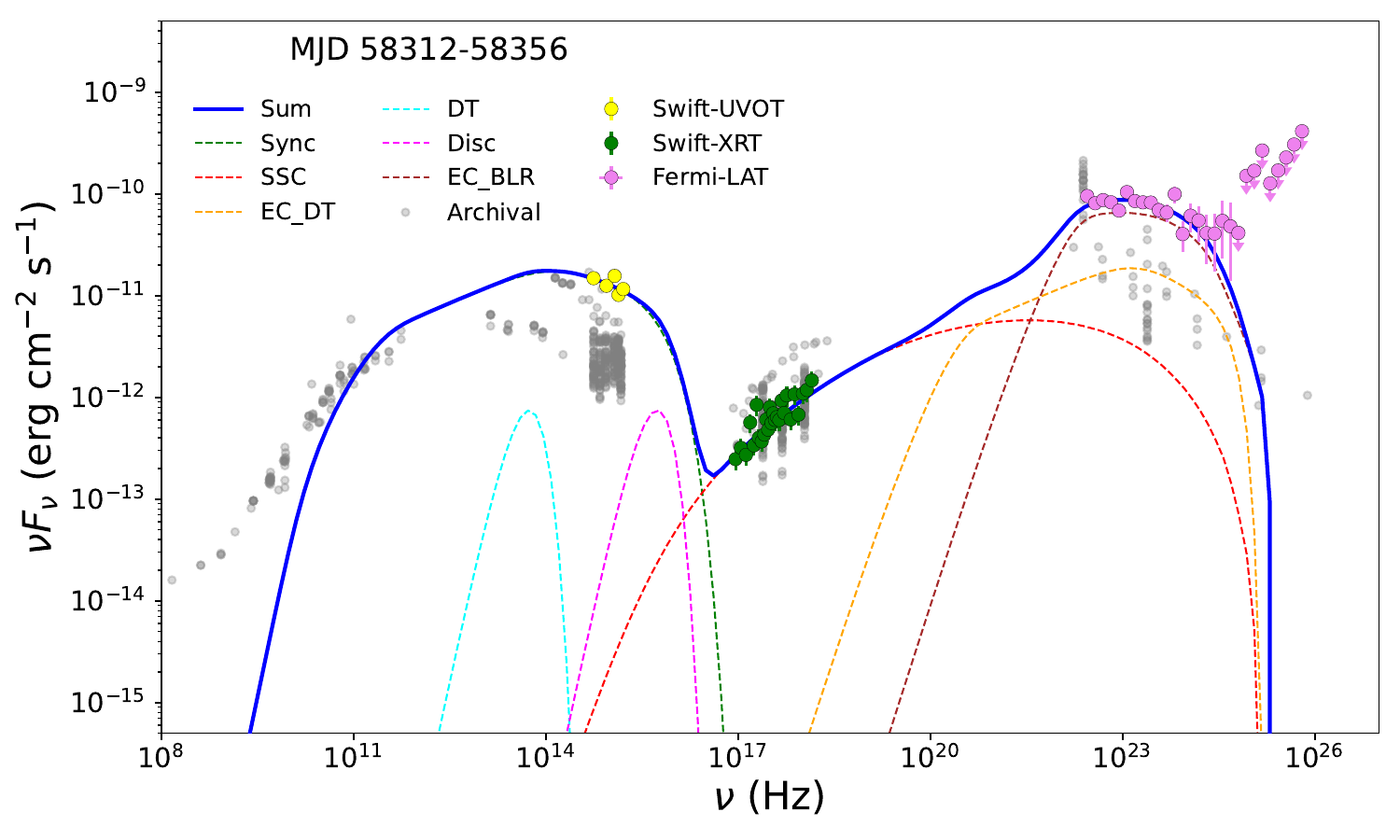}
    \caption{The broadband SED modeling of PKS 0035-252 with one-zone leptonic model. }
    \label{Fig-JETSET_PKS0035-252}    
\end{figure*}

\begin{figure*}
    \centering
    \includegraphics[width=0.89\textwidth]{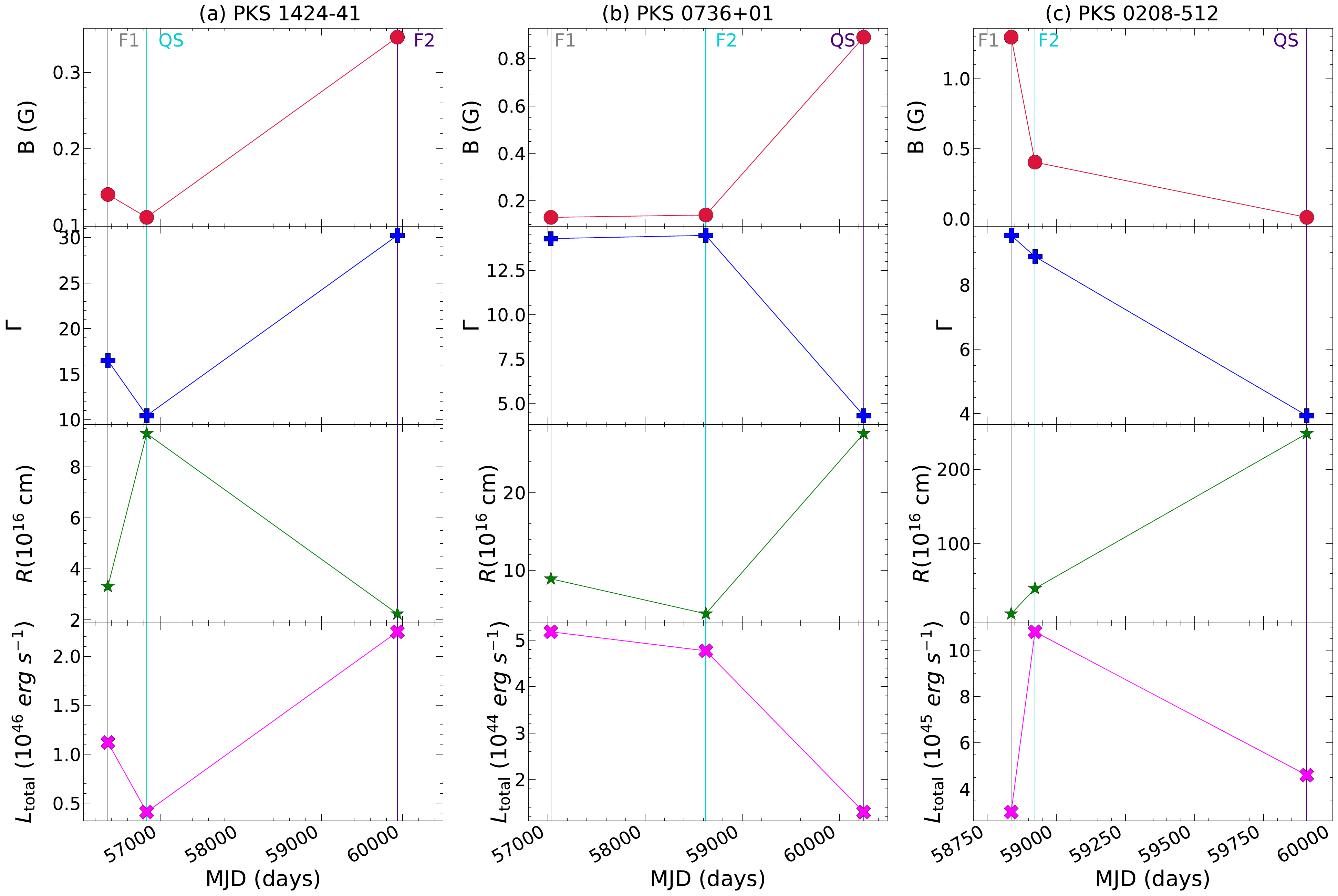}
    \caption{The variation of jet parameters in different flux states. }
    \label{Fig-JETSET_parameters_variations}    
\end{figure*}

\section{Simultaneous broadband SED modeling}\label{sec:sec4_SED_modeling}

The emission mechanisms of blazars can be comprehensively understood through the study of simultaneous multi-wavelength Spectral Energy Distributions (SEDs). These SEDs typically exhibit two broad humps, which are well explained by the one-zone Synchrotron Self-Compton (SSC) model within the leptonic framework. In this scenario, relativistic leptons--primarily electrons and positrons--interact with the magnetic field in a compact emission region, producing synchrotron radiation that spans from radio to soft X-ray frequencies. This synchrotron emission forms the first hump of the SED. The second hump, extending from X-rays to gamma-rays, arises from inverse Compton (IC) scattering processes, wherein the same population of relativistic electrons upscatter low-energy photons to higher energies. In the SSC models \citep{ghisellini1993high, maraschi1992jet}, the seed photons for IC scattering are the synchrotron photons previously emitted by the electrons themselves. In case of External Compton (EC) models \citep{dermer1993model, sikora1994comptonization}, the seed photons originate from external photon fields, which include:
\begin{itemize}
    \item Direct optical emission from the accretion disk,
    \item Reprocessed ultraviolet-optical emission from the Broad-Line Region (BLR) \citep{donea2003radiation}, and
    \item Infrared emission from the Dusty Torus (DT).
\end{itemize}

External photons involved in the External Compton (EC) process can originate from multiple sources. The accretion disc emits direct optical/UV photons characterized by a luminosity $L_{disc}$ and temperature $T_{disc}$. A fraction of this disc emission is reprocessed by surrounding structures. The Broad-Line Region (BLR), located at a distance $R_{BLR}$ from the central engine, reprocesses a portion of the disc radiation into ultraviolet photons. Similarly, the dusty torus, situated at a distance $R_{DT}$ and with a characteristic temperature $T_{DT}$, re-emits infrared photons. The fractions of the disc luminosity reprocessed by the BLR and the torus are denoted by $\tau_{BLR }$ and $\tau_{DT}$, respectively. These photon fields serve as seed photons for inverse Compton scattering in the EC process.\par

In this study, we performed the broadband SED modeling of different flux states of blazars PKS 0736+01, PKS 14124-41, PKS 0208-512, and PKS 0035-252 using publicly available code JetSet\footnote{\url{https://jetset.readthedocs.io/en/1.3.0/}} \citep{tramacere2009swift, tramacere2011stochastic, tramacere2020jetset}. JETSET (version 1.3.0) fits numerical models to the observed data to determine the optimal jet parameters. We accumulated the simultaneous observations from Fermi-LAT, Swift-XRT, and -UVOT of different flux states of blazars to perform broadband SED modeling. In the one-zone leptonic emission model, the spherical blob of radius $R$ is located at a distance of $R_H$ from the central black hole, which is moving down the jet with the bulk Lorentz factor $\Gamma$ at a small angle to the observer, is permeated with the magnetic field $B$. It is postulated that this spherical region is injected with non-thermally accelerated electrons that follow a distribution characterized by a broken PL. The broken PL distribution of electrons is defined as follows:

\begin{equation}
N(\gamma)\, d\gamma =
\left\{
\begin{aligned}
    &K\gamma^{-p} d\gamma, \ \ \ \ \ \ \ \ \  \ \ \ \ \ \gamma_{min} < \gamma < \gamma_{break} \\
    &K\gamma_{break}^{p_1 - p}\gamma^{-p_1}d\gamma, \ \ \ \ \gamma_{break} < \gamma < \gamma_{max}
\end{aligned}
\right.
\end{equation}

where $\gamma$ is the electron Lorentz factor, $ N$ is the electron density in units of $cm^{-3}$, and $K$ is the normalization constant. $p$ and $p_1$ are the low- and high-energy spectral; slopes, and $\gamma_{min}$, $\gamma_{max}$, and $\gamma_{break}$ are the Lorentz factors corresponding to the low-energy cut-off, high-energy cut-off, and turn over energy, respectively. In general, broadband SED modeling involves a large number of parameters, which can often lead to model degeneracy. To minimize degeneracy and improve the reliability of the fit, it is crucial to reduce the number of free parameters. Therefore, in our modeling approach, several key parameters were fixed, including the $z$, $\theta$, $L_{disc}$, $T_{disc}$, $n$(ratio of cold protons to relativistic electrons), and $T_{DT}$. The inclusion of $L_{disc}$ allows us to estimate the characteristic distances of external photon fields. Specifically, the inner radius of the Broad-Line Region (BLR) is computed using the relation:

\begin{equation}
R_{\text{BLR-in}} = 3 \times 10^{17} \times \sqrt{\frac{L_{\text{disc}}}{10^{46}}} , \text{cm}
\label{eq:R_BLR_in}
\end{equation}

The outer radius of the BLR is then taken as:
\begin{equation}
    R_{BLR-out} = 1.1\times R_{BLR-in}
    \label{eq:R_BLR_out}
\end{equation}
Similarly, the radius of the dusty torus is estimated by:

\begin{equation}
R_{\text{DT}} = 2 \times 10^{19} \times \sqrt{\frac{L_{\text{disc}}}{10^{46}}} , \text{cm}
\label{eq:R_DT}
\end{equation}

Based on these relations, $R_{BLR-in}, R_{BLR-out}$ and $R_{DT}$ are estimated and kept fixed during the modeling. However, the size of the emitting region ($R$) and its location along the jet ($R_H$) are treated as free parameters. The findings from broadband SED modeling of different states of sources are illustrated below.\par

%It is important to note that the viewing angle $\theta$, which was fixed in our model, was independently determined through quasi-periodic oscillations analysis of the light curves for all sources, as discussed in \textbf{PAPER  \MakeUppercase{\romannumeral 1}}. 

\subsection{PKS 1424-41}\label{sec:PKS1424_SED_modeling}
The multi-wavelength observations of PKS 1424–41, as discussed in Section 2, enable the construction of broadband SEDs corresponding to distinct flux states of the source. These states are characterized based on temporal variations observed in $\gamma$-rays. The broadband SEDs were generated as follows:
\begin{enumerate}[label=(\alph*)]
    \item Flare 1: A high-flux state identified during the period MJD 56299–56411, wherein the $\gamma$-ray flux exceeded the long-term average and 3$\sigma$ flux threshold, as shown in Figure~\ref{Fig-LC_PKS1424}. This enhanced activity was further confirmed using Bayesian block analysis and supported by simultaneous multi-band data.
    \item Flare 2: The brightest $\gamma$-ray flare ever detected from this source by the Fermi-LAT instrument, with a peak flux of approximately \( \sim 4 \times 10^{-6} \ \mathrm{ph \ cm^{-2} \ s^{-1}} \), observed between MJD 59782–60089, see Figure~\ref{Fig-LC_PKS1424}. This period also coincided with enhanced flux levels in both Swift-XRT and UVOT bands, indicating a significant multi-wavelength outburst.
    \item Quiescent state: A low-flux period between MJD 56751–56920, during which the $\gamma$-ray emission was consistently below the observed average flux level as shown in Figure~\ref{Fig-LC_PKS1424}.
\end{enumerate}

As mentioned earlier, several physical parameters were fixed during the SED modeling using the JetSet code to reduce model degeneracy and ensure reliable fits. 
%The viewing angle of the jet was set to 1.34$^{\circ}$, a value derived from modeling the $\gamma$-ray light curve during a time interval in which a quasi-periodic oscillation (QPO) was detected. This QPO was interpreted within the framework of a supermassive binary black hole (SMBBH) scenario, which emerged as a plausible explanation for the observed periodic modulation. From this modeling, the viewing angle of the jet was estimated to be approximately 1.34$^{\circ}$, and since this interval encompasses Flare 1 and quiescent state, we fixed the angle accordingly. For further details, we refer the reader to Paper \MakeUppercase{\romannumeral 1}. 
The redshift $z$ and viewing angle $\theta$ are fixed at $z=1.52$ and 3.05, respectively, \citep{abhir2021multi}. A key objective of the modeling is to investigate whether the jet is particle-dominated or magnetically dominated. To assess the contribution of cold protons to the total jet power, we fixed the ratio of cold protons to relativistic electrons at 0.1, following the approach adopted in \citep{ghisellini2012electron}.
The accretion disc parameters were also fixed, with a temperature $T_{disc} = 1\times 10^5$ K and the luminosity $L_{disc} = 1\times 10^{46} \ \mathrm{erg/s}$. Based on this disc luminosity, we derived and fixed the characteristic radii of the external photon fields: $3.0\times 10^{17}$, $3.3\times 10^{17}$, and $2.0\times 10^{19}$ cm, using the relations as given by Equations~\ref{eq:R_BLR_in}, \ref{eq:R_BLR_out}, and \ref{eq:R_DT}. We optimize the other parameters such as magnetic field, particle distribution slopes, particle Lorentz factors, the BLR and DT optical depths, size and location of emission region, and most importantly, the Doppler boosting parameter to achieve a best-fit model for the SED.

In the broadband SED modeling of PKS 1424–41, the contributions from external photon fields were incorporated, as both the broad-line region (BLR) and the dusty torus provide a significant supply of seed photons necessary for $\gamma$-ray production via external Compton (EC) processes. Notably, we find that higher photon densities are required during flaring episodes compared to the quiescent state, indicating enhanced external field interactions during active phases.\par

For all three flux states—Flare 1, Flare 2, and the Quiescent period—the infrared (IR), optical, and ultraviolet (UV) spectra are well reproduced by a synchrotron component. The corresponding magnetic field strengths derived from the fits are 0.14, 0.34, and 0.11 Gauss, respectively. These values are consistent with those reported during the interval MJD 56299–56412 by \citet{abhir2021multi}. 
%In contrast, \citet{aleksic2011magic} modeled the SED, including the very-high-energy (VHE) $\gamma$-ray counterpart during the lower flux phase (MJD 55269–55305), and found magnetic fields in the range of 0.006–0.033 Gauss.\par

An important factor that likely contributes to the enhanced $\gamma$-ray flux during Flare 1 and Flare 2 is the Doppler boosting, $\delta \sim \Gamma$. The Lorentz factors ($\Gamma$) derived for Flare 1 and Flare 2 are 16.49 and 30.24, respectively, while the quiescent state shows a significantly lower value of 10.44. This suggests that variations in the Doppler (or Lorentz) factor play a key role in driving the observed high-energy flares. Notably, \citet{abhir2021multi} assumed a fixed value of $\Gamma = 20$ during their modeling.\par

Furthermore, it is widely believed that short-term, bright $\gamma$-ray flares originate from compact emission regions. Supporting this idea, we find that the sizes of the emitting regions during Flare 1 and Flare 2 are $3.31 \times 10^{16}$ cm and $2.24 \times 10^{16}$ cm, respectively, while the quiescent state corresponds to a significantly larger region of $9.3 \times 10^{16}$ cm.\par

From the SEDs modeling, the derived energy densities of BLR ($U_{BLR}^{'}$), DT ($U_{DT}^{'}$), electrons ($U_{e}^{'}$), and magnetic field ($U_{B}^{'}$), are 0.43, $9.6\times 10^{-3}$, 0.25, and $8.04\times 10^{-4}$ in units of $\mathrm{erg \ cm^{-3}}$ for Flare 1, respectively. For Flare 2, these are 2.59, 0.019, 0.088, and $4.76\times 10^{-3}$ in  $\mathrm{erg \ cm^{-3}}$ and for quite flux state, these are 0.014, $1.63\times 10^{-3}$, $1.9\times 10^{-3}$, and $9.09\times 10^{-4}$ in $\mathrm{erg \ cm^{-3}}$, respectively. The photon density from the broad-line region (BLR) is found to be significantly higher during the flaring states compared to the quiescent phase. This indicates a more efficient supply of seed photons for inverse-Compton scattering, leading to enhanced $\gamma$-ray production during active episodes.\par

In our modeling, all three spectral energy distributions (SEDs) are well described by a one-zone leptonic scenario, as shown in Figure~\ref{Fig-JETSET_PKS1424}, where the emission region is situated within or in close proximity to the BLR. Both the BLR and DT contribute a sufficient number of external seed photons to support the observed high-energy emission through external Compton processes. Our results further indicate that the jet is particle-dominated, with a greater share of its total power carried by relativistic electrons rather than by the magnetic field.

\subsection{PKS 0736+01}\label{sec:PKS0736_SED_modeling}
The broadband SEDs of the blazar PKS 0736+01 across different flux states have been constructed, as described in Sect.~\ref{sec:data_reduction_sec2}. These states are summarized below:

\begin{enumerate}[label=(\alph*)]
\item Flare 1: A high-flux state observed during the period MJD 56982–57080, identified through Bayesian block analysis. This interval is characterized by elevated emission across multiple wavelengths, with significantly enhanced flux detected in both the Swift-XRT and UVOT bands, indicating a prominent multi-wavelength flare, see Figure~\ref{Fig-LC_PKS0736}.

\item Flare 2: The most intense $\gamma$-ray flare ever recorded from this source by Fermi satellite, reaching a peak flux of approximately  \( \sim 1.6 \times 10^{-6} \ \mathrm{ph \ cm^{-2} \ s^{-1}} \), during MJD 58623–58630. While this flare was accompanied by strong activity in the Swift-UVOT bands, the Swift-XRT flux was comparatively lower than that observed during Flare 1, as visible in Figure~\ref{Fig-LC_PKS0736}.

\item Sub-flare: A relatively low-flux state spanning MJD 60000–60500, during which the $\gamma$-ray emission surpassed the average flux and remained close to the $3\sigma$ flux level. This phase represents a subdued activity period across the observed wavebands. That's why we defined it as a sub-flare state, see Figure~\ref{Fig-LC_PKS0736}. 
\end{enumerate}

Using the JetSet code within the leptonic emission framework as discussed above, we modeled all three flux states of the source by keeping several physical parameters fixed. These include the viewing angle, set to  $\theta=1.8^{\circ}\pm 0.3^{\circ}$as estimated by \citet{pushkarev2017mojave}, the accretion disc temperature $T_{disc} = 1\times 10^{5} \ K$ and the disc luminosity $L_{disc} = 1\times 10^{45}$ erg/s, adopted from \citet{abdalla2020hess}, allow us to constrain the characteristic sizes of the external photon fields involved in inverse-Compton processes. Specifically, we estimate the inner and outer radii of the broad-line region as $R_{\mathrm{BLR, \ in}} \sim 9.4\times 10^{16}$cm and $R_{\mathrm{BLR, \ out}} \sim 1.04\times 10^{17}$cm respectively, while the radius of the dusty torus is estimated to be $R_{\mathrm{DT}} \sim 6.3\times 10^{18}$cm. The derived magnetic field ($B$) values are 0.13, 0.14, and 0.89, corresponding to Flare 1, Flare 2, and sub-flare. The flux enhancements observed during Flare 1 and Flare 2 are likely driven by increased bulk Lorentz factors, which are found to be $\Gamma=14.29$ and $\Gamma=14.47$, respectively. In contrast, during the sub-flare, the Lorentz factor is significantly lower, with $\Gamma=4.13$. For comparison, \citet{abdalla2020hess} reported a slightly higher value of $\Gamma=17.7$, which exceeds the values derived in our analysis. The sizes of the emitting regions corresponding to Flare 1, Flare 2, and the sub-flare are estimated to be approximately 
$\sim 8.9\times 10^{16} \ cm$, $\sim 4.4\times 10^{16} \ cm$, and $\sim 2.76\times 10^{17} \ cm$, respectively. The corresponding locations of these regions along the jet are 
$\sim 1.44\times 10^{17} \ cm$, $\sim 1.68\times 10^{17} \ cm$, and $\sim 1.32\times 10^{17} \ cm$. The energy densities derived from the SED modeling for each state are summarized in Table~\ref{tab:SEDresults}. The fitted SEDs of this source are shown in Figure~\ref{Fig-JETSET_PKS0736}.

\subsection{PKS 0208-512}\label{sec:PKS0208_SED_modeling}
The broadband SEDs of the blazar PKS 0208–512 have been modeled across different flux states. A brief description of each state is provided below:

\begin{enumerate}[label=(\alph*)]
\item Flare 1: A flaring event observed in $\gamma$-ray emission during the period MJD 58834–58841, accompanied by enhanced multi-wavelength activity detected in the Swift-XRT and -UVOT bands, Figure~\ref{Fig-LC_PKS0208}.

\item Flare 2: The maximum $\gamma$-ray flux ever recorded from this source, with a peak flux of approximately \( \sim 1.5 \times 10^{-6} \ \mathrm{ph \ cm^{-2} \ s^{-1}} \), see Figure~\ref{Fig-LC_PKS0736}, occurring during MJD 58917–58929. This flare was also associated with significant emission in both Swift-XRT and UVOT bands.

\item Quiescent: A relatively low-flux state observed between MJD 59868–59944, representing a period of low activity across all observed wavebands, see Figure~\ref{Fig-LC_PKS0208}.
\end{enumerate}

We followed the modeling approach described above to analyze the different flux states of PKS 0208–512, including two flaring states and one low-flux (quiescent) state, see Figure~\ref{Fig-JETSET_PKS0208}. The results of this analysis are presented below. The broadband SED modeling was performed by fixing several key parameters, as done for other sources in our sample. These include the viewing angle $\theta = 2^{\circ}$ \citep{khatoon2022temporal}, disc temperature $T_{disc}=1\times 10^5 K$, and disc luminosity $L_{disc}=1\times 10^{46} $erg/s. Using equations~\ref{eq:R_BLR_in}, ~\ref{eq:R_BLR_out}, and ~\ref{eq:R_DT}, we estimated the characteristic scales of the external photon fields: 
$R_{BLR, \ in}$, $R_{BLR, \ out}$, and $R_{DT}$. The magnetic field strengths derived for Flare 1, Flare 2, and the quiescent state are found to be 1.296, 0.404, and 0.01 Gauss, respectively. \citet{khatoon2022temporal} previously reported magnetic field values exceeding 1.38 Gauss. In our study, the bulk Lorentz factors corresponding to the flaring episodes are $\Gamma=9.54$ and $\Gamma=8.88$, which are slightly higher than those reported by \citet{khatoon2022temporal}. For the quiescent state, we find a significantly low Lorentz factor of $\Gamma=3.94$, which likely contributes to the reduced flux level.

\subsection{PKS 0035-252}\label{sec:PKS0035_SED_modeling}
We constructed a broadband spectral energy distribution (SED) using multi-wavelength observations collected during the period MJD 58312–58356. During this interval, the maximum observed $\gamma$-ray flux was approximately $\sim 7 \times 10^{-7} \ \mathrm{ph \ cm^{-2} \ s^{-1}}$, indicating a relatively low-flux state as compared to other bright sources. Following a consistent modeling approach, we fixed several parameters: the viewing angle $\theta = 2^{\circ}$, accretion disc temperature $T_{\mathrm{disc}} = 1 \times 10^5$ K, and disc luminosity $L_{\mathrm{disc}} = 1 \times 10^{45} \ \mathrm{erg,s^{-1}}$ and estimated the sizes of external photon fields. The observed magnetic field strength ($B$) and bulk Lorentz factor ($\Gamma$) were found to be 0.22 G and 4.21, respectively. The resulting energy densities and other derived physical parameters are summarized in Table~\ref{tab:SEDresults}. The modeled SED is shown in Figure~\ref{Fig-JETSET_PKS0035-252}.

\section{Jet power}\label{sec:JET_power}
We have also estimated the total jet luminosity, which is defined as \citep{celotti2008power}:

\begin{align}
L_{\mathrm{total}} &= L_{\mathrm{e}} + L_{\mathrm{p}} + L_{\mathrm{B}} + L_{\mathrm{r}} \nonumber \\
                   &= \pi R^2 \Gamma^2 c (U_{\mathrm{e}} + U_{\mathrm{p}} + U_{\mathrm{B}} + U_{\mathrm{r}})
\label{eq:total_jet_power}
\end{align}

where $L_{\mathrm{e}}$, $L_{\mathrm{p}}$, $L_{\mathrm{B}}$, and $L_{\mathrm{r}}$ represent the power carried by electrons, protons, magnetic fields, and radiation, respectively, while $U_{\mathrm{e}}$, $U_{\mathrm{p}}$, $U_{\mathrm{B}}$, and $U_{\mathrm{r}}$ denote their corresponding energy densities. The derived values of these parameters for each flux state are presented in Table~\ref{tab:SEDresults}. For PKS 1424-41, the electron and magnetic field luminosities are notably high during both flaring states. In particular, the ratios $L_{\mathrm{e}}/L_{\mathrm{B}} \sim 325$ for Flare 1 and $\sim 19$ for Flare 2 indicate a particle-dominated jet. In contrast, during the quiescent state, the ratio $L_{\mathrm{e}}/L_{\mathrm{B}} \sim 2.1$ suggests that the system approaches equipartition, but still particle dominated. As expected, the total jet power is marginally higher during flaring episodes compared to the low-flux state. 
%Furthermore, during flaring, the total jet luminosity exceeds the disk luminosity, whereas in the quiescent phase, it remains slightly below the disk luminosity \citep{ghisellini2014power}.

From the SED modeling of PKS 0736+01, the derived $L_{\mathrm{e}}/L_{\mathrm{B}}$ ratios for Flare 1, Flare 2, and the quiescent state are 2.83, 10.32, and 0.36, respectively. These values indicate a mildly particle-dominated jet during the flaring episodes and a magnetically dominated jet in the quiescent phase. For PKS 0208-512, the electron-to-magnetic field luminosity ratios are found to be 0.066 for Flare 1, 0.093 for Flare 2, and 82 for the quiescent state. This clearly suggests that the jet is magnetic field dominated during high-activity states, while it becomes strongly particle dominated during the low-flux state. For PKS 0035–252, the electron and magnetic field energy densities are comparable, with values of 0.21 and 0.57 erg/cm, respectively; however, it is dominated by the magnetic field.

\section{Minimum Doppler factor}\label{sec:min_Doppler_factor}

The minimum Doppler factor $\delta_{\mathrm{min}}$ can be estimated numerically using a $\delta$-function approximation for the $\gamma\gamma$ opacity constraint and detected high-energy $\gamma-$ray photons \citep{dondi1995gamma, ackermann2010fermi}. The expression of $\delta_{\mathrm{min}}$ is given as

\begin{equation}
    \delta_{\mathrm{min}}\sim \left[ \frac{\sigma_{\mathrm{T}} \ d_{\mathrm{L}}^2 \ (1+z)^2 \ \varepsilon \ F_{\mathrm{X-ray}}}{4 \ t_{\mathrm{var}} \ m_e \ c^4}  \right]^{1/6} \ \ \ ,
\end{equation}

where $\sigma_{\mathrm{T}}$ is the Thomson scattering cross-section for electron ($6.65\times 10^{-25} \ \mathrm{cm^2}$), $d_{\mathrm{L}}$ is the cosmology distance parameter $d_L$\footnote{Throughout this paper, we assume a standard $\Lambda \ \mathrm{CDM}$ cosmology with $H_0 = 71 \ \mathrm{km \ s^{-1} \ Mpc^{-1}}$, $ \Omega_m = 0.3$, and $\Omega_{\Lambda} = 0.7$ and derived cosmological distance $d_L$ using Astropy \textit{FlatLambdaCDM} package} for the source estimated using Astropy package\footnote{\url{https://docs.astropy.org/en/stable/api/astropy.cosmology.FlatLambdaCDM.html}}, $z$ is the redshift, $F_{X-ray}$ is the X-ray flux in the energy range (0.3 - 10.0 keV), $t_{\mathrm{var}}$ is the shortest variability timescale, and $\varepsilon$ is the energy of the highest-energy photon observed scaled by $m_ec^2$(0.511 MeV). In this analysis, we assumed the optical depth for $\gamma$-ray photon absorption via pair production to be $\tau_{\gamma\gamma} = 1$ for the highest-energy detected photon. The shortest variability timescale ($t_{\mathrm{var}}$) was determined for those blazars exhibiting peak $\gamma$-ray fluxes exceeding $1\times10^{-6} \ \mathrm{ph \ cm^{-2} \ s^{-1}}$, as described in Sect.\ref{sec:doubling}. For PKS 1424-41, the fastest variability was observed during Flare 2 with $t_{\mathrm{var}} \sim 3$ h (see Table\ref{tab:Fastest_variability}), a luminosity distance of $d_{\mathrm{L}} = 10.93$ Gpc at redshift $z = 1.52$, X-ray flux $F_{\mathrm{X-ray}} = 4.56 \times 10^{-12} \ \mathrm{erg \ cm^{-2} \ s^{-1}}$, and the highest detected photon energy $E_{\mathrm{max}} = 62.24$ GeV. Using these parameters, we estimated the minimum Doppler factor as $\delta_{\mathrm{min}} \sim 20.93$.

Similarly, for PKS 0736+01 during Flare 2, with $t_{\mathrm{var}} \sim 2.73$ h, $d_{\mathrm{L}} = 912.57$ Mpc, $z = 0.19$, $F_{\mathrm{X-ray}} = 6.85 \times 10^{-12} \ \mathrm{erg \ cm^{-2} \ s^{-1}}$, and $E_{\mathrm{max}} = 6.3$ GeV, we derived $\delta_{\mathrm{min}} \sim 5.29$. For PKS 0208-512 during Flare 1, the observed values are $t_{\mathrm{var}} \sim 9.83$ h, $d_{\mathrm{L}} = 6.53$ Gpc, $z = 1.003$, $F_{\mathrm{X-ray}} = 1.20 \times 10^{-12} \ \mathrm{erg \ cm^{-2} \ s^{-1}}$, and $E_{\mathrm{max}} = 6.319$ GeV, resulting in $\delta_{\mathrm{min}} \sim 7.36$.

The derived values of $\delta_{\mathrm{min}}$ indicate that a minimum Doppler factor is required for the $\gamma$-ray flux to reach the observed levels. The Lorentz factors obtained from the SED modeling are found to exceed these lower bounds, thereby providing important constraints on both the location of the emission region and the bulk Lorentz factor of the jet (see Sect.~\ref{sec:constraint_r_Gamma}).

\section{Constraining the location of gamma-ray emission site}\label{sec:constraint_r_Gamma}

\begin{figure*}
    \centering
    \includegraphics[width=0.49\textwidth]{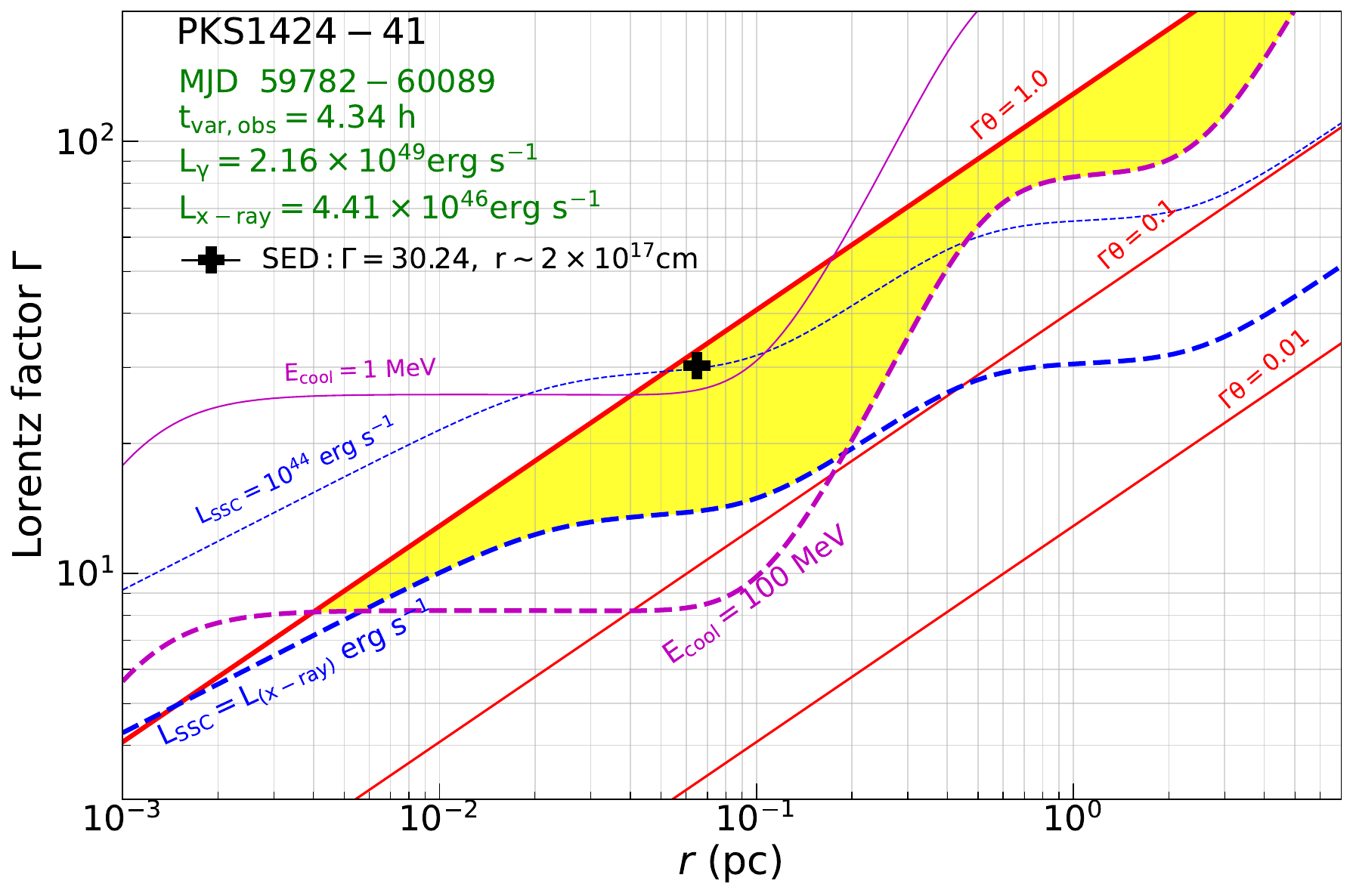}\hspace{1pt}
    \includegraphics[width=0.49\textwidth]{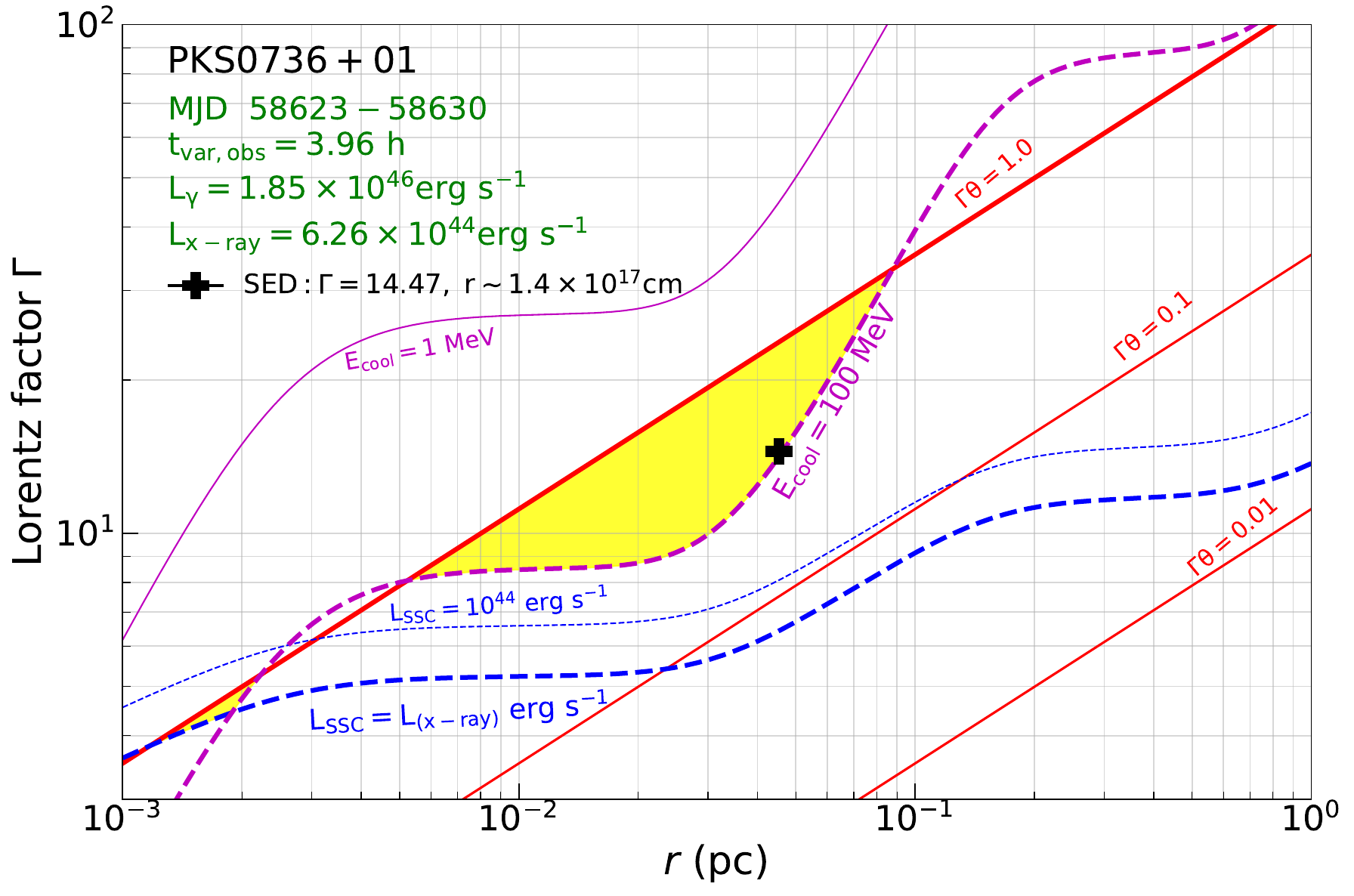}\vspace{1pt}
    \includegraphics[width=0.49\textwidth]{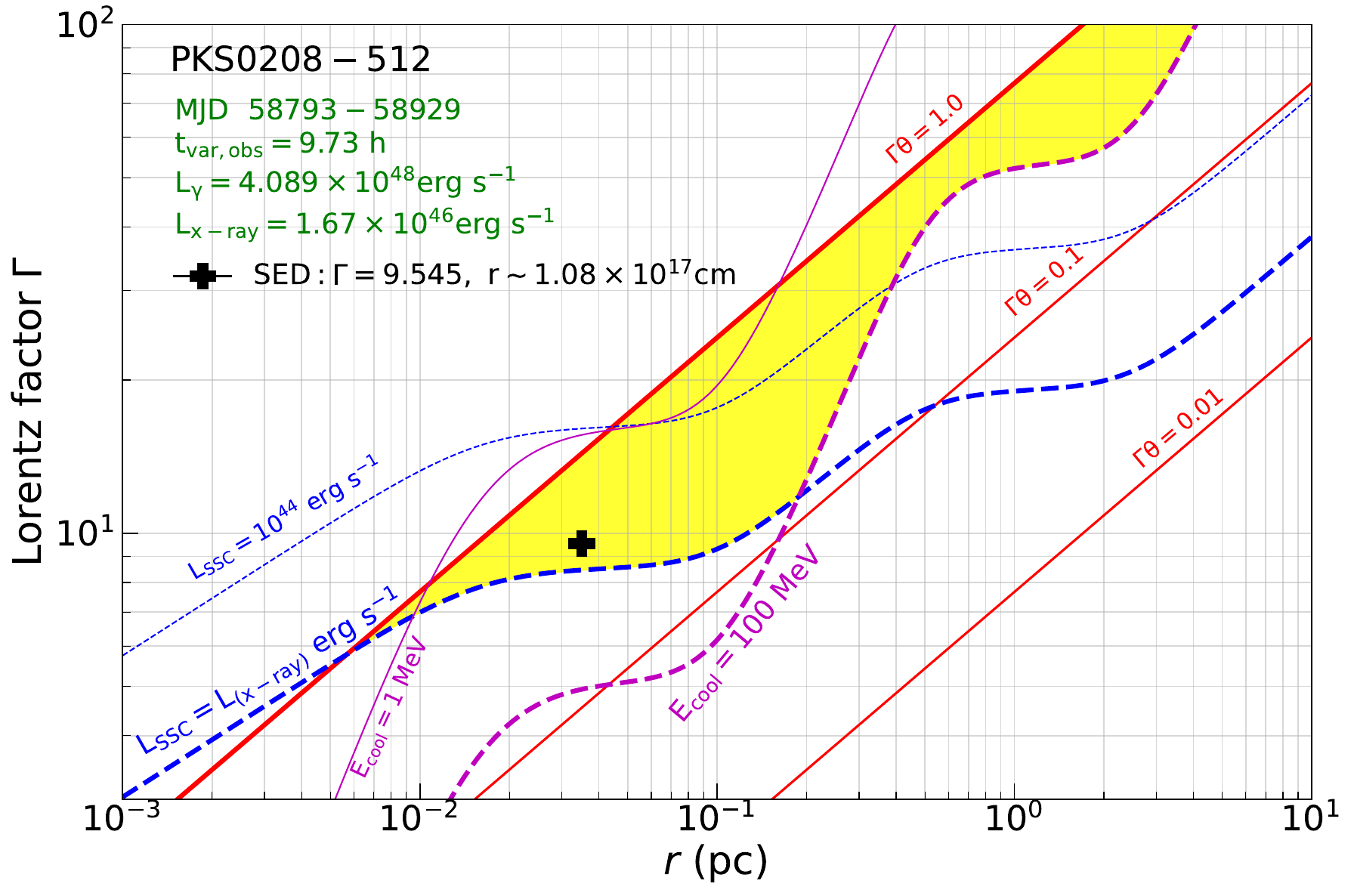}
    \caption{Constraints on the parameter space of the location ($r$) and bulk Lorentz factor ($\Gamma$) of the emitting region responsible for the bright $\gamma$-ray flares of blazars, determined using the shortest variability timescale (Table~\ref{tab:Fastest_variability}) within the respective flare durations. For PKS 1424-41, the brightest flare was observed during MJD 59782–60089; for PKS 0736+01, during MJD 58623–58630; and for PKS 0208-512, during MJD 58793–58929. In all three panels: The solid red lines represent the jet collimation constraint ($\Gamma\theta < 1$). The blue dashed lines indicate the synchrotron self-Compton (SSC) luminosity constraint ($L_{\mathrm{SSC}} < L_{\mathrm{X}}$) and the magenta lines correspond to the cooling condition $E_{\mathrm{cool,obs}} < 100$ MeV. The yellow shaded region defines the allowed parameter space for the emission site ($r$) and $\Gamma$. The black “+” symbols within the shaded region denote the values of $r$ and $\Gamma$ derived from broadband SED modeling for the respective bright flares, Table~\ref{tab:SEDresults}.  }
    \label{Fig-Lorentz_factor_r_relation}    
\end{figure*}

The localization of $\gamma$-ray emission sites within blazar jets remains a long-standing and critical problem in understanding the mechanisms behind $\gamma$-ray production under favorable environmental conditions. Previous studies, such as \citet{chatterjee2013implications} and \citet{nalewajko2014constraining}, have investigated $\gamma$-ray flares in blazars and established constraints on the distance of the emission region from the central supermassive black hole (SMBH), as well as the bulk Lorentz factor $\Gamma$. 
%of the $\gamma$-ray emitting region in luminous blazars.

Several theoretical models have been proposed to explain energy dissipation and particle acceleration within blazar jets. However, the precise localization of the $\gamma$-ray emission sites remains unresolved, with estimates spanning several orders of magnitude.

In this work, we explore the parameter space defined by the location $r$ and the Lorentz factor $\Gamma$ of the emitting regions responsible for major $\gamma$-ray flares in blazars. Our analysis incorporates various observable quantities, including the $\gamma$-ray luminosity ($L_{\gamma}$), the shortest observed variability timescale in the $\gamma$-ray band ($t_{\mathrm{var, \ obs}}$), synchrotron luminosity ($L_{\mathrm{syn}}$), X-ray luminosity ($L_X$), and accretion disc luminosity ($L_d$). Additionally, we consider key assumptions such as the Doppler-to-Lorentz factor ratio ($\mathcal{D}/\Gamma$) and the covering factors of external radiation fields ($\xi_{\mathrm{BLR}}, \ \xi_{\mathrm{DT}}$). These inputs allow us to place constraints in the $(r, \Gamma)$ parameter space based on several physical considerations, including the jet collimation condition ($\Gamma \theta$), synchrotron self-Compton (SSC) luminosity ($L_{\mathrm{SSC}}$), and the observed energy of external radiation Comptonization (ERC) photons corresponding to the threshold for efficient electron cooling ($E_{\mathrm{cool,\ obs}}$).\par

We consider an emitting region located at a distance $r$ from the central engine, viewed at an angle $\theta_{\mathrm{obs}}$ relative to the observer's line of sight. The bulk Lorentz factor of the region is given by $\Gamma = \left(1 - \beta^2\right)^{-1/2}$, where $\beta = v/c$ is the dimensionless jet velocity. The Doppler factor of the observed radiation is defined as: $\mathcal{D} = \left[ \Gamma \left( 1 - \beta \ \mathrm{cos}\theta_{\mathrm{obs}}  \right)   \right]^{-1}$. The characteristic size $R$ of the emitting region is related to the co-moving variability timescale, expressed as: $R \simeq c \ t_{var}^{'}$, where $t_{var}^{'} = \mathcal{D} \ t_{var, \ obs} \ / (1+z) $, where $z$ is the source redshift and $t_{\mathrm{var,\ obs}}$ is the shortest observed variability timescale.

Additionally, the size of the emitting region can be approximated as $R \simeq \theta r$, where $\theta$ is the opening angle of the emitting region, distinct from the jet opening angle $\theta_{\mathrm{j}}$, with the physical constraint $\theta \leq \theta_{\mathrm{j}}$.

In blazars, the Doppler factor $\mathcal{D}$ is typically of the same order as the Lorentz factor $\Gamma$, although the ratio $\mathcal{D}/\Gamma$ introduces significant uncertainty in constraining the distance $r$ as a function of $\Gamma$. It's important to note that the Lorentz factor of the emitting region, $\Gamma$, may not necessarily coincide with the bulk Lorentz factor of the jet, $\Gamma_{\mathrm{j}}$. However, for simplicity, we often assume $\Gamma \simeq \Gamma_{\mathrm{j}}$.

For a compact emitting region with a viewing angle $\theta_{\mathrm{obs}} \simeq 1/\Gamma$, we typically find $\mathcal{D}/\Gamma \simeq 1$. In the extreme case of $\theta_{\mathrm{obs}} \simeq 0$, this ratio increases to $\mathcal{D}/\Gamma \simeq 2$. In reality, the emitting regions within a blazar jet may span a range of viewing angles, leading to a broad range of possible $\mathcal{D}/\Gamma$ values. Putting now all these assumptions and relations, we constraint the emission region $r$ and the Lorentz factor $\Gamma$.

\subsection{Collimation Constraint}

For convenience, we define the collimation efficiency parameter as the product $\Gamma \theta$, which combines the effects of jet collimation and relativistic beaming. We can now write an expression for the Lorentz factor as a function of $ \Gamma \theta$:

\begin{equation}
    \Gamma \left( r, \ \Gamma \theta \right)\simeq \left( \frac{\mathcal{D}}{\Gamma}  \right)^{-1/2} \left[ \frac{(1+z) \ \Gamma \theta \ r}{c \ t_{\mathrm{var, \ obs}}} \right]^{1/2}
    \label{eq:collimation}
\end{equation}

The are strong theoretical and observational indications that $\Gamma \theta <1$ for AGN jets \citep{pushkarev2017mojave, komissarov2009magnetic}, references are therein.

\subsection{SSC Constraint}

In addition to constraint on $r$ and $\Gamma$ based on the collimation parameter $\Gamma \theta$, a further constraint can be placed based on the SSC luminosity $L_{\mathrm{SSC}}$. In $\gamma$-ray emission via comptonization of external radiation by the relativistic electrons, the observable luminosities - the peak luminosity of the ERC component $L_{\mathrm{ERC}}$, $L_{\mathrm{syn}}$, and $L_{\mathrm{SSC}}$ - can be related to the energy densities of external radiation in co-moving frame $u_{\mathrm{ext}}^{'}$, magnetic field $u_{\mathrm{B}}^{'} = B^{'2}/8\pi$, and synchrotron radiation $u_{\mathrm{syn}}^{'} \simeq L_{\mathrm{syn}}/(4\pi c \ \mathcal{D}^4 R^2)$, respectively. The SSC luminosity component is estimated utilizing the relation $L_{\mathrm{SSC}}/L_{\mathrm{syn}}\simeq g_{\mathrm{SSC}} (u_{\mathrm{syn}}^{'}/u_{\mathrm{B}}^{'})$, where $g_{\mathrm{SSC}} \simeq 3/4$ is a correction factor \citep{nalewajko2014constraining}. The compton dominance parameter $q = L_{\gamma}/L_{syn} = g_{\mathrm{ERC}}(\mathcal{D}/\Gamma)^2 (u_{\mathrm{ext}}^{'}/u_{\mathrm{B}}^{'}))$ , where $g_{\mathrm{ERC}} \simeq 1/2$ is a correction factor mainly due to Klein-Nishina effects and co-moving energy density of external radiation $u_{\mathrm{ext}}^{'}\simeq \zeta(r) \Gamma^2 L_{\mathrm{d}} / (3 \pi c r^2) $, here  $L_{\mathrm{d}}$ represents the accretion disc luminosity and $\zeta(r)$ is related to the composition of external radiation fields, involving the contributions from the BLR, the dusty torus, and the accretion disc. The $\zeta$ is defined as a function of $r$ given as:

\begin{equation}
    \zeta(r)\simeq \frac{0.4 \ \xi_{\mathrm{BLR}} \ (r/r_{\mathrm{BLR}})^2}{1+(r/r_{\mathrm{BLR}})^4} + \frac{0.4 \ \xi_{\mathrm{IR}} \ (r/r_{\mathrm{IR}})^2}{1+(r/r_{\mathrm{IR}})^4} + \frac{0.21 R_g}{r}
    \label{eq:zeta}
\end{equation}
where $\xi_{\mathrm{BLR}}$ and $\xi_{\mathrm{IR}}$ are the covering factors of the BLR of characteristic radius $r_{\mathrm{BLR}}$ and the dusty torus with characteristic radius $r_{\mathrm{IR}}$, and $R_g$ represents the gravitational radius of the SMBH. We used the following scaling laws to estimate the $r_{\mathrm{BLR}}\simeq 0.1 \ \left(\frac{L_{\mathrm{d}}}{10^{46} \ \mathrm{erg \ s^{-1}}}\right)^{1/2}$ pc and $r_{\mathrm{IR}}\simeq 2.5 \ \left(\frac{L_{\mathrm{d}}}{10^{46} \ \mathrm{erg \ s^{-1}}}\right)^{1/2}$ pc \citep{sikora2009constraining}. Putting all relations mentioned above, we obtain a constraint on $\Gamma$ \citep{chatterjee2013implications, nalewajko2014constraining}:

\begin{align}
\Gamma (r, L_{\mathrm{SSC}}) &\simeq \left[ 3 \left(  \frac{g_{\mathrm{SSC}}}{g_{\mathrm{ERC}}}  \right) 
\left(  \frac{L_{\mathrm{syn}}}{L_{\mathrm{SSC}}}  \right) 
\left(  \frac{L_{\gamma}}{\zeta(r) L_{\mathrm{d}}}  \right) \right]^{1/8} \nonumber \\
&\quad \times \left( \frac{\mathcal{D}}{\Gamma} \right)^{-1} 
\left[ \frac{(1+z)r}{2 \ c \ t_{\mathrm{var, \ obs}}} \right]^{1/4}
\label{eq:SSC}
\end{align}

In blazars, high-energy electrons in the jet first emit synchrotron radiation in a broad  range. These same electrons can then scatter the synchrotron photons up to X-ray or gamma-ray energies. However, in many blazars, the observed X-ray emission is too hard to be explained by SSC alone. This suggests that another mechanism, likely External Compton (EC), might be dominating the X-ray emission. Thus, our $SSC$ constraint is defined as $L_{\mathrm{SSC}} \lesssim L_{\mathrm{X}}$.

\subsection{Cooling Constraint}

Blazars exhibit rapid and intense gamma-ray flares with nearly time-symmetric profiles, suggesting that the high-energy electrons responsible for this emission cool very efficiently. In luminous blazars, this radiative cooling is primarily governed by the external radiation Compton (ERC) process, with a characteristic cooling timescale given by $t_{\mathrm{cool}}^{'}(\gamma) \simeq 3 m_e c / (4 \sigma_{\mathrm{T}} \gamma u_{\mathrm{ext}}^{'})$, where $\gamma$ denotes the electron Lorentz factor (note that this symbol specifically refers to electrons when used in this context). For flares with time-symmetric shapes, the cooling timescale must be comparable to or shorter than the observed variability timescale, i.e., $t_{\mathrm{cool}}^{'}(\gamma) \lesssim t_{\mathrm{var}}^{'}$. By equating these timescales, one can define a characteristic electron Lorentz factor $\gamma_{\mathrm{cool}}$, such that $t_{\mathrm{cool}}^{'}(\gamma_{\mathrm{cool}}) \simeq t_{\mathrm{var}}^{'}$ This corresponds to an observed ERC photon energy: $E_{\mathrm{cool,\ obs}} \simeq \mathcal{D} \Gamma \gamma_{\mathrm{cool}}^{2} E_{\mathrm{ext}}(r) / (1+z)$. By combining these expressions, we obtain a constraint on the bulk Lorentz factor  $\Gamma$, which depends on the variability timescale, external photon field, and location of the emission region.

\begin{align}
\Gamma (r, \ E_{\mathrm{cool, \ obs}}) &\simeq \left( \frac{\mathcal{D}}{\Gamma} \right)^{-1/4} 
\left[ \frac{9 \pi m_e c^2 r^2}{4 \sigma_{\mathrm{T}}\zeta(r) L_{\mathrm{d}} t_{\mathrm{var, \ obs}}} \right]^{1/2} \nonumber \\
&\quad \times \left[ \frac{(1+z) E_{\mathrm{ext}}(r)}{E_{\mathrm{cool, \ obs}}} \right]^{1/4}
\label{eq:cooling}
\end{align}

Since Fermi-LAT observations are typically analyzed over the energy range of 100 MeV to 300 GeV, the cooling constraint is imposed by requiring the observed ERC photon energy to satisfy $E_{\mathrm{cool, \ obs}} \lesssim 100$ MeV. 

We refer to \citet{nalewajko2014constraining} for a comprehensive discussion on the constraints related to the location $r$ and Lorentz factor $\Gamma$ of the gamma-ray emitting region. In our analysis, we explored the parameter space in the ($r, \Gamma$) plane by applying three key constraint relations: (\MakeUppercase{\romannumeral 1}) an upper limit on the collimation parameter, $\Gamma \theta \lesssim 1$; (\MakeUppercase{\romannumeral 2}) an upper limit on the synchrotron self-Compton (SSC) luminosity, $L_{\mathrm{SSC}} \lesssim L_{\mathrm{X}}$; and (\MakeUppercase{\romannumeral 3}) a constraint on the efficient electron cooling condition, requiring the observed ERC photon energy to satisfy $E_{\mathrm{cool, \ obs}} \lesssim 100$ MeV. \par 

We applied these constraints to blazars in our sample that exhibited prominent gamma-ray flares with contemporaneous multi-wavelength coverage. The findings are discussed below.

\begin{itemize}
    \item \textit{PKS 1424-41} ($z = 1.522$, $d_L = 10.93$ Gpc): The brightest flare, we referred as Flare 2 in this paper, ever detected from this source by the Fermi-LAT during the period MJD 59782-60089. To estimate the shortest variability timescale, we analyzed the light curve using the doubling/halving method described in Section~\ref{sec:doubling}. We identified a minimum doubling timescale of $t_{\mathrm{var, \ obs}} \sim 3.0\pm 0.9$ h at $\sim$MJD 59935. The corresponding gamma-ray luminosity was estimated to be $L_{\gamma} \sim 2.16\times 10^{49} \  \mathrm{erg \ s^{-1}}$. This flare was accompanied by simultaneous multi-wavelength observations, including X-ray and UVOT data. From the X-ray data, we measured a flux of $F_{\mathrm{X}} = 4.049\times 10^{-12} \ \mathrm{erg \ cm^{-2} \ s^{-1}}$ and spectral index of 1.70. Following that, we derived the X-ray luminosity $L_{\mathrm{X}}\sim 4.41\times 10^{46} \ \mathrm{erg \ s^{-1}}$ \citep{worrall2009x}. The $L_{\mathrm{syn}}$ was estimated from SED modeling to be $3.33\times 10^{44} \ \mathrm{erg \ s^{-1}}$. For the external photon fields, we adopted a disk luminosity of $L_d\sim 1\times 10^{46} \ \mathrm{erg \ s^{-1}}$, also used in SED modeling, and characteristic sizes of $r_{\mathrm{BLR}}\sim 0.1$ pc and $r_{\mathrm{IR}}\sim 2.5$ pc for the BLR and IR torus, respectively. Using the expressions for collimation (\ref{eq:collimation}), SSC luminosity (\ref{eq:SSC}), and cooling constraints (\ref{eq:cooling}), we mapped the allowed region in ($r$, $\Gamma$) parameter space. The top left panel of Figure~\ref{Fig-Lorentz_factor_r_relation} shows the yellow-shaded region representing the parameter space allowed by the three constraints: $\Gamma\theta <1$, $L_{\mathrm{SSC}} \lesssim L_{\mathrm{X}}$, and $E_{\mathrm{cool, \ obs}} \lesssim 100 $ MeV, assuming $\xi_{\mathrm{BLR}} \simeq \xi_{\mathrm{IR}} \simeq 0.1$, $\mathcal{D}/\Gamma = 1$, as described in \citep{nalewajko2014constraining}. We adopted a black hole mass of $M_{\mathrm{BH}}\simeq 4.5\times 10^9 \ M_{\odot}$ \citep{abhir2021multi}.
    
   In our SED modeling using JetSet, we treated the bulk Lorentz factor $\Gamma$ as a free parameter. The best-fit values of the emitting region location and Lorentz factor were found to be $r\sim 2\times 10^{17}$ cm and $\Gamma\sim$30.24, respectively, which fall within the constrained region. \citet{nalewajko2014constraining} further explored the sensitivity of these constraints by tuning the physical parameters. Notably, while previous studies assumed variability timescales on the order of a few days, our analysis reveals a much shorter timescale in $\gamma$-rays of order of a few hours, indicating that it likely originated from a highly compact region located at a distance within BLR. 

   \item \textit{PKS 0736+01} ($z = 0.189$, $d_L = 912.57 $ Mpc): This blazar exhibits repetitive flaring events with a characteristic timescale of $\sim 4$ years, as recently reported by \citep{sharma2025exploring}. The most prominent flare occurred during the period MJD 58623–58630, reaching its peak flux. A detailed analysis of the $\gamma$-ray light curve revealed the shortest variability timescale to be $2.73 \pm 0.63$ h, observed around MJD 58619. The derived luminosities are: $L_{\gamma} \sim 1.85\times 10^{46} \ \mathrm{erg \ s^{-1}}$, $L_{\mathrm{syn}} \sim 4.37\times 10^{42} \ \mathrm{erg \ s^{-1}}$, and $L_{\mathrm{X}} \sim 6.26\times 10^{44} \ \mathrm{erg \ s^{-1}}$. We adopted a disk luminosity of $L_d \sim 1\times 10^{45} \ \mathrm{erg \ s^{-1}}$ for this study. The black hole mass has been estimated in the literature to lie in the range $10^8$–$10^{8.7} \ M_{\odot}$ \citep{wandel1991eddington, mclure2001black, woo2002active, marchesini2004transition, dai2007correlation, xiong2014intrinsic, abdalla2020hess, zhang2024fundamental}. In our analysis, we adopted a black hole mass of $M_{\mathrm{BH}} \sim 10^8 \ M_{\odot}$. As in previous cases, we assumed $\mathcal{D}/\Gamma = 1$ and $\xi = 0.1$ for both the BLR and IR radiation fields, with characteristic radii of $r_{\mathrm{BLR}} \simeq 0.031$ pc and $r_{\mathrm{IR}} \simeq 0.79$ pc, respectively. The location of the $\gamma$-ray emitting region and the bulk Lorentz factor, derived from SED modeling of Flare 2 using multi-wavelength observations, are $r \sim 1.4\times 10^{17}$ cm and $\Gamma \sim 14.47$, respectively. The finding lies within the yellow shaded region in the corresponding panel of Figure~\ref{Fig-Lorentz_factor_r_relation} (top right panel), consistent with the physical constraints discussed. The corresponding location of $\gamma$-ray flare with the Lorentz factor from our analysis is consistent with the findings reported by \citep{abdalla2020hess}, which were carried out during the time frame from MJD 57066 - 57080, referred to as Flare 1 in our study. This suggests that the flares occurred under similar environmental conditions and were produced via similar underlying physical mechanisms. 

   \item \textit{PKS 0208-512} ($z = 1.003$, $d_L = 6.53 $ Gpc): The blazar exhibits repetitive flaring events with an approximate recurrence interval of $\sim 2.3$ years \citep{sharma2025exploring}. We derived constraints on the physical parameters $r$ and $\Gamma$ for the brightest flare, which occurred around MJD 58794. The estimated luminosities are: $L_{\gamma} \sim 4.08\times 10^{48} \ \mathrm{erg \ s^{-1}}$, $L_{\mathrm{syn}} \sim 1.20\times 10^{44} \ \mathrm{erg \ s^{-1}}$, and $L_{\mathrm{X}} \sim 1.47\times 10^{46} \ \mathrm{erg \ s^{-1}}$. The shortest observed variability timescale during this flare is $9.83 \pm 2.65$ days.
   For this analysis, we adopted the following parameters: $L_d \sim 1\times 10^{46} \ \mathrm{erg \ s^{-1}}$, $\xi \simeq 0.1$, $r_{\mathrm{BLR}} \simeq 0.1$ pc, $r_{\mathrm{IR}} \simeq 2.5$ pc, $\mathcal{D}/\Gamma = 1$, and a black hole mass of $M_{\mathrm{BH}} \simeq 1.6\times 10^{9} \ M_{\odot}$ \citep{nalewajko2014constraining}. The SED modeling of Flare 2 revealed the location of the $\gamma$-ray emitting region and the bulk Lorentz factor to be $r \sim 1.08\times 10^{17}$ cm and $\Gamma \sim 9.54$, respectively. The finding corresponds to these physical parameters falling within the yellow shaded region in the bottom panel of Figure~\ref{Fig-Lorentz_factor_r_relation}, confirming consistency with the applied constraints.
\end{itemize}

\section{Discussion and conclusion}\label{sec:discussion}
In this work, we present the results of a comprehensive multi-wavelength analysis of the blazars PKS 1424-41, PKS 0736+01, PKS 0208-512, and PKS 0035-252. The study utilizes Fermi-LAT data collected between August 5, 2008, and April 1, 2025, along with Swift-XRT and UVOT observations obtained simultaneously. The analysis includes temporal and cross-correlation studies, as well as an estimation of the shortest variability timescale using the doubling/halving method. The shortest timescales are then used to constrain the size and location of the $\gamma$-ray emission regions from the central SMBH. Additionally, we performed simultaneous broadband spectral energy distribution (SED) modeling, which allowed us to derive constraints on the emission region's distance $r$ and the bulk Lorentz factor $\Gamma$. A detailed discussion of these results is provided in the sections below.\par

We conducted a fractional variability analysis across multiple energy bands, with the results illustrated in Figure~\ref{Fig-Fvar} and summarized in Table~\ref{tab:Fvar}. For PKS 1424$-$41, the optical–UV and $\gamma$-ray bands exhibit comparable, yet significantly higher, fractional variability amplitudes relative to the X-ray band. A noticeable dip in the X-ray variability contributes to a characteristic double-hump structure in the $F_{\mathrm{var}}$ distribution. Such a feature has also been reported in other blazars \citep{abdo2010gamma, rani2017probing, prince2022multiwavelength, sharma2024probing, tantry2025study}, and is thought to reflect differences in the emission mechanisms dominating each spectral regime. For PKS 0736+01, the highest fractional variability amplitude is observed in the $\gamma$-ray band. The optical–UV bands show slightly greater variability than the X-rays, yet the overall $F_{\mathrm{var}}$ profile retains a double-hump morphology, similar to that typically observed in the broadband spectral energy distributions (SEDs) of blazars. A similar pattern in the variation of fractional variability amplitudes has been observed for both PKS 0208–512 and PKS 0035–252.\par

The symmetric profiles of both prominent and sub-prominent flares in the $\gamma$-ray light curves were characterized using a sum-of-exponentials function, as described in Sect.\ref{sec:doubling}. The findings are summarized in Table~\ref{tab:flare_fitting_results} and the correlation between rise and decay timescales is illustrated in Figure~\ref{Fig-Rise_decay_timescale}.\par

In addition, the shortest variability timescales were estimated using the doubling/halving method. The fastest $\gamma$-ray variability timescales observed for PKS 1424-41, PKS 0736+01, and PKS 0208-512 are $3.0\pm0.9$ h, $2.73\pm0.63$ h, and $9.83\pm2.65$ h, respectively. Based on these timescales, we derived the corresponding locations and sizes of the $\gamma$-ray emitting regions. For PKS 1424-41, the emission region is located at a distance of $\sim6.05\times10^{16}$ cm from the central engine, with a characteristic size of $\sim1.97\times10^{15}$ cm. In the case of PKS 0736+01, the emission site is found at $\sim1.43\times10^{16}$ cm, with an emitting region radius of $\sim4.21\times10^{15}$ cm. While for PKS 0208-512, the emission region is located farther out at $\sim7.2\times10^{17}$ cm, with a size scale of $\sim1.3\times10^{16}$ cm. The timescales are tabulated in Table~\ref{tab:Fastest_variability}.\par

We further examined the correlation between fluxes and photon indices of the $\gamma$-ray emissions, see Figure~\ref{Fig-Index_flux_correlation}. This analysis was carried out using Spearman’s rank correlation method, and the results are summarized in Table~\ref{tab:correlation_results}. To gain deeper insights into the origin and interplay of emissions across different energy bands, we also conducted a cross-correlation analysis between $\gamma$-ray and optical-UV light curves. For this, we employed the interpolated cross-correlation function (ICCF) technique, which is particularly effective in estimating time lags between unevenly sampled time series. A detailed methodology and interpretation of these results are presented in Sect.\ref{sec:ICCF} \citep{2018ascl.soft05032S}, \citep{sharma2025searching}, with the derived time lags listed in Table\ref{tab:ICCF}. The corresponding cross-correlation profiles are visually illustrated in Figures~\ref{Fig-CCF_PKS1424-41}.\par

We carried out broadband SED modeling of blazars PKS 1424-41, PKS 0736+01, PKS 0208-512, and PKS 0035-252 across multiple flux states, including flaring and quiescent states. For PKS 1424-41 in particular, two prominent flaring periods were identified during MJD 56299–56411 and MJD 59782–60089, along with a quiescent phase between MJD 56751–56920, based on a Bayesian block analysis. A detailed discussion of the SED modeling for this source is provided in Sect.\ref{sec:PKS1424_SED_modeling}, with key parameter variations illustrated in Figure\ref{Fig-JETSET_parameters_variations} and summarized in Table~\ref{tab:SEDresults}.

For PKS 1424-41, the temporal evolution of the magnetic field strength, as shown in the first panel in the first column of Figure~\ref{Fig-JETSET_parameters_variations}, reveals a noticeable increasing trend during flaring states, which is indicative of enhanced synchrotron emission. The estimated magnetic field values are consistent with those reported by \citet{abhir2021multi}. The second panel of first column of  Figure~\ref{Fig-JETSET_parameters_variations} presents the variation in the jet's bulk Lorentz factor ($\Gamma$), showing significantly higher values during flaring episodes compared to the quiescent state, consistent with expectations for relativistic jet behavior during enhanced activity. It is worth noting that \citet{abhir2021multi} assumed a fixed $\Gamma$ in their modeling. The third panel of the first column displays changes in the size of the emission region. Here, a larger emission region is observed during the quiescent phase, while the flaring states correspond to more compact zones, implying that the flares likely originate from smaller, more localized regions within the jet. The localization of the emission regions based on broadband SED modeling suggests that they lie within or near the BLR, as illustrated in Figure~\ref{Fig-JETSET_PKS1424}. This implies that the relativistic electrons responsible for the high-energy $\gamma$-ray emission interact with a significant number of BLR photons via inverse Compton scattering. For Flare 1, the $\gamma$-ray emission region appears to be located closer to the central supermassive black hole (SMBH) or the accretion disk compared to the other flux states, which is consistent with the observed higher disk photon energy density ($U_{\mathrm{disk}}$) during this period. In contrast, the emission regions during Flare 2 and the quiescent state are located farther from the central engine, likely closer to the BLR. This is supported by the dominance of $U_{\mathrm{BLR}}$ over $U_{\mathrm{DT}}$ in these states (see Table~\ref{tab:SEDresults}).

We also estimated the total jet power across different flux states and present its variation in the bottom panel of first column of Figure~\ref{Fig-JETSET_parameters_variations}. The highest jet luminosity was observed during Flare 2, which corresponds to the brightest flare ever detected from PKS 1424-41, while the lowest jet luminosity was found in the quiescent state. To assess the power output of the jet relative to the Eddington limit, we use the standard relation for Eddington luminosity:

\begin{equation}
L_{\mathrm{Edd}} = 10^{38} \ \frac{M}{M_{\odot}} \ \mathrm{erg \ s^{-1}},
\label{eq:eddinton}
\end{equation}

where M is the mass of the central SMBH. Adopting a black hole mass of $M \sim 4.5\times 10^{9} \ M_{\odot}$ \citep{abhir2021multi}, we estimate $L_{\mathrm{Edd}} \approx 4.5\times10^{47} \ \mathrm{erg \ s^{-1}}$. The total jet luminosity corresponds to $\sim0.91\%$ of $L_{\mathrm{Edd}}$ in the quiescent state, and increases to $\sim2.5\%$ and $\sim5.0\%$ during Flare 1 and Flare 2, respectively. The analysis suggests that the flaring activity is primarily driven by variations in the magnetic field strength, bulk Lorentz factor, size of the emission region, and changes in the external photon fields, particularly $U_{\mathrm{BLR}}$, $U_{\mathrm{DT}}$, and $U_{\mathrm{disc}}$.\par

For PKS 0736+01, the temporal evolution of key jet parameters across different flux states is illustrated in the second column of Figure~\ref{Fig-JETSET_parameters_variations}. The magnetic field strength is found to be higher during the quiescent state compared to the flaring episodes, suggesting a more magnetically dominated environment in the low-activity phase. In contrast, the bulk Lorentz factors ($\Gamma$) during Flare 1 and Flare 2 are nearly identical, and both are significantly higher than in the quiescent state, as expected. \citealt{abdalla2020hess} reported a high magnetic field during the period MJD 57066–57082, with comparable Lorentz factor with observed ones in flaring activity . As observed in other sources, the flares in PKS 0736+01 appear to originate from relatively compact emission regions. The localization of the $\gamma$-ray emission zones, based on SED modeling, indicates that the emission regions during Flare 1 and Flare 2 lie beyond the BLR but within the dusty torus, as shown in Figure~\ref{Fig-JETSET_PKS0736} and tabulated in Table~\ref{tab:SEDresults}. In these flaring states, the external Compton (EC) process is predominantly driven by photons from the dusty torus, contributing significantly to the high-energy $\gamma$-ray emission. Conversely, during the quiescent state, the emission region shifts slightly closer to the BLR, making the BLR photon field more influential in the EC process relative to the torus. The estimated total jet power during the flaring states is found to be comparable between the two flares, whereas it drops by approximately an order of magnitude during the quiescent state. We also calculated the Eddington luminosity using Equation~\ref{eq:eddinton}, assuming a black hole mass of $M \sim 10^8 \ M_{\odot}$, which yields $L_{\mathrm{Edd}} \approx 1\times10^{46} \ \mathrm{erg \ s^{-1}}$.  Comparing this with the jet power estimates, we found that the total jet luminosity in the quiescent state corresponds to approximately $\sim 1.3\%$ of $L_{\mathrm{Edd}}$, while during Flare 1 and Flare 2, it increases to about $\sim 5.18\%$  and $\sim 4.77\%$ of $L_{\mathrm{Edd}}$, respectively.\par

For PKS 0208-512, the temporal evolution of jet parameters across different flux states is shown in the third column of Figure~\ref{Fig-JETSET_parameters_variations}. The magnetic field strength is found to be higher during the flaring states compared to the quiescent state, a similar trend is observed in the bulk Lorentz factors. \citep{khatoon2022temporal} also reported elevated magnetic field values during the period $\sim$ MJD 58800–59000, and the bulk Lorentz factors reported are consistent with our findings. The size of the emission region is larger during the quiescent state, whereas it becomes more compact during flaring events. The location of the emission regions during flaring states is found to lie within the radius of the broad-line region (BLR), where relativistic electrons interact with a dense field of BLR photons to produce high-energy $\gamma$-ray emissions. In contrast, during the quiescent state, the emission region lies outside the BLR but within the dusty torus (DT). Notably, during this low-flux state, the high-energy emission is primarily produced by the same population of relativistic electrons responsible for the synchrotron radiation, as seen in the bottom panel of Figure~\ref{Fig-JETSET_PKS0208}. While modeling, we included contributions from both the BLR and DT; however, their role in high-energy emission was found to be minimal in the quiescent state. To evaluate the jet power in the context of the Eddington luminosity, we estimated $\approx6.91\times 10^{46} \ \mathrm{erg \ s^{-1}}$, using a black hole mass of $M\sim 6.91\times10^{8} \ M_{\odot}$ \citep{khatoon2022temporal}. The total jet luminosity in the quiescent state corresponds to $\sim6.65\%$ of $L_{\mathrm{Edd}}$, while it is $\sim4.35\%$ and $\sim15.62\%$ of $L_{\mathrm{Edd}}$ in Flare 1 and Flare 2, respectively.

For PKS 0035-252, the key jet parameters derived from SED modeling are summarized in Table~\ref{tab:SEDresults}. The location of the emission region is found to lie within the broad-line region (BLR), suggesting that BLR photons significantly contributed to the production of high-energy $\gamma$-ray emission, as illustrated in Figure~\ref{Fig-JETSET_PKS0035-252}. The estimated Eddington luminosity is $L_{\mathrm{Edd}}\approx 1.52\times 10^{45} \ \mathrm{erg \ s^{-1}}$, using the black hole mass of $M\sim 1.51\times 10^7 \ M_{\odot}$ \citep{pei2022estimation}, which is slightly higher than the total jet luminosity inferred from the modeling.

In this work, we aimed to constrain the $\gamma$-ray emission sites ($r$) along the jets relative to the central SMBH, as well as the bulk Lorentz factor $\Gamma$. These constraints were derived based on several key parameters, including the collimation factor $\Gamma\theta$, the SSC luminosity $L_{\mathrm{SSC}}$, and the observed photon energy corresponding to the efficient cooling threshold, $E_{\mathrm{cool, \ obs}}$. To estimate $r$ and $\Gamma$, we utilized multiple observational quantities: the $\gamma$-ray variability timescale $t_{\mathrm{var, obs}}$, $\gamma$-ray luminosity $L_{\gamma}$, synchrotron luminosity $L_{\mathrm{syn}}$, X-ray luminosity $L_{\mathrm{X}}$, and the accretion disk luminosity $L_{\mathrm{d}}$. In addition, we adopted assumptions for the Doppler-to-Lorentz factor ratio $\mathcal{D}/\Gamma$ and covering factors of the external photon fields ($\xi_{\mathrm{BLR}}, \ \xi_{\mathrm{IR}}$).

We performed SED modeling of the brightest flares from luminous blazars, including PKS 1424-41, PKS 0736+01, and PKS 0208-512 in this study, exhibiting well-sampled multi-wavelength data and estimated the parameters required to place constraints on $r$ and $\Gamma$. The most stringent constraints can be achieved under the following conditions: $\Gamma\theta \lesssim 1$, $L_{\mathrm{SSC}} < L_{\mathrm{X}}$, and $E_{\mathrm{cool, obs}} \lesssim 100$ MeV.

Using essential quantities derived from SED modeling and incorporating reasonable assumptions, we constrained the emission location $r$ and Lorentz factor $\Gamma$, as illustrated in Figure~\ref{Fig-Lorentz_factor_r_relation}. For an emission region located at a distance $r$ from the central SMBH and moving with Lorentz factor $\Gamma$, the expected variability timescale, $\sim \frac{r}{\Gamma^2 c}$, corresponds to several hours. This timescale is consistent with the shortest variability timescale detected by Fermi-LAT \citep{tavecchio2010constraining, saito2013very, raiteri2013awakening, pandey2022detection} and those estimated using flux doubling/halving analyses in this study. We present a comprehensive temporal and spectral analysis of blazars using multi-wavelength observations from \textit{Fermi}-LAT and \textit{Swift}-XRT/UVOT. Through broadband spectral modeling of the brightest flares, we constrained the emission region and Lorentz factor in the ($r$, $\Gamma$) parameter space.

%\documentclass{article}

% Define the \aap command to expand to "Astronomy & Astrophysics"
\newcommand{\aap}{Astronomy \& Astrophysics}
\newcommand{\apj}{The Astrophysical Journal}
\newcommand{\ssr}{Space Science Reviews}
\newcommand{\mnras}{Monthly Notices of the Royal Astronomical Society}
\newcommand{\apjl}{The Astrophysical Journal Letters}
\newcommand{\pasp}{Publications of the Astronomical Society of the Pacific}
%\begin{document}

%\cite{example}

% Your bibliography and document content here

%\end{document}

%$\textunderscore$
\section{Acknowledgements}
A. Sharma is grateful to Prof. Sakuntala Chatterjee at S.N. Bose National Centre for Basic Sciences for providing the necessary support to conduct this research.
 
\paragraph{\textbf{Data Availability Statement}}
This research utilizes publicly available data of blazars in our sample obtained from the Fermi-LAT data server provided by NASA Goddard Space Flight Center (GSFC): \url{https://fermi.gsfc.nasa.gov/ssc/data/access/} and Swift-XRT and UVOT observations from \url{https://heasarc.gsfc.nasa.gov/w3browse/swift/swiftmastr.html}.

\bibliographystyle{elsarticle-harv}

\newpage
\bibliography{example}



%% For example:

%% \bibitem[Aladro et al.(2015)]{Aladro15} Aladro, R., Martín, S., Riquelme, D., et al. 2015, \aas, 579, A101
\appendix

\section{Cross-correlation analysis}
\begin{figure*}
    \centering
    \includegraphics[width=0.99\textwidth]{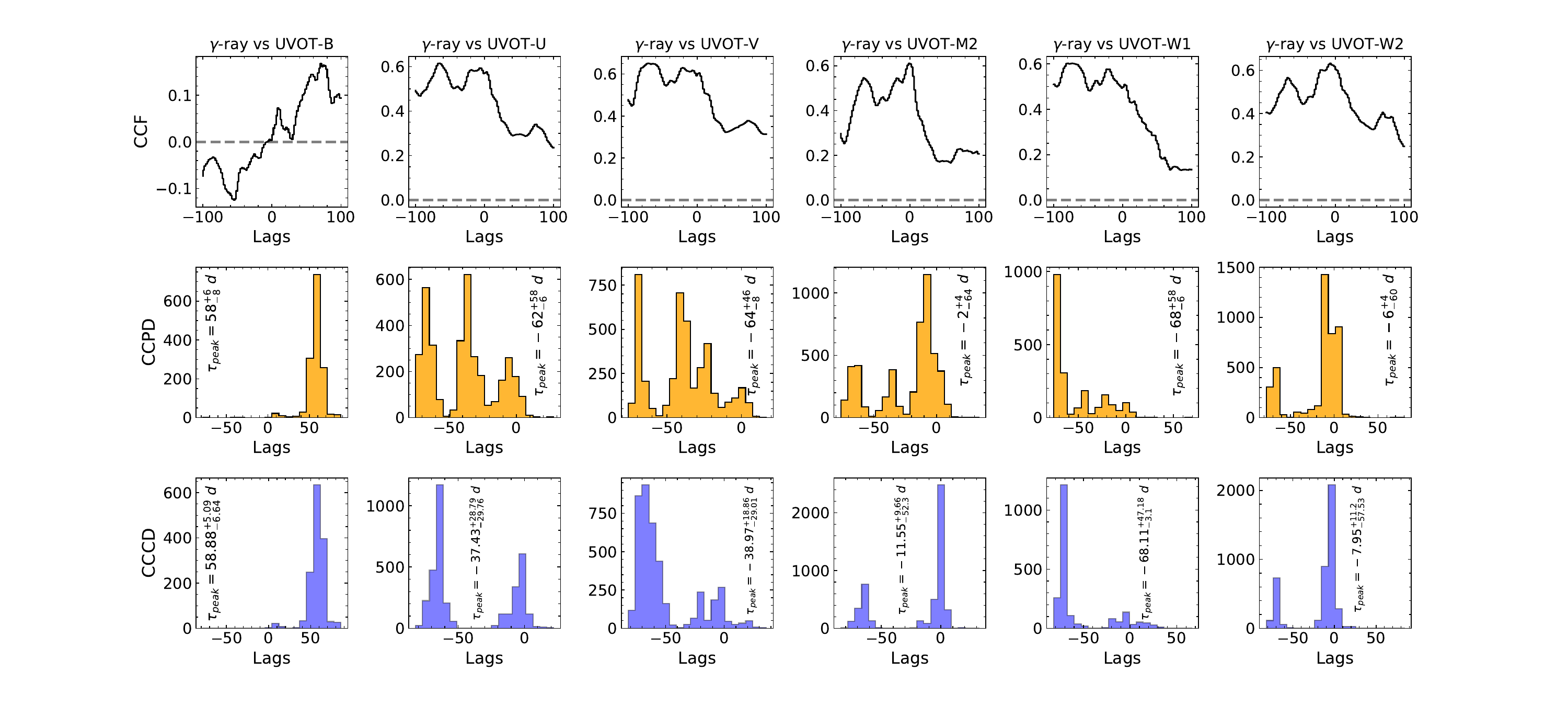}
    \caption{Cross-correlation analysis between $\gamma$-ray and Swift-UVOT flux variations of blazar PKS 1424-41. The upper row shows ICCF (solid black curve). The middle and bottom panels represent the cross-correlation centroid distribution (CCCD) and cross-correlation peak distribution (CCPD) in orange and blue, respectively.  }
    \label{Fig-CCF_PKS1424-41}    
\end{figure*}

\begin{figure*}
    \centering
    \includegraphics[width=0.99\textwidth]{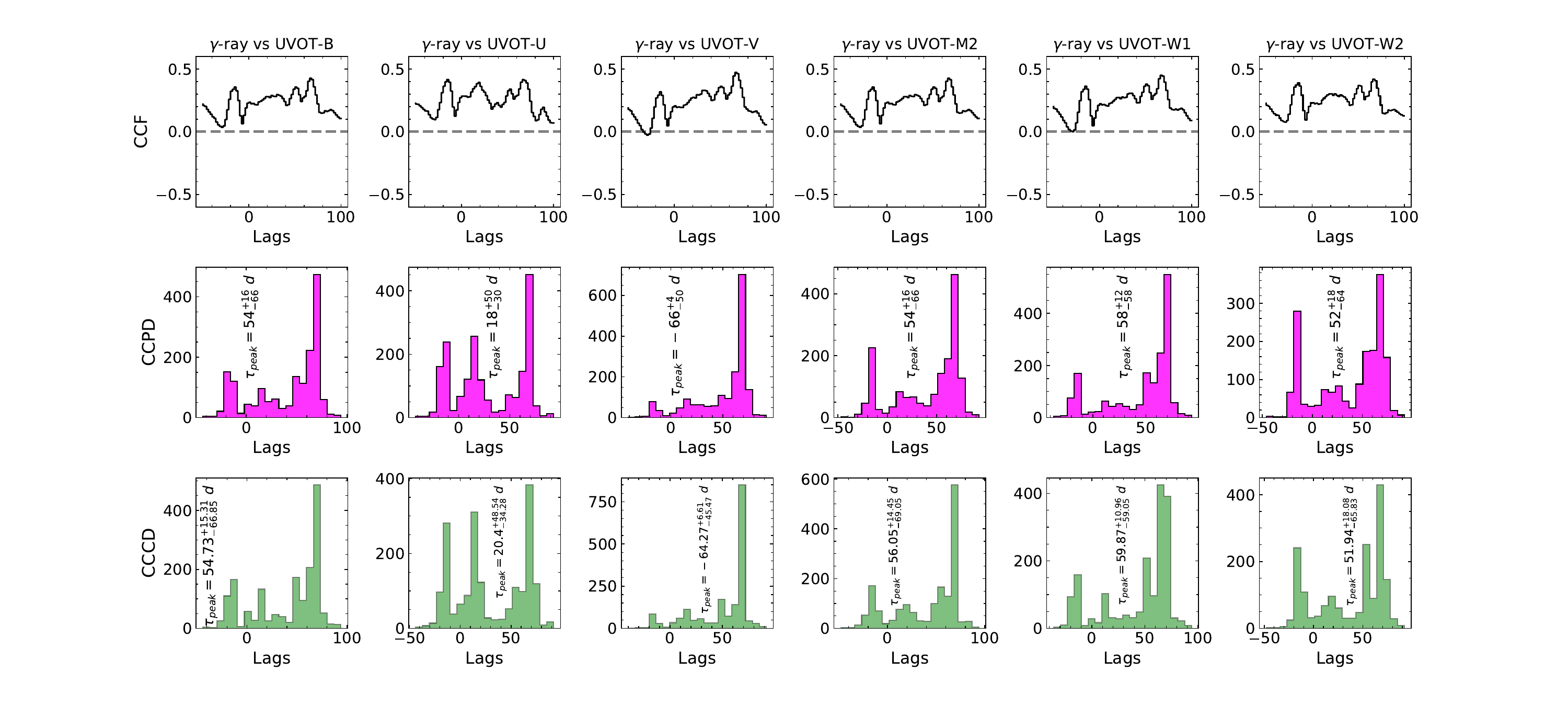}
    \caption{Cross-correlation analysis between $\gamma$-ray and Swift-UVOT flux variations of PKS 0736+01. The upper row shows ICCF (solid black curve) between $\gamma$-ray light curve and all UVOT filters, while middle and bottom panels exhibit the CCCD in magenta and CCPD in green. }
    \label{Fig-CCF_PKS0736}    
\end{figure*}

\begin{figure*}
    \centering
    \includegraphics[width=0.99\textwidth]{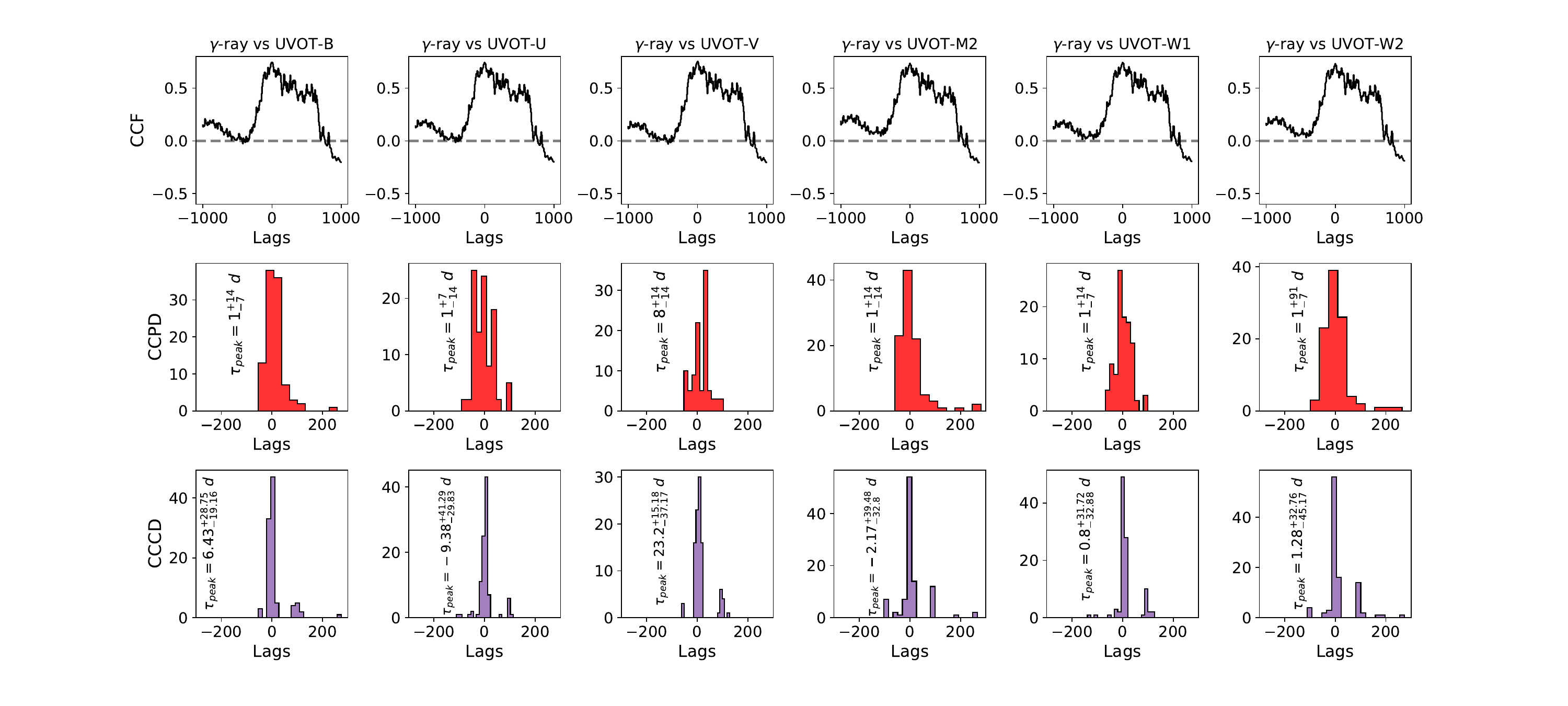}
    \caption{Cross-correlation analysis between $\gamma$-ray and Swift-UVOT flux variations of PKS 0208-512. The upper row shows ICCF in black. The middle and bottom panels represent the CCCD in red and CCPD in indigo.}
    \label{Fig-CCF_PKS0208}    
\end{figure*}

We performed cross-correlation analysis using ICCF method discussed in Sect.~\ref{sec:correlation} and observed results are shown in Figure~\ref{Fig-CCF_PKS1424-41}, ~\ref{Fig-CCF_PKS0736}, and ~\ref{Fig-CCF_PKS0208}. 

\section{Broadband SED}
A table of observed parameters and findings from broadband SEDs modeling of sources is included here, Table~\ref{tab:SEDresults}. 

\begin{table*}
%\tablewidth{50pt}
\setlength{\extrarowheight}{5.5pt}
\setlength{\tabcolsep}{4pt}
\centering
\caption{Table of input parameters for the SSC and EC Models used to reproduce the observed SEDs of two flaring and one quiescent states of blazars PKS 1424-41 and PKS 0736+01.}
\begin{tabular}{lcccccccccc}
\hline
\hline
 & \multicolumn{3}{c}{PKS 1424-41} & \multicolumn{3}{c}{PKS 0736+01} & \multicolumn{3}{c}{PKS 0208-512} & \multicolumn{1}{c}{PKS 0035-252} \\
\cline{2-4} \cline{5-7} \cline{8-10} \cline{11-11}
Parameter symbol & F1 & F2 & Q & F1 & F2 & Q & F1 & F2 & Q & \\
[+2pt]
\hline
$\gamma_{min}$ [$10^1$] & 2.69  & 0.67 &  30.66 & 42.06 & 34.40 & 16.22 & 10.8 & 23.02 & 27.8 & 17.9\\
$\gamma_{max}$ [$10^5$] & 38.17  & 1.01 & 0.109 & 749.91 & 2.01 & 2.73 & 1.04 & 0.99 & 3.29 & 0.33\\
$\gamma_{break}$ [$10^3$] & 11.52  & 5.17 & 10.2 & 8.44 & 6.25 & 2.67 & 3.86 & 1.80 & 51.2 & 4.95 \\
N [1/$cm^3$]  & $1.07\times 10^{3}$  & $1.93\times 10^{3}$ & 37.7 & 36.82 & 16.25 & 77.84 & 20.18 & 9.62 & 0.58 & 1.88\\
$p$ & 1.75  & 1.92 & 2.91 & 3.83 & 3.11 & 2.83 & 2.53 & 1.47 & 2.54 & 2.44\\
$p_1$  & 7.89  & 7.11 & 8.69 & 8.11 & 6.68 & 5.28 & 6.48  & 4.90 & 2.38 & 3.37\\
$T_{DT}^*$ [$10^3$K]   & 1.0  & 1.0 & 1.0 & 1.0 & 1.0 & 1.0 & 1.0 & 1.0 & 1.0 & 1.0\\
$R_{DT}^* [10^{19}$ cm]   & 2.0  & 2.0 & 2.0 & 0.63 & 0.63 & 0.63 & 2.0 & 2.0 & 0.63 & 0.63\\
$\tau_{DT}$ [$10^{-3}$]   & 534.5  & 136 & 114 & 6.95 & 140.9 & 1000 & 1.47 & 738.5 & 1.50$\times 10^{-7}$ & 988.2\\
$R_{BLR, \ in}^*$ [$10^{17}$cm]   & 3.0  & 3.0 & 3.0 & 0.94 & 0.94 & 0.94 & 3.0 & 3.0 & 0.94 & 0.94\\
$R_{BLR, \ out}^*$ [$10^{17}$cm]   & 3.3  & 3.3 & 3.3 & 1.04 & 1.04 & 1.04 & 3.3 & 3.3 & 1.04 & 1.04\\
$\tau_{BLR}$ [$10^{-2}$]   & 0.79  & 0.60 & 3.65 & 0.008 & 19.14 & 2.77 & 8.29 & 0.102 & 5.04$\times 10^{-6}$ & 0.31\\
$L_{Disk}^*$ [$10^{46} \ erg/cm$]   & 1.0  & 1.0 & 1.0 & 0.1 & 0.1 & 0.1 & 1.0 & 1.0 & 0.1 & 0.1\\
$T_{Disk}^*$ [$10^5$ K]   & 1.0  & 1.0 & 1.0 & 1.0 & 1.0 & 1.0 & 1.0 & 1.0 & 1.0 & 1.0\\
R [$10^{17}$ cm]   & 0.331  & 0.224 & 0.93 & 0.89 & 0.44 & 2.76 & 0.56 & 3.96 & 24.8 & 4.21\\
$R_H$ [$10^{17}$ cm]   & 0.911  & 2 & 3.2 & 1.44 & 1.68 & 1.32 & 1.08 & 1.006 & 10.0 & 1.0\\
B [gauss]    & 0.14  & 0.346 & 0.11 & 0.13 & 0.14 & 0.89 & 1.296 & 0.404 & 0.01 & 0.22\\
$\Gamma$  & 16.49  & 30.24 & 10.44 & 14.29 & 14.47 & 4.13 & 9.54 & 8.88 & 3.94 & 4.21\\
$\theta^*$  & 1.34  & 1.34 & 1.34 & 1.8 & 1.8 & 1.8 & 2.0 & 2.0 & 0.8 & 2.0\\
$z_{\mathrm{cosm}}^*$  & 1.52  & 1.52 & 1.52 & 0.19 & 0.19 & 0.19 & 1.003 & 1.003 & 1.003 & 0.49\\
$N_{\mathrm{H\_cold\_to\_rel\_e}}^*$  & 0.1  & 0.1 & 0.1 & 0.1 & 0.1 & 0.1 & 0.1 & 0.1 & 0.1 & 0.1\\
[+5pt]
\hline
\hline
\multicolumn{11}{c}{Energy Densities}\\
\hline
$U_{BLR}$  [$10^{-2} \ erg/cm^3$]    & 43.6  & 259  & 1.41 & 0.005 & 0.0002 & 24.4 & 310.35 & 3.30 & 1.83$\times 10^{-10}$ & 4.05\\
%$U_{Disk}$ [$10^{-2} \ erg/cm^3$]  & 464  & 161.9 & 0.65 & 19.16 & 0.58 & 8.87 & 236.53 & 254.6 & 6.17$\times 10^{-3}$ & 9.74\\
$U_{DT}$ [$10^{-3} \ erg/cm^3$]  & 9.66  & 19.1 & 1.65 & 0.18 & 3.94 & 2.25 & 0.017 & 7.75 & 3.07$\times 10^{-10}$ & 2.30\\
$U_e$ [$10^{-4} \ erg/cm^3$]  & 2573  & 882.8 & 19.0 & 18.32 & 83.61 & 210.2 & 44.6 & 6.05 & 3.67 & 7.56\\
$U_B$ [$10^{-4} \ erg/cm^3$]  & 8.02  & 47.6 & 9.09 & 7.27 & 8.02 & 566.6 & 668.97 & 64.9 & 0.045 & 19.9\\
\hline
\hline
\multicolumn{11}{c}{Luminosity}\\
\hline
$L_e$ [$10^{45} erg/s$]   & 3.58  & 4.40 & 1.47 & 0.34 & 0.32 & 0.025 & 0.122 & 0.704 & 3.20 & 0.21\\
$L_B$ [$10^{45} erg/s$]  & 0.011  & 0.23 & 0.70 & 0.12 & 0.031 & 0.069 & 1.84 & 7.55 & 0.039 & 0.57\\
$L_p$ [$10^{45} erg/s$]  & 2.25  & 14.4 & 0.43  & 0.087 & 0.095 & 0.014 & 0.083 & 0.168 & 0.76  & 0.081\\
$L_r$ [$10^{45} erg/s$]  & 5.38  & 3.40 & 1.53 & 0.0082 & 0.023 & 0.025 & 0.966 &  2.44 & 0.57 & 0.58\\
$L_{jet}$ [$10^{46} erg/s$]  & 1.12  & 2.25 & 0.41 & 0.0518 & 0.0477 & 0.013 & 0.301 & 1.08 & 0.46 & 0.146\\
Success fraction & 0.57 & 0.61 & 0.59 & 0.61 & 0.64 & 0.60 & 0.62 & 0.61 & 0.67 & 0.61\\
red. chisq & 1.38 & 0.62 & 0.64 & 2.85 & 2.34 & 2.22 & 2.47 & 1.39 & 1.93 & 2.05\\
[+5pt]
\hline
\end{tabular}

\label{tab:SEDresults}
\end{table*}

%%\end{thebibliography}

\end{document}